\documentclass[times, twocolumn, trackchanges]{aastex63}
\usepackage{inputenc}
\graphicspath{{./}{figures/}}

\newcommand{\mjup}{\ensuremath{M_{\mathrm{Jup}}}}
\newcommand{\nobject}{1054\;}
\newcommand{\Ks}{\mbox{$K_S$\;}}     
\newcommand{\fldg}{\hbox{\sc fld-g }}
\newcommand{\intg}{\hbox{\sc int-g }}
\newcommand{\vlg}{\hbox{\sc vl-g }}
\newcommand{\Gaia}{{\sl Gaia}}

\begin{document}

\title{The Hawaii Infrared Parallax Program. VI. The Fundamental Properties of 1000+ Ultracool Dwarfs and Planetary-mass Objects Using Optical to Mid-IR SEDs and Comparison to {\sc BT-Settl} and {\sc ATMO} 2020 Model Atmospheres}

\correspondingauthor{Aniket Sanghi}
\email{asanghi@caltech.edu}

\author[0000-0002-1838-4757]{Aniket Sanghi}
\altaffiliation{NSF Graduate Research Fellow.}
\affiliation{Department of Astronomy, California Institute of Technology, 1200 E. California Boulevard, Pasadena, CA 91125, USA}
\affiliation{Institute for Astronomy, University of Hawaii, 2680 Woodlawn Drive, Honolulu, HI 96822, USA}

\author[0000-0003-2232-7664]{Michael C. Liu}
\affiliation{Institute for Astronomy, University of Hawaii, 2680 Woodlawn Drive, Honolulu, HI 96822, USA}

\author[0000-0003-0562-1511]{William M. Best}
\affiliation{The University of Texas at Austin, Department of Astronomy, 2515 Speedway, C1400, Austin, TX 78712, USA}

\author[0000-0001-9823-1445]{Trent J. Dupuy}
\affiliation{Institute for Astronomy, University of Edinburgh, Royal Observatory, Blackford Hill, Edinburgh, EH9 3HJ, UK}

\author[0000-0001-5016-3359]{Robert J. Siverd}
\affiliation{Institute for Astronomy, University of Hawaii, 2680 Woodlawn Drive, Honolulu, HI 96822, USA}

\author[0000-0002-3726-4881]{Zhoujian Zhang}
\affiliation{University of California, Santa Cruz, 1156 High St. Santa Cruz, CA 95064, USA}

\author[0000-0002-6903-9080]{Spencer A. Hurt}
\affiliation{Department of Physics and Astronomy, University of Wyoming, Laramie, WY 82071, USA}

\author[0000-0002-7965-2815]{Eugene A. Magnier}
\affiliation{Institute for Astronomy, University of Hawaii, 2680 Woodlawn Drive, Honolulu, HI 96822, USA}

\author[0000-0003-0208-6146]{Kimberly M. Aller}
\affiliation{Institute for Astronomy, University of Hawaii, 2680 Woodlawn Drive, Honolulu, HI 96822, USA}

\author[0000-0003-2440-7350]{Niall R. Deacon}
\affiliation{Max Planck Institute for Astronomy, Konigstuhl 17, D-69117 Heidelberg, Germany}

\shorttitle{Fundamental Parameters of 1000+ Ultracool Dwarfs and Comparison to Atmospheric Models}

\shortauthors{Sanghi et al.}

\begin{abstract}
We derive the bolometric luminosities ($L_{\mathrm{bol}}$) of 865 field-age and 189 young ultracool dwarfs (spectral types M6--T9, including 40 new discoveries presented here) by directly integrating flux-calibrated optical to mid-IR spectral energy distributions (SEDs). The SEDs consist of low-resolution ($R\sim$ 150) near-IR (0.8--2.5 $\mu$m) spectra (including new spectra for 97 objects), optical photometry from the Pan-STARRS1 survey, and mid-IR photometry from the CatWISE2020 survey and Spitzer/IRAC. Our $L_{\mathrm{bol}}$ calculations benefit from recent advances in parallaxes from {\em Gaia}, Spitzer, and UKIRT, as well as new parallaxes for 19 objects from CFHT and Pan-STARRS1 presented here. Coupling our $L_{\mathrm{bol}}$ measurements with a new uniform age analysis for all objects, we estimate substellar masses, radii, surface gravities, and effective temperatures ($T_{\mathrm{eff}}$) using evolutionary models. We construct empirical relationships for $L_{\mathrm{bol}}$ and $T_{\mathrm{eff}}$ as functions of spectral type and absolute magnitude, determine bolometric corrections in optical and infrared bandpasses, and study the correlation between evolutionary model-derived surface gravities and near-IR gravity classes. Our sample enables a detailed characterization of {\sc BT-Settl} and {\sc ATMO} 2020 atmospheric model systematics as a function of spectral type and position in the near-IR color-magnitude diagram. We find the greatest discrepancies between atmospheric and evolutionary model-derived $T_{\mathrm{eff}}$ (up to 800 K) and radii (up to 2.0 $R_{\mathrm{Jup}}$) at the M/L transition boundary. With \nobject objects, this work constitutes the largest sample to date of ultracool dwarfs with determinations of their fundamental parameters.
\end{abstract}


\section{Introduction}
\label{sec:intro}
Brown dwarfs are substellar objects more massive than gas-giant planets \cite[$\gtrsim$ 13\;\mjup;][]{2011ApJ...727...57S} but less massive than stars \cite[$\lesssim$ 70~\mjup;][]{2017ApJS..231...15D}. All brown dwarfs fuse deuterium \citep[e.g.][]{1996ApJ...460..993S} and those between 60--75 \mjup\;will also burn lithium \citep{1998ASPC..134..394B}. Brown dwarfs cool through their lifetimes, progressing through a wide temperature range spanning the M, L, T and Y spectral types \citep[e.g.][]{1997ApJ...491..856B, 2005ARAA..43..195K, 2011ApJ...743...50C}. This continuous cooling means that for a given luminosity and effective temperature, and hence spectral type, the age and mass of a brown dwarf are not uniquely determined. Such an observational degeneracy makes it challenging to use colors, magnitudes, and spectral types to determine the physical properties of brown dwarfs, such as mass, age, and radius. However, the luminosities and temperatures for a population of brown dwarfs do characterize their evolutionary history and provide unique information about the star-formation history of our Galaxy \citep{2021AJ....161...42B, 2021ApJS..253....7K}. Additionally, young brown dwarfs, identified by their unusually red near- to mid-infrared colors as well as lower surface gravities, serve as crucial templates for understanding the atmospheric properties (e.g., clouds) of giant exoplanets given their similar effective temperatures and ages \citep[e.g.][]{2013ApJ...777L..20L, 2016ApJS..225...10F, 2016ApJ...833...96L}. A large, well-defined sample of brown dwarfs with accurate and precise luminosity measurements would enhance our ability to test models of substellar formation, evolution, and atmospheres.

A proven approach to determining the physical properties of ultracool dwarfs ($\gtrsim$ M6) is to assemble and integrate broadband spectral energy distributions (SEDs) consisting of flux-calibrated optical, near-infrared, and mid-infrared spectra and/or photometry to obtain bolometric fluxes \citep[e.g.][]{2015ApJ...810..158F}. Given parallax measurements, we can then derive the corresponding bolometric luminosities ($L_{\mathrm{bol}}$). The bolometric luminosities can be used together with age estimates and evolutionary models to compute the masses and radii of ultracool dwarfs. Finally, combining the bolometric luminosities with evolutionary model-derived radii and applying the Stefan-Boltzmann Law enables the calculation of effective temperatures. 

This approach should yield less model-dependent results than deriving the bolometric luminosities, effective temperatures, and surface gravities simultaneously from atmospheric model spectra fits, and then obtaining the masses and radii. Bolometric luminosites derived by atmospheric model fits can introduce systematic uncertainties since such measurements are dependent upon the fidelity of the fitting procedures adopted, the sampling of the model grid, and the differing input physics between various atmospheric models \citep[e.g.][]{2009ApJ...702..154S, 2021ApJ...921...95Z}. \citet{2015ApJ...810..158F} adopted the SED integration strategy, combining optical, near-infrared, and mid-infrared spectra and photometry to construct spectral energy distributions for 145 field-age and 53 young ultracool dwarfs and thereby derive their fundamental physical properties. The uncertainties in the distances were the largest contributor to the uncertainty in their bolometric luminosities. Thus, high-precision parallax measurements are a necessity for accurate and precise determinations of bolometric luminosities. Similarly, since the uncertainty in age is the dominant contributor to the uncertainty in ultracool dwarf masses, radii, surface gravities, and effective temperatures, well-constrained ages are a necessity for accurate and precise determinations of these physical properties. The limited availability of such measurements has constrained the sample size of objects in previous work.

\tabletypesize{\scriptsize}
\begin{deluxetable*}{l}
\tablecolumns{1}
\tablecaption{Comprehensive Reference List for Table of Ultracool Fundamental Properties Data\label{tab:ref}}
\tablehead{\colhead{References}}
\startdata
\multicolumn{1}{m{\textwidth}}{\citet{1998ApJ...509..836L}; 
\citet{2020AJ....159..257B}; 
\citet{2016ApJ...821..120A}; 
\citet{2005AA...441..653C}; 
\citet{2005AA...440.1061L}; 
\citet{2013MNRAS.431.3222L}; 
\citet{2001AA...374L..12S}; 
\citet{2002AJ....124.1170D}; 
\citet{1997MNRAS.284..507T}; 
\citet{2006AJ....131.1163S}; 
\citet{2013ApJ...777L..20L}; 
\citet{1999AA...351L...5E}; 
\citet{2006ApJ...651..502P}; 
\citet{2010MNRAS.405.1140G}; 
\citet{2013ApJ...779..172G}; 
\citet{2013ApJS..205....6M}; 
\citet{2006ApJ...639.1120K}; 
\citet{2002MNRAS.336L..49S}; 
\citet{2004AJ....128.1733G}; 
\citet{2013MNRAS.433..457B}; 
\citet{2004AA...425..519S}; 
\citet{2018MNRAS.481.3548S}; 
\citet{2003IAUS..211..197W}; 
\citet{2006PASP..118..671R}; 
\citet{2004AA...427L...1F}; 
\citet{2005AN....326..974S}; 
\citet{2017ApJ...837...95B}; 
\citet{2014ApJ...792L..17G}; 
\citet{1991MNRAS.252P..61I}; 
\citet{1979ApJ...233..226L}; 
\citet{2012AJ....144...94G}; 
\citet{2005AJ....130.2347E}; 
\citet{2006MNRAS.373..705R}; 
\citet{2009ApJ...695.1517L}; 
\citet{2016yCat.2343....0E}; 
\citet{2012AJ....144..148G}; 
\citet{2015ApJ...814..118B}; 
\citet{2007AA...468..163D}; 
\citet{1997AA...327L..25D}; 
\citet{2013AJ....146..161M}; 
\citet{2011AJ....141..203A}; 
\citet{2011AJ....142...77D}; 
\citet{1993ApJ...414..279T}; 
\citet{2010AA...517A..53M}; 
\citet{2004AJ....128.2460H}; 
\citet{2011AJ....142..171G}; 
\citet{2017AJ....154...69S}; 
\citet{2011ApJ...736L..34G}; 
\citet{2007ApJ...658..617B}; 
\citet{2016PhDT.......189A}; 
\citet{2015AJ....149..158S}; 
\citet{1988Natur.336..656B}; 
\citet{2012ApJ...760..152L}; 
\citet{2012MNRAS.422.1922P}; 
\citet{2006MNRAS.373L..31M}; 
\citet{2008AJ....135..785W}; 
\citet{1995AJ....109..797K}; 
\citet{1991ApJS...77..417K}; 
\citet{1999ApJ...522L..61S}; 
\citet{2012MNRAS.427.3280F}; 
\citet{2000AA...353..958S}; 
\citet{2002AJ....123..458S}; 
\citet{2003ApJ...589L..51T}; 
\citet{2007ApJ...655..522L}; 
\citet{2013ApJ...777...84B}; 
\citet{2006ApJ...645..676L}; 
\citet{2007ApJ...658..557B}; 
\citet{2016ApJ...822L...1S}; 
\citet{2004ApJ...614L..73B}; 
\citet{2015ApJ...799..154M}; 
\citet{2003AJ....125..850B}; 
\citet{1976PhDT........14B}; 
\citet{1996MNRAS.281..644T}; 
\citet{2017MNRAS.465.4723K}; 
\citet{2007ApJ...654..570L}; 
\citet{2016ApJS..225...10F}; 
\citet{2014ApJ...785L..14G}; 
\citet{2000AJ....119..928F}; 
\citet{1999AAS..135...41D}; 
\citet{2010AJ....139.1808S}; 
\citet{2022arXiv220800070G}; 
\citet{2019ApJS..240...19K}; 
\citet{2006AJ....131.1007B}; 
\citet{2014ApJ...794..143B}; 
\citet{2017MNRAS.467.1126D}; 
\citet{2008ApJ...689L..53B}; 
\citet{2017AJ....154..112K}; 
\citet{2000ApJ...531L..57B}; 
\citet{2002AA...389L..20L}; 
\citet{2007AA...474..653V}; 
\citet{2012ApJ...757..100D}; 
\citet{2012ApJ...746....3C}; 
\citet{2011ApJS..197...19K}; 
\citet{2002ApJ...564..452L}; 
\citet{2009AJ....137.3345C}; 
\citet{2001AA...380..590P}; 
\citet{2016Natur.533..221G}; 
\citet{2014MNRAS.445.3908L}; 
\citet{2007MNRAS.374..445K}; 
\citet{2007AJ....134.1162L}; 
\citet{2003AJ....126.2421C}; 
\citet{2018ApJS..234....1B}; 
\citet{2012ApJ...752...56F}; 
\citet{2016arXiv161205560C}; 
\citet{2016ApJ...833...96L}; 
\citet{2006AJ....131.2722C}; 
\citet{2015ApJS..219...33G}; 
\citet{2013ApJ...774...55B}; 
\citet{2012ApJ...755...94D}; 
\citet{2012ApJ...753..142B}; 
\citet{2009AA...497..619Z}; 
\citet{2003AJ....125..343L}; 
\citet{2016ApJ...826...73P}; 
\citet{2019ApJ...883..205B}; 
\citet{2021ApJ...916...53Z}; 
\citet{2008MNRAS.383..831P}; 
\citet{2013AA...557A..43B}; 
\citet{1991ApJ...367L..59R}; 
\citet{1992ApJS...82..351L}; 
\citet{2008ApJ...686..528L}; 
\citet{2008AA...481..661B}; 
\citet{1986ApJ...305..784G}; 
\citet{1988MNRAS.234..177H}; 
\citet{1990ApJ...354L..29H}; 
\citet{2007ApJ...659..655B}; 
\citet{2010MNRAS.408L..56L}; 
\citet{2017ApJS..228...18G}; 
\citet{2017AJ....154..262M}; 
\citet{2013Msngr.154...35M}; 
\citet{2015ApJ...802...37B}; 
\citet{2008AJ....136.1290R}; 
\citet{2004AJ....127.3553K}; 
\citet{2021ApJ...911....7Z}; 
\citet{2023AJ....166..103S}; 
\citet{2014ApJ...787..126L}; 
\citet{2010ApJ...710.1627L}; 
\citet{2008ApJ...681..579B}; 
\citet{2012AJ....144..180M}; 
\citet{2013ApJ...776..126C}; 
\citet{2002ApJ...564..466G}; 
\citet{2004AJ....127.2856B}; 
\citet{2002MNRAS.329..543D}; 
\citet{2021AJ....161...42B}; 
\citet{2005AA...435..363D}; 
\citet{1993AJ....105.1169T}; 
\citet{2003ApJ...594..510B}; 
\citet{2012ApJ...748...74L}; 
\citet{2001AJ....121.3235K}; 
\citet{2007MNRAS.378..901F}; 
\citet{2011ApJ...743...50C}; 
\citet{1961AJ.....66..528V}; 
\citet{2009AA...493L..27S}; 
\citet{2008ApJ...676.1281M}; 
\citet{2017AJ....154..147D}; 
\citet{2000AJ....120..447K}; 
\citet{2004PASP..116..362C}; 
\citet{2000ApJ...538..363B}; 
\citet{2003tmc..book.....C}; 
\citet{2012yCat.2316....0U}; 
\citet{2008ApJ...674..451B}; 
\citet{2002ApJ...564..421B}; 
\citet{2015ApJS..216....7B}; 
\citet{2013ApJ...772...79A}; 
\citet{1981MNRAS.196p..15R}; 
\citet{1999ApJ...519..802K}; 
\citet{2009AJ....137.4109L}; 
\citet{2015MNRAS.450.2486C}; 
\citet{2003AJ....126..975T}; 
\citet{2009AA...494..949S}; 
\citet{2006AJ....132..891R}; 
\citet{2000ApJ...536L..35L}; 
\citet{1994ApJ...436..262M}; 
\citet{1985MNRAS.213..257G}; 
\citet{2015ApJ...799..203G}; 
\citet{2006AJ....132.2360H}; 
\citet{2007AJ....133..439C}; 
\citet{2013AA...560A..52M}; 
\citet{2016ApJ...820...32B}; 
\citet{2017AJ....153..196S}; 
\citet{2006ApJ...639.1095B}; 
\citet{2001AJ....122.1989W}; 
\citet{2002ApJ...575..484G}; 
\citet{2002AJ....123.3409H}; 
\citet{1997PASP..109..849G}; 
\citet{1993AJ....105.1045T}; 
\citet{2003AA...403..929K}; 
\citet{2004AA...416L..17K}; 
\citet{2010AA...524A..38M}; 
\citet{2010ApJ...715L.165R}; 
\citet{2015AJ....150..182K}; 
\citet{2012ApJS..201...19D}; 
\citet{1971lpms.book.....G}; 
\citet{2007MNRAS.379.1599L}; 
\citet{2006MNRAS.373..781L}; 
\citet{2019MNRAS.487.1149G}; 
\citet{2000AJ....120.1100B}; 
\citet{2005PASP..117..676R}; 
\citet{2015MNRAS.446.3878M}; 
\citet{2004AJ....127.2948V}; 
\citet{2014ApJ...783..122K}; 
\citet{1998Sci...282.1309R}; 
\citet{2014AJ....147..113C}; 
\citet{2013MNRAS.431.2745G}; 
\citet{2006ApJ...639.1114R}; 
\citet{2010ApJ...715.1408Z}; 
\citet{1991AJ....102.1180S}; 
\citet{2023AA...674A...1G}; 
\citet{2018ApJ...862..138G}; 
\citet{1997AJ....113.1421K}; 
\citet{2016AJ....152...24W}; 
\citet{2007AJ....133.2320S}; 
\citet{2006ApJ...637.1067B}; 
\citet{2014AJ....147...34S}; 
\citet{1995AJ....110.1838R}; 
\citet{2008ApJ...689.1295K}; 
\citet{2013Sci...341.1492D}; 
\citet{2014ApJ...792..119D}; 
\citet{2015MNRAS.449.3651M}; 
\citet{2005ApJS..159..141V}; 
\citet{1997ApJ...476..311K}; 
\citet{2018AJ....156...57D}; 
\citet{2005AJ....129.1483L}; 
\citet{2009ApJ...703..399L}; 
\citet{2008ApJ...689..471R}; 
\citet{2009AJ....137....1F}; 
\citet{2016ApJ...817..112S}; 
\citet{2012AA...548A..53L}; 
\citet{2010ApJS..190..100K}; 
\citet{2010AA...515A..92S}; 
\citet{2006ApJ...651.1166M}; 
\citet{2016AA...589A..49S}; 
\citet{2018MNRAS.473.5113D}; 
\citet{2015AJ....150..179G}; 
\citet{2006AJ....132.2074M}; 
\citet{2014MNRAS.443.2327S}; 
\citet{2010ApJ...715..561A}; 
\citet{2010ApJ...710.1142B}; 
\citet{2010MNRAS.406.1885B}; 
\citet{2009AJ....137..304S}; 
\citet{2000AJ....119..369R}; 
\citet{2011ApJ...732...56G}; 
\citet{2016ApJ...830..144R}; 
\citet{2003AJ....126.1526B}; 
\citet{2018AA...619L...8R}; 
\citet{2005AA...430L..49S}; 
\citet{2011AA...532L...5S}; 
\citet{2014ApJ...781....4L}; 
\citet{2013MNRAS.433.2054S}; 
\citet{2015AA...574A.118P}; 
\citet{2005AJ....130.2326T}; 
\citet{2014yCat.2328....0C}; 
\citet{2016ApJS..224...36K}; 
\citet{2010ApJ...725.1405B}; 
\citet{2006MNRAS.366L..40P}; 
\citet{2003ApJ...586L.149S}; 
\citet{2007ApJ...655.1079L}; 
\citet{2012yCat.2314....0L}; 
\citet{2017ASInC..14....7B}; 
\citet{2007ApJ...669L..97L}; 
\citet{2013AJ....145....2F}; 
\citet{2021yCat.2367....0M}; 
\citet{2000AJ....120.1085G}; 
\citet{2006PASP..118..659L}; 
\citet{2011MNRAS.414L..90B}; 
\citet{2021ApJ...921...95Z}; 
\citet{2018ApJS..236...28T}; 
\citet{2001AJ....121.2185G}; 
\citet{2012ApJ...753..156K}; 
\citet{2003AJ....125.3302G}; 
\citet{2003AJ....126.2487B}; 
\citet{1998AA...338.1066T}; 
\citet{1999AJ....118.2466M}; 
\citet{2021ApJS..257...45H}; 
\citet{1999ApJ...522L..65B}; 
\citet{2013PASP..125..809T}; 
\citet{2021ApJS..253....7K}; 
\citet{1983ApJ...274..245P}; 
\citet{2006ApJ...651L..57A}; 
\citet{2004AJ....128..463R}; 
\citet{1994ApJS...94..749K}; 
\citet{1991AJ....101..662B}; 
\citet{2004AA...421..643R}; 
\citet{2021ApJS..253....8M}; 
\citet{2011ApJ...740L..32L}; 
\citet{2000ApJ...531L..61T}; 
\citet{2010AJ....139..176F};
\citet{2011AA...530A.138C};
Best et al. (AAS Journals, submitted); 
Hurt et al. (AAS Journals, submitted);
\emph{The UltracoolSheet} (version 2.0.0, in preparation);
This Work.}
\enddata
\tablecomments{References for ultracool dwarf discoveries, spectral types (optical/IR), gravity classifications (optical/IR), parallaxes, SpeX prism spectra, ages, Pan-STARRS 1 $grizy$ photometry, 2MASS $JHK_S$ photometry, MKO $JHK$ photometry, WISE \emph{W1}, \emph{W2}, \emph{W3}, and \emph{W4} photometry, and Spitzer Channel 1 and Channel 2 photometry presented in the Table of Ultracool Fundamental Properties associated with this work (see \S\ref{sec:intro}) are included above.}
\end{deluxetable*}

The first generation of ultracool dwarf parallax programs focused on obtaining measurements for small samples of objects across the spectral type range M, L, and T as they were newly discovered \citep{2002AJ....124.1170D, 2003AJ....126..975T, 2004AJ....127.2948V}. With the rapid acceleration in the number of ultracool dwarf discoveries catalyzed by the advent of large digital sky surveys, the latest generation of ultracool parallax programs have focused their efforts on distinct classes of objects and large volume-limited samples. The Hawaii Infrared Parallax Program has observed L/T transition dwarfs \citep{2012ApJS..201...19D}, young ultracool dwarfs \citep{2016ApJ...833...96L}, and ultracool binaries \citep{2012ApJS..201...19D, 2017ApJS..231...15D} with WIRCam on the Canada–France–Hawaii Telescope (CFHT), and most recently obtained infrared parallax measurements for 348 L and T dwarfs with WFCAM on the United Kingdom Infrared Telescope \citep{2020AJ....159..257B}. Several programs have targeted late-T and Y dwarfs, the coldest objects in the ultracool dwarf temperature sequence \citep{2013Sci...341.1492D, 2014ApJ...783...68B, 2014ApJ...796...39T, 2018ApJ...867..109M, 2019ApJS..240...19K, 2021ApJS..253....7K}. Therefore, a large-scale analysis aimed at deriving ultracool dwarf properties by taking advantage of the tremendous advances in precision and availability of trigonometric parallax measurements, particularly for later-type objects, is timely. 

Age dating stars and brown dwarfs is a notoriously difficult task, but steady progress has been made over the past decade. The simultaneous study of a large ensemble of stars has proven to be a successful avenue to obtaining new age measurements. The discovery of coeval, kinematically comoving associations of young stars and brown dwarfs in the solar neighborhood called young moving groups (YMGs) has been instrumental to the effort \citep[e.g.][]{2004ARA&A..42..685Z, 2014ApJ...783..121G, 2016ApJ...821..120A, 2016ApJS..225...10F, 2016ApJ...833...96L, 2018ApJ...856...23G, 2022arXiv220800070G}. Group members generally span a wide range of masses and, thus, can be used to age date each group using lithium depletion boundaries and isochrone fitting \citep[e.g.][]{2015MNRAS.454..593B, 2015ApJ...808...23H}. For the field population of ultracool dwarfs, \citet[][]{2017ApJS..231...15D} empirically determined an age distribution consistent with the Besançon model of the solar neighborhood \citep{2003A&A...409..523R} based on cooling ages of 20 ultracool dwarfs in binaries derived from high-precision dynamical mass measurements. In addition, various techniques have been developed geared towards identifying signs of youth in individual objects. Activity-based indicators such as chromospheric H$\alpha$ emission \citep{2008AJ....135..785W, 2021AJ....161..277K}, coronal X-ray emission \citep{2012MNRAS.422.2024J, 2017MNRAS.471.1012B}, and UV emission \citep{2011ApJ...727....6S, 2013ApJ...774..101R} are found to be well-correlated with stellar ages. Tracing lithium absorption features in ultracool dwarf spectra and perfoming a comparison with predictions of lithium-burning timescales from evolutionary models can provide valuable insights into the age of young dwarfs \citep{2008ApJ...689.1295K, 2016MNRAS.455.3345B, 2022ApJ...935...15Z}. For mid-M to L dwarfs, there are a number of gravity-sensitive features in the near-infrared such as FeH (0.99, 1.20, 1.55 $\mu$m), VO (1.06 $\mu$m), Na I (1.138 $\mu$m), K I (1.169, 1.177, 1.244, 1.253 $\mu$m), and the shape of the \emph{H}-band continuum (1.46--1.68 $\mu$m) that can be used to uncover low gravity (young) objects \citep{2013ApJ...772...79A, 2018AJ....155...34C}. Similarly, at optical wavelengths, weak Na I (8183, 8195 \AA), Cs I (8521, 8943 \AA), Rb I doublet lines (7800, 7948 \AA), weak pressure-broadened K I wings (7665, 7699 \AA), weak FeH (8692 \AA), and strong VO (7300--7550 \AA\;and 7850--8000 \AA) in early type dwarfs act as spectroscopic signatures of low-gravity (youth) \citep{1998ApJ...507L..41M, 2009AJ....137.3345C, 2010ApJS..190..100K}.

The large-scale effort to leverage advances in parallax and age measurements to empirically determine the fundamental properties of ultracool dwarfs has been greatly facilitated by the development of \emph{The UltracoolSheet} (version 2.0.0, in preparation). \emph{The UltracoolSheet} is a catalog of 3000+ ultracool dwarfs and directly imaged exoplanets, including optical to mid-infrared photometry from Pan-STARRS1 \citep{2016arXiv161205560C}, 2MASS \citep{2006AJ....131.1163S}, UKIDSS/UHS \citep{2007MNRAS.379.1599L, 2018MNRAS.473.5113D}, and CatWISE \citep{2020ApJS..247...69E, 2021ApJS..253....8M}, absolute astrometry, proper motions, parallaxes, multiplicity (including orbit determinations), and spectroscopic classifications. A major benefit offered by \emph{The UltracoolSheet} for deriving bolometric luminosities is that it not only compiles high-precision parallax values from the latest surveys (e.g., \emph{Gaia DR3}) and literature measurements \citep[e.g.][]{2020AJ....159..257B, 2021ApJS..253....7K}, but also compares the various sources to readily make available the most precise measurement available to-date for each listed object. Additionally, it provides photometric distance estimates for objects without parallax measurements. In short, \emph{The UltracoolSheet} is a valuable resource well matched to the goals of this work.

\begin{figure*}
    \label{fig:unpub-spl}
        \centering
        \includegraphics[scale=0.8]{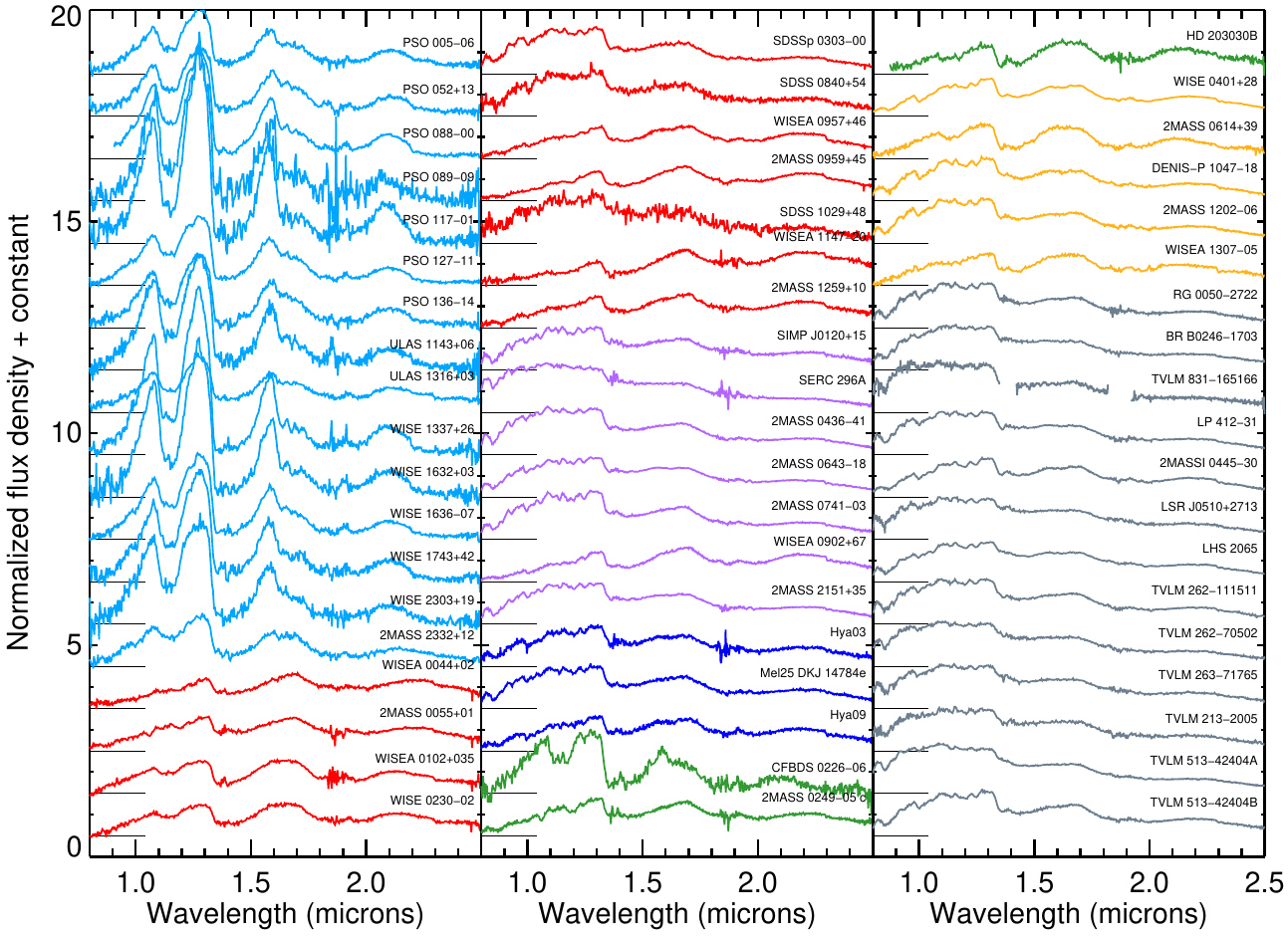}
        \caption{Near-infrared (0.8--2.5 $\mu$m) SpeX prism spectra for the 57 known objects presented in Table~\ref{table:spex-known}. From top to bottom and left to right, plotting order follows the table order for objects, namely by category (which is colorized in the plot) and then by right ascension. Light blue corresponds to T dwarfs, red corresponds to red L dwarfs, purple corresponds to candidate low-gravity objects, dark blue corresponds to Hyades members, green corresponds to wide substellar/planetary-mass companions, yellow corresponds to solar neighborhood candidates, and gray corresponds to M/L transition dwarfs. The horizontal line to the left of each spectrum shows the zero flux level.}
\end{figure*}

In this paper, we derive the fundamental physical properties of \nobject ultracool dwarfs by directly integrating flux-calibrated SEDs consisting of low-resolution near-infrared SpeX prism spectra and optical and mid-infrared photometry. This work increases the sample of ultracool dwarfs with semi-empirical physical parameter measurements by nearly five-fold, and takes advantage of significant growth in parallax data for ultracool dwarfs. Additionally, we construct and analyze empirical relationships for bolometric luminosity and effective temperature as functions of spectral type and absolute magnitude, determine bolometric corrections across the optical to mid-infrared filters, benchmark the near-infrared gravity classifications against evolutionary model-derived surface gravity measurements, and study systematic offsets between fundamental parameters derived with evolutionary (BHAC15 and SM08) and atmospheric ({\sc BT-Settl} and {\sc ATMO} 2020) models. We prepare a Table of Ultracool Fundamental Properties\footnote{Zenodo Data Publication: \url{https://doi.org/10.5281/zenodo.8315643}} that is a subset of \emph{The UltracoolSheet} (version 2.0.0, in preparation) tied uniquely to this work. It is a comprehensive compilation of the astrometric, photometric, spectroscopic, and age properties of all objects in our sample. Importantly, it contains the fundamental parameters calculated in this work for all objects in our sample. References for all data used in this work, compiled in the Table of Ultracool Fundamental Properties, are included in Table \ref{tab:ref}.

\begin{figure*}
    \label{fig:unpub-obj}
        \centering
        \includegraphics[scale=0.8]{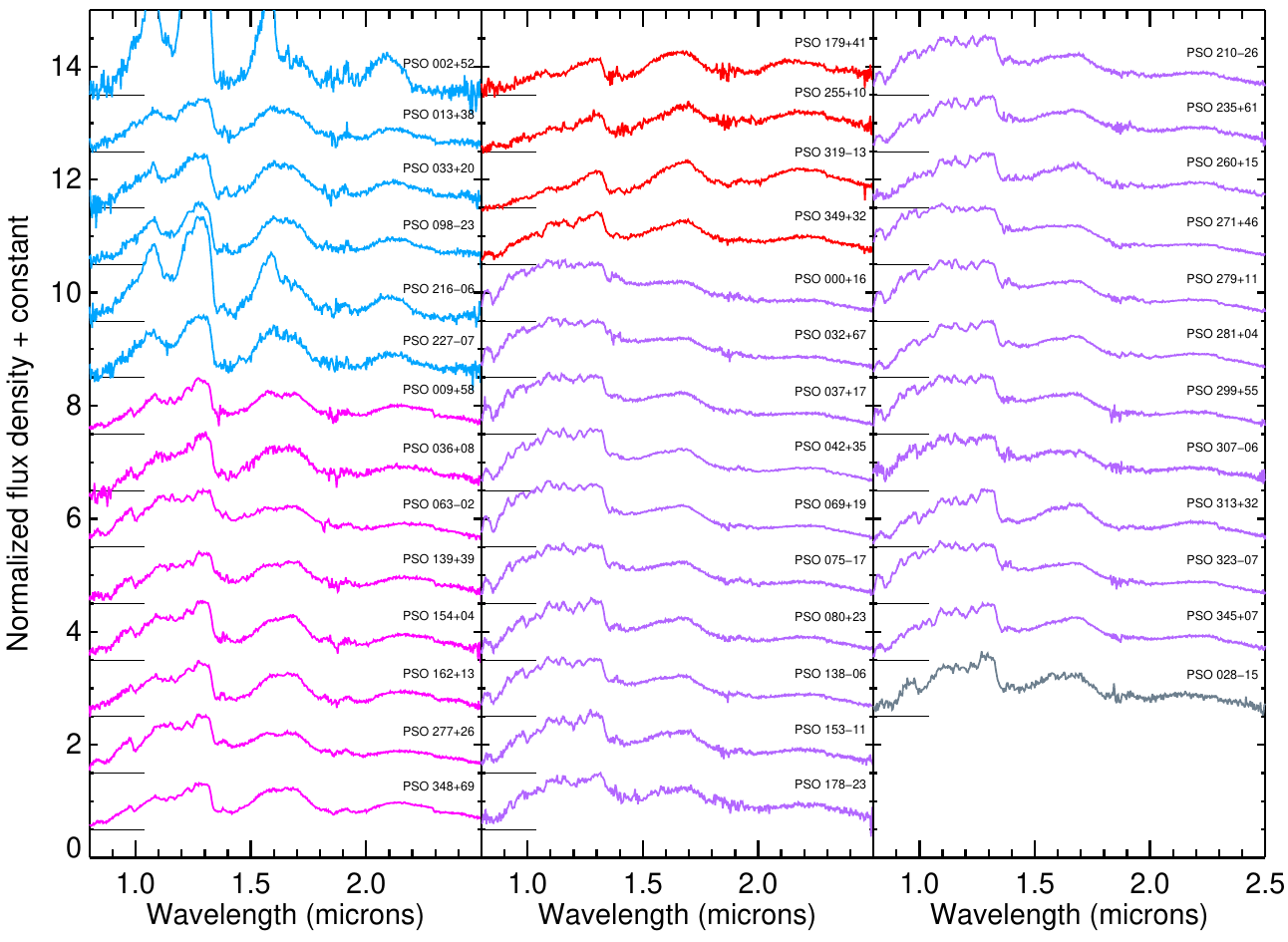}
        \caption{Near-infrared (0.8--2.5 $\mu$m) SpeX prism spectra for the 40 new discoveries presented in Table~\ref{table:spex-discoveries}. From top to bottom and left to right, plotting order follows the table order for objects, namely by category (which is colorized in the plot) and then by right ascension. Light blue corersponds to T dwarfs, magenta corresponds to L dwarfs, red corresponds to red L dwarfs, purple corresponds to low-gravity objects, and gray corresponds to M/L transitions dwarfs. The horizontal line to the left of each spectrum shows the zero flux level.}
\end{figure*}

This paper is organized as follows. Section \ref{sec:obs} details the construction of our ultracool dwarf sample and presents newly discovered objects, new object spectra, and new parallax measurements. Section \ref{sec:SED} describes the spectra and photometry used to assemble and flux-calibrate multiwavelength ultracool dwarf spectral energy distributions (SEDs). Section \ref{sec:model_desc} details our procedures to estimate flux in regions where no spectra are available using atmospheric model fitting. Sections \ref{sec:lbol} and \ref{sec:ages} discuss the calculation of bolometric luminosities and assignment of ages for the \nobject ultracool dwarfs in our sample. Section \ref{sec:evo-sample} summarizes the techniques used to derive the masses, radii, surface gravities, and effective temperatures for objects in our ultracool dwarf sample. Section \ref{sec:Discussion} presents calculations of bolometric corrections, analysis of the relationship between gravity classes and derived surface gravities, and a detailed look at atmospheric model-derived parameters in comparison with evolutionary model-derived parameters. Finally, Section \ref{sec:concl} presents our conclusions. Appendix \ref{sec:abs-spt} presents revised absolute magnitude-spectral type relations for young ultracool dwarfs based on all available objects in \emph{The UltracoolSheet}.

\begin{figure*}
        \centering
        \includegraphics[scale=1.01]{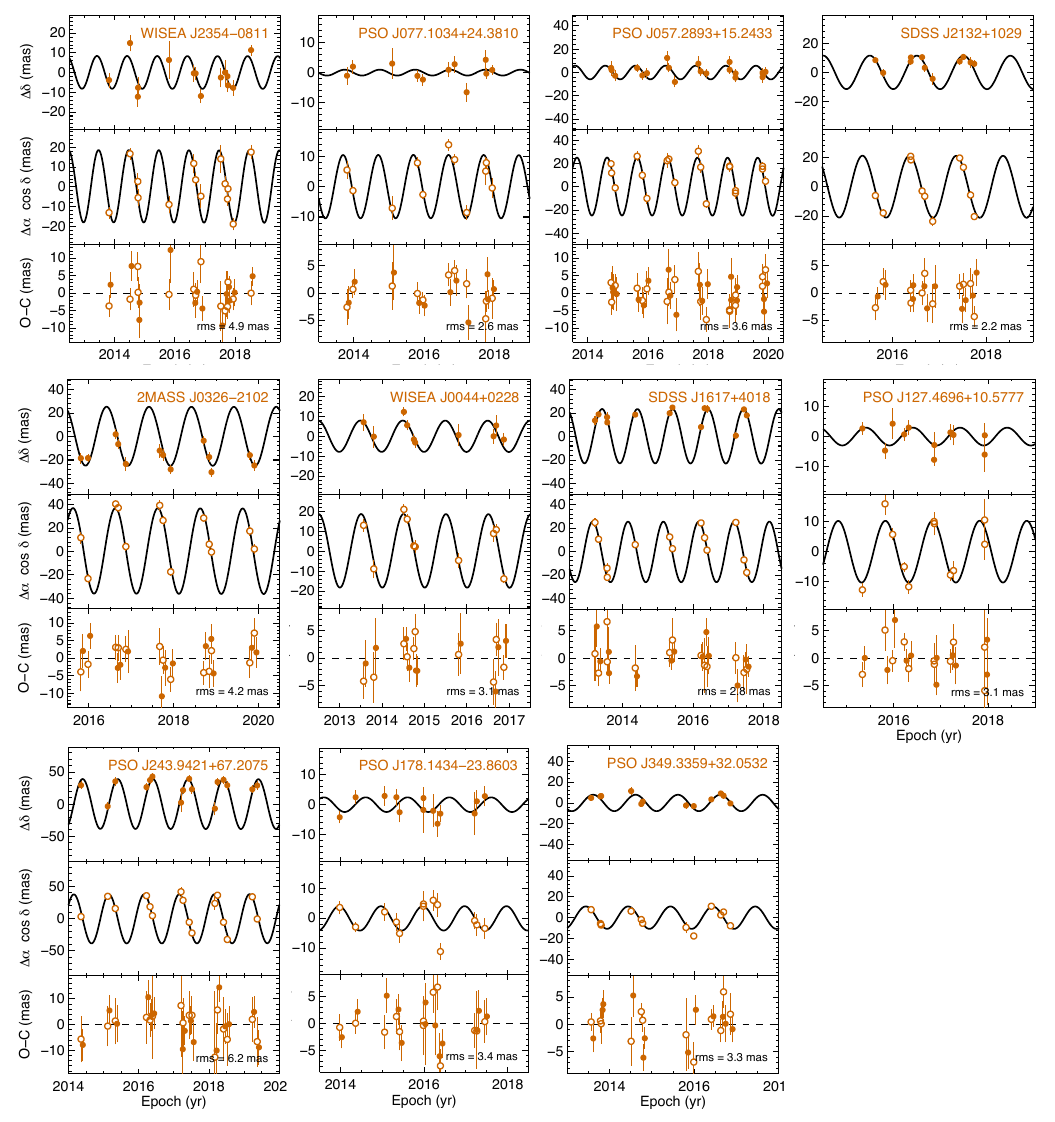}
        \caption{Parallax solutions for 11 objects with new CFHT/WIRCam astrometry. For each object, three panels are shown: Decl. measurements with the proper motion and zero point subtracted (top); R.A. measurements with the proper motion and zero point subtracted (middle); and the residuals of both R.A. (open symbols) and Decl. (solid symbols) plotted with the best-fit parallax and proper motion solution subtracted (bottom). The standard deviation of the residuals of both R.A. and Decl. combined is quoted as the rms in the bottom panel. The top and middle subpanels show the motion expected from the best-fit parallax (solid line). Objects have varying observational time baselines, which is reflected in the varying x-axis ranges.}
        \label{fig:parallax}
\end{figure*}

\section{Sample \& Additional Observations}
\label{sec:obs}

\subsection{Construction of Ultracool Dwarf Sample}
Our sample is constructed using \emph{The UltracoolSheet} (version 2.0.0, in preparation), selecting all objects with parallaxes or photometric distances and SpeX prism spectra readily available to us. We exclude unresolved binaries and multiples, subdwarfs, and members of active star-forming regions (e.g. Taurus, Chamaeleon, Lupus, IC348, etc.). \emph{The UltracoolSheet} columns relevant to the above criteria are \texttt{multiple\_unresolved\_in\_this\_table}, \texttt{flag}, and \texttt{age\_category}, respectively. In total, \nobject objects in \emph{The UltracoolSheet} satisfy the above criteria (including the newly discovered ultracool dwarfs discussed below).

\subsection{New Objects and Spectra}
We obtained low-resolution near-IR (0.8--2.5~\micron) spectroscopy for 97 objects in our sample using the NASA Infrared Telescope Facility (IRTF) on the summit of Mauna Kea, Hawaii. We used the facility spectrograph SpeX \citep{2003PASP..115..362R} in prism mode with either the 0.5\arcsec\ or 0.8\arcsec\ slit, which provided an average spectral resolution ($R\equiv\Delta\lambda/\lambda$) of $\approx$150 or $\approx$100, respectively. We dithered in an ABBA pattern on the science target to enable sky subtraction during the reduction process. For each epoch, we obtained calibration frames (arcs and flats) and observed an A0~V star contemporaneously for telluric calibration. All spectra were reduced using the Spextool software package \citep{2003PASP..115..389V,2004PASP..116..362C}.

These spectra were obtained for a variety of programs conducted by us over the past decade, either for (1) spectroscopic characterization of known objects (Figure ~\ref{fig:unpub-spl} and Table~\ref{table:spex-known}) or (2) discovery of new ultracool dwarfs (Figure~\ref{fig:unpub-obj} and Table~\ref{table:spex-discoveries}). For (1), we obtained spectra of known objects that had potential indications of low-gravity spectra from previous work \citep[e.g.][]{2013ApJ...772...79A} or objects that could serve as benchmark objects by virtue of being members of known stellar associations (the Hyades and Pleiades) or being wide substellar/planetary-mass companions to stars \citep[e.g.][]{2006ApJ...651.1166M,2018AJ....156...57D}. Finally, as a bad-weather program, we obtained spectra of objects with pre-\Gaia\ parallaxes and with spectral types at the M/L transition. For (2), the discoveries result from various multi-catalog photometric searches relying on the Pan-STARRS~1 optical survey as a staring point, including searches for T~dwarfs \citep[e.g.][]{2011AJ....142...77D, 2011ApJ...740L..32L}, very red (low-gravity) L~dwarfs \citep[e.g.][]{2013ApJ...777L..20L}, young moving group members \citep[e.g.][]{2016ApJ...821..120A, 2016PhDT.......189A}, L/T transition dwarfs and objects in the solar neighborhood \citep[e.g.][]{2015ApJ...814..118B}. For spectral typing and gravity classification, we use the near-IR classification system described by \citet{2013ApJ...772...79A} and \citet{2016ApJ...821..120A} for late-M and L~dwarfs. For T dwarfs, we follow the near-IR system described by \citet{2006ApJ...637.1067B}. Among the new discoveries, we find that PSO J080.3940+23.0999 is a new candidate AB Doradus Moving Group member with an 82.5\% BANYAN~$\Sigma$ probability \citep{2018ApJ...856...23G}.

\subsection{New Parallaxes}
For eleven objects --- WISEA J004403.39+022810.6, 2MASS J03264225-2102057, PSO J057.2893+15.2433, PSO J077.1034+24.3810, PSO J127.4696+10.5777, SDSS J161731.65+401859.7, SDSS J213240.36+102949.4, WISEA J235422.31-081129.7, PSO J178.1434-23.8603, PSO J349.3359+32.0532, and PSO J243.9421+67.2075 --- we present new parallax measurements (Figure~\ref{fig:parallax}). They are available in the Table of Ultracool Fundamental Properties associated with this paper (see \S\ref{sec:intro}). We monitored these objects with the facility infrared camera WIRCam \citep{2004SPIE.5492..978P} on the Canada-France-Hawaii Telescope (CFHT). In the $J$ band, we measured their $(x, y)$ positions and those of many reference stars. Using our custom pipeline \citep{2012ApJS..201...19D, 2015ApJ...805...56D}, we reduced these individual measurements into high-precision multi-epoch relative astrometry, with the absolute calibration provided by low-proper-motion 2MASS stars \citep{2003tmc..book.....C}. We derived the relative parallax and proper motion for the targets using our standard MCMC approach and then, in order to be consistent with our many previously published CFHT parallaxes, converted to an absolute reference frame using the Besançon galaxy model to simulate the distances of the reference stars \citep{2003A&A...409..523R}.

We also present new Pan-STARRS\,1 (PS1) parallaxes for eight objects in our sample: 2MASS J03001631+2130205, 2MASSW J0309088-194938, SDSSp J032817.38+003257.2, 2MASSW J0355419+225702, 2MASSW J0918382+213406, 2MASS J13120707+3937445, SDSS J151603.03+025928.9, and SDSS J154849.02+172235.4. They are available in the Table of Ultracool Fundamental Properties associated with this paper (see \S\ref{sec:intro}). These objects were observed by the PS1 telescope from 2009--2015 as part of the Pan-STARRS $3\pi$ Steradian Survey \citep{2016arXiv161205560C} in the $grizy$ filters. The $3\pi$ Survey was well-suited to parallaxes as every survey region was observed at opposition as well as evening and morning twilight. Astrometric calibration and automated calculation of parallaxes and proper motions are described by \cite{2020ApJS..251....6M}. The resulting astrometry is tied to the Gaia DR1 inertial system, with a correction for the proper motion bias introduced by Galactic rotation and solar motion.

\startlongtable
\begin{deluxetable*}{lccccc}
\centering
\tablecaption{IRTF/SpeX Observing Log: Objects from the Literature \label{table:spex-known}}
\tabletypesize{\scriptsize}
\tablewidth{0pt}
\tablehead{
 \colhead{Object} &
 \colhead{UT Date} &
 \colhead{Slit (\arcsec)} &
 \colhead{$<$Airmass$>$} &
 \colhead{$T_{\mathrm{int}}$ (s)} &
 \colhead{$<$S/N$> (Y,J,H,H_{\mathrm{blue}},K,K_{\mathrm{blue}})$} 
}

\startdata                                                                                                  
\cutinhead{T dwarfs}                                                                                                
PSO J005.6302-06.8669        &   2012-01-20   &  0.8\arcsec &    1.49 &    120.0   &   16,  37,  20,  20,  16,  24 \\  
PSO J052.2746+13.3754        &   2012-09-24   &  0.8\arcsec &    1.24 &    420.0   &   18,  45,  23,  25,   9,  18 \\   
PSO J088.5709-00.1430        &   2012-09-24   &  0.8\arcsec &    1.08 &   1440.0   &   34, 103,  33,  46,  11,  35 \\   
PSO J089.1751-09.4513        &   2011-04-20   &  0.5\arcsec &    2.10 &    240.0   &    7,  18,   3,   5,   1,   3 \\   
PSO J117.0600-01.6779        &   2012-04-20   &  0.8\arcsec &    1.16 &    240.0   &   15,  28,   6,  12,   2,  10 \\  
PSO J127.5648-11.1861        &   2012-04-20   &  0.8\arcsec &    1.24 &    240.0   &   30,  66,  30,  34,  11,  27 \\   
PSO J136.5380-14.3267        &   2012-04-20   &  0.8\arcsec &    1.35 &    240.0   &   17,  34,  15,  17,   6,  13 \\   
ULAS J114340.47+061358.9     &   2015-05-26   &  0.8\arcsec &    1.27 &   1554.1   &   16,  27,   7,  10,   3,   7 \\   
ULAS J131610.13+031205.5     &   2012-04-20   &  0.8\arcsec &    1.05 &    240.0   &   23,  50,  31,  32,  30,  36 \\   
WISE J133750.46+263648.6     &   2012-01-19   &  0.8\arcsec &    1.03 &    120.0   &   28,  68,  15,  23,   4,  14 \\   
WISE J163236.47+032927.3     &   2012-08-10   &  0.8\arcsec &    1.29 &     60.0   &   19,  42,  10,  19,   3,   9 \\   
WISE J163645.56-074325.1     &   2012-04-20   &  0.8\arcsec &    1.13 &    240.0   &   28,  62,  26,  31,   9,  21 \\   
WISE J174303.71+421150.0     &   2012-07-05   &  0.8\arcsec &    1.62 &    240.0   &   15,  37,  10,  14,   4,  10 \\  
WISE J230356.79+191432.9     &   2012-01-22   &  0.8\arcsec &    1.64 &    120.0   &   12,  26,   8,  11,   3,   8 \\   
2MASS J23322678+1234530      &   2012-02-01   &  0.8\arcsec &    1.95 &    120.0   &   14,  31,  18,  16,  18,  20 \\   
\cutinhead{Red L Dwarfs}                                                                                           
WISEA J004403.39+022810.6    &   2012-10-14   &  0.8\arcsec &    1.06 &   1320.0   &   11,  31,  34,  29,  46,  45 \\   
2MASS J00550564+0134365      &   2012-10-25   &  0.8\arcsec &    1.06 &    840.0   &   25,  62,  48,  43,  61,  61 \\   
WISEA J010202.11+035541.4    &   2012-10-25   &  0.8\arcsec &    1.04 &    840.0   &   27,  52,  46,  41,  44,  50 \\  
WISE J023038.90-022554.0     &   2012-10-28   &  0.8\arcsec &    1.22 &   1080.0   &   15,  34,  31,  28,  34,  37 \\   
SDSSp J030321.24-000938.2    &   2012-11-08   &  0.8\arcsec &    1.06 &    840.0   &   51,  85,  48,  43,  42,  43 \\   
SDSS J084016.42+543002.1     &   2012-04-30   &  0.8\arcsec &    1.34 &    120.0   &   12,  20,  12,  11,  14,  14 \\   
WISEA J095729.41+462413.5    &   2012-11-08   &  0.8\arcsec &    1.31 &    840.0   &   25,  61,  57,  51,  69,  69 \\  
2MASS J09593276+4523309      &   2012-04-20   &  0.8\arcsec &    1.18 &    180.0   &   25,  68,  67,  58,  86,  87 \\   
SDSS J102947.68+483412.2     &   2012-04-30   &  0.8\arcsec &    1.63 &    120.0   &    8,  14,   7,   7,   8,   7 \\   
WISEA J114724.10-204021.3    &   2014-01-17   &  0.8\arcsec &    1.33 &   1080.0   &    9,  28,  37,  32,  46,  48 \\   
2MASS J12594167+1001380      &   2012-01-20   &  0.8\arcsec &    1.02 &    120.0   &   13,  42,  43,  37,  51,  52 \\   
\cutinhead{Candidate Low-Gravity Objects}                                                                          
SIMP J01205253+1518277       &   2015-09-25   &  0.5\arcsec &    1.01 &   1075.9   &   49,  64,  49,  47,  43,  43 \\   
SERC 296A                    &   2015-11-24   &  0.5\arcsec &    1.94 &   1912.7   &  118, 133, 108, 104,  71,  80 \\  
2MASS J04362788-4114465      &   2016-02-03   &  0.5\arcsec &    2.06 &    418.4   &  102, 117,  83,  76,  65,  71 \\   
2MASS J06431685-1843375      &   2015-01-20   &  0.5\arcsec &    1.28 &    418.4   &  110, 158, 130, 153, 110, 117 \\   
2MASS J07410404-0359495      &   2016-03-25   &  0.5\arcsec &    1.39 &    836.8   &   83, 113,  79,  71,  72,  73 \\   
WISEA J090258.99+670833.1    &   2016-02-05   &  0.5\arcsec &    1.47 &   2271.3   &   24,  52,  62,  50,  75,  75 \\   
2MASS J21512797+3547206      &   2015-08-07   &  0.5\arcsec &    1.07 &    266.9   &  101, 135, 118, 111,  99, 100 \\   
\cutinhead{Hyades Members}                                                                                         
Hya03                        &   2014-11-17   &  0.8\arcsec &    1.03 &     87.6   &   19,  35,  31,  33,  28,  27 \\  
Cl\* Melotte 25 DKJ 14784e   &   2016-03-21   &  0.5\arcsec &    1.25 &   1554.1   &   41,  52,  33,  29,  44,  41 \\   
Hya09                        &   2016-02-03   &  0.5\arcsec &    1.41 &   1315.0   &   12,  25,  20,  17,  28,  28 \\   
\cutinhead{Wide Substellar/Planetary-Mass Companions}                                                              
CFBDS J022644.65-062522.0    &   2011-12-02   &  0.8\arcsec &    1.13 &    120.0   &    9,  20,   8,   9,   3,   5 \\   
2MASS J0249-0557 c           &   2015-09-25   &  0.5\arcsec &    1.11 &   1554.1   &   20,  35,  33,  28,  38,  37 \\   
HD 203030B                   &   2012-07-07   &  0.5\arcsec &    1.08 &    240.0   &    8,  19,  22,  20,  20,  23 \\   
\cutinhead{Solar Neighborhood Candidates}                                                                          
WISE J040137.21+284951.7     &   2013-12-11   &  0.8\arcsec &    1.04 &    420.0   &  117, 261, 242, 234, 189, 198 \\   
2MASS J06143818+3950357      &   2011-10-14   &  0.8\arcsec &    1.11 &    120.0   &   22,  44,  39,  38,  32,  38 \\   
DENIS-P J104731.1-181558     &   2012-04-30   &  0.8\arcsec &    1.38 &     15.0   &   49,  98,  65,  58,  62,  63 \\   
2MASS J12022564-0629026      &   2015-05-19   &  0.8\arcsec &    1.12 &    133.4   &   67,  89,  68,  62,  58,  59 \\   
WISEA J130729.56-055815.4    &   2011-03-31   &  0.5\arcsec &    1.18 &    120.0   &   23,  53,  48,  45,  45,  48 \\   
\cutinhead{M/L Transition Dwarfs}                                                                                  
RG 0050-2722                 &   2014-11-17   &  0.8\arcsec &    1.47 &    107.0   &   48,  76,  77,  75,  52,  53 \\ 
BR B0246-1703                &   2014-01-19   &  0.8\arcsec &    1.50 &     35.0   &   37,  79,  68,  68,  40,  43 \\  
TVLM 831-165166              &   2014-10-15   &  0.8\arcsec &    1.07 &     87.6   &   20,  26,  23,  24,  15,  16 \\  
LP 412-31                    &   2013-11-06   &  0.8\arcsec &    1.00 &     50.0   &   45, 104,  95,  94,  58,  61 \\  
2MASSI J0445538-304820       &   2016-02-03   &  0.8\arcsec &    1.57 &    418.4   &   66, 103,  83,  75,  74,  76 \\  
LSR J0510+2713               &   2014-01-19   &  0.8\arcsec &    1.42 &     11.0   &   24,  55,  52,  52,  28,  30 \\   
LHS 2065                     &   2014-01-19   &  0.8\arcsec &    1.11 &     21.0   &   50, 127, 124, 124,  77,  81 \\   
TVLM 262-111511              &   2014-01-18   &  0.8\arcsec &    1.21 &    105.0   &   68, 125,  99,  97,  62,  65 \\  
TVLM 262-70502               &   2014-01-18   &  0.8\arcsec &    1.22 &     70.0   &   46,  85,  71,  71,  42,  44 \\  
TVLM 263-71765               &   2014-01-18   &  0.8\arcsec &    1.19 &     35.0   &   32,  70,  62,  62,  35,  37 \\   
TVLM 213-2005                &   2014-01-19   &  0.8\arcsec &    1.20 &     35.0   &   17,  38,  32,  32,  18,  19 \\  
TVLM 513-42404A              &   2014-07-29   &  0.8\arcsec &    1.02 &    836.8   &  128, 131, 108,  93, 101, 102 \\   
TVLM 513-42404B              &   2014-07-29   &  0.8\arcsec &    1.03 &   1315.0   &   76, 100,  67,  55,  66,  69 \\    
\enddata

\tablecomments{The last column gives the median S/N of the spectrum for
  the standard IR bandpasses and also for two wavelength regions free of
  methane absorption ($H_{\mathrm{blue}}$ = 1.49--1.63~\micron\ and $K_{\mathrm{blue}}$ =
  2.03--2.20~\micron), which are relevant for T-dwarf spectra.}

\end{deluxetable*} 
\startlongtable
\begin{deluxetable*}{lccccc}
\tablecaption{IRTF/SpeX Observing Log: New Discoveries \label{table:spex-discoveries}}
\centering
\tabletypesize{\scriptsize}
\tablewidth{0pt}
\tablehead{
 \colhead{Object} &
 \colhead{UT Date} &
 \colhead{Slit (\arcsec)} &
 \colhead{$<$Airmass$>$} &
 \colhead{$T_{\mathrm{int}}$ (s)} &
 \colhead{$<$S/N$> (Y,J,H,H_{\mathrm{blue}},K,K_{\mathrm{blue}})$} 
}

\startdata                                                    
\cutinhead{T dwarfs}                                                                                                                                                                     
PSO J002.0878+52.0687 &  2012-01-22   &   0.8\arcsec &    1.52 &     120.0 &   12,  32,   9,  13,   3,  11 \\   
PSO J013.7740+38.2804 &  2011-10-14   &   0.8\arcsec &    1.13 &     120.0 &   16,  36,  23,  23,  15,  21 \\   
PSO J033.2936+20.4493 &  2011-10-14   &   0.8\arcsec &    1.11 &     120.0 &   11,  24,  20,  21,  16,  21 \\  
PSO J098.2822-23.2845 &  2011-10-14   &   0.8\arcsec &    1.37 &     120.0 &   17,  36,  23,  23,  18,  21 \\   
PSO J216.4707-06.2849 &  2012-04-20   &   0.8\arcsec &    1.12 &     240.0 &   18,  45,  18,  22,   6,  14 \\   
PSO J227.1576-07.9608 &  2012-04-20   &   0.8\arcsec &    1.18 &     240.0 &   12,  23,  15,  15,   7,  12 \\   
\cutinhead{L Dwarfs}                                                                                                                                                                   
PSO J009.8334+58.5781 &  2015-11-28   &   0.5\arcsec &    1.33 &     836.8 &   27,  56,  42,  39,  48,  53 \\   
PSO J036.9069+08.5025 &  2012-01-20   &   0.8\arcsec &    1.16 &     120.0 &    9,  31,  23,  21,  19,  21 \\  
PSO J063.1519-02.8904 &  2012-01-22   &   0.8\arcsec &    1.08 &     120.0 &   40,  79,  53,  47,  46,  49 \\  
PSO J139.5493+39.0380 &  2012-01-19   &   0.8\arcsec &    1.08 &     120.0 &   20,  50,  40,  38,  34,  37 \\   
PSO J154.9622+04.8279 &  2012-01-19   &   0.8\arcsec &    1.09 &     120.0 &   22,  52,  35,  33,  29,  32 \\   
PSO J162.0614+13.9756 &  2012-01-19   &   0.8\arcsec &    1.01 &     120.0 &   27,  59,  46,  45,  36,  39 \\   
PSO J277.3873+26.0116 &  2015-07-15   &   0.8\arcsec &    1.01 &     207.6 &   41,  69,  57,  54,  47,  49 \\   
PSO J348.5125+69.6598 &  2015-12-08   &   0.5\arcsec &    1.55 &    1793.2 &   50,  87,  76,  69,  86,  92 \\   
\cutinhead{Red L Dwarfs}                                                                                                                                                               
PSO J179.2489+41.5121 &  2014-01-17   &   0.8\arcsec &    1.08 &    1080.0 &   10,  25,  28,  25,  29,  33 \\  
PSO J255.0913+10.3263 &  2013-09-23   &   0.8\arcsec &    1.59 &     240.0 &    7,  20,  19,  16,  30,  31 \\   
PSO J319.7576-13.0982 &  2013-09-22   &   0.8\arcsec &    1.38 &     240.0 &   11,  38,  40,  33,  52,  54 \\   
PSO J349.3359+32.0532 &  2012-10-14   &   0.8\arcsec &    1.03 &    2280.0 &   25,  53,  38,  34,  41,  40 \\   
\cutinhead{Low-Gravity Objects}                                                                                                                                                        
PSO J000.2794+16.6237 &  2015-11-03   &   0.5\arcsec &    1.26 &     418.4 &   58,  60,  42,  41,  34,  34 \\  
PSO J032.2211+67.4179 &  2015-12-24   &   0.5\arcsec &    1.48 &    2032.3 &   65,  75,  56,  51,  47,  49 \\  
PSO J037.4616+17.2761 &  2015-09-30   &   0.5\arcsec &    1.05 &     537.9 &   35,  52,  53,  49,  43,  43 \\   
PSO J042.8405+35.3726 &  2013-09-23   &   0.5\arcsec &    1.08 &     120.0 &   73, 132,  93,  86,  77,  77 \\   
PSO J069.4987+19.6346 &  2015-01-21   &   0.5\arcsec &    1.29 &     629.2 &   83, 111,  93,  89,  77,  74 \\   
PSO J075.4182-17.8327 &  2015-11-02   &   0.5\arcsec &    1.32 &     418.4 &   47,  58,  41,  38,  38,  38 \\   
PSO J080.3940+23.0999 &  2015-10-14   &   0.5\arcsec &    1.03 &     207.6 &   42,  61,  51,  47,  42,  42 \\  
PSO J138.1887-06.9862 &  2015-12-24   &   0.5\arcsec &    1.17 &     418.4 &   55,  73,  58,  54,  52,  52 \\   
PSO J153.7969-11.3631 &  2015-05-20   &   0.5\arcsec &    1.28 &    2510.4 &   28,  39,  26,  23,  23,  24 \\   
PSO J178.1434-23.8603 &  2013-04-04   &   0.8\arcsec &    1.39 &    2400.0 &   22,  39,  22,  20,  22,  22 \\   
PSO J210.6211-26.9300 &  2016-05-29   &   0.5\arcsec &    1.47 &    1554.1 &   44,  54,  40,  36,  37,  37 \\  
PSO J235.3149+61.5071 &  2015-07-06   &   0.5\arcsec &    1.34 &    1434.5 &   43,  59,  47,  42,  40,  41 \\   
PSO J260.1789+15.3091 &  2013-04-19   &   0.5\arcsec &    1.01 &     180.0 &   24,  50,  32,  28,  36,  34 \\  
PSO J271.8353+46.9326 &  2014-08-01   &   0.5\arcsec &    1.17 &     836.8 &   72, 127,  89,  83,  68,  78 \\   
PSO J279.6187+11.6036 &  2015-09-25   &   0.5\arcsec &    1.02 &     418.4 &   87, 111,  87,  78,  75,  79 \\   
PSO J281.4296+04.2296 &  2015-07-15   &   0.8\arcsec &    1.07 &     207.6 &   77, 107,  88,  81,  75,  75 \\   
PSO J299.5032+55.0722 &  2015-09-25   &   0.5\arcsec &    1.23 &     537.9 &   55,  73,  57,  52,  47,  50 \\   
PSO J307.7981-06.6854 &  2015-07-04   &   0.5\arcsec &    1.14 &    1075.9 &   19,  29,  26,  24,  22,  22 \\   
PSO J313.1067+32.9117 &  2015-12-08   &   0.5\arcsec &    1.44 &    2032.3 &   36,  57,  41,  34,  43,  44 \\   
PSO J323.4300-07.5353 &  2015-08-07   &   0.5\arcsec &    1.17 &     266.9 &   91, 108,  88,  83,  71,  72 \\   
PSO J345.2949+07.7665 &  2015-07-04   &   0.5\arcsec &    1.06 &    1315.0 &   43,  70,  68,  62,  53,  55 \\   
\cutinhead{M Dwarfs}                                                                                          
PSO J028.0715-15.5717 &  2012-10-15   &   0.8\arcsec &    1.28 &    2520.0 &   13,  30,  17,  16,  14,  14 \\   
\enddata

\tablecomments{The last column gives the median S/N of the spectrum for the standard IR bandpasses
  and also for two wavelength regions free of methane absorption
  ($H_{\mathrm{blue}}$ = 1.49--1.63~\micron\ and $K_{\mathrm{blue}}$ = 2.03--2.20~\micron).}

\end{deluxetable*}

\begin{figure*}
    \centering
    \includegraphics[scale=0.045]{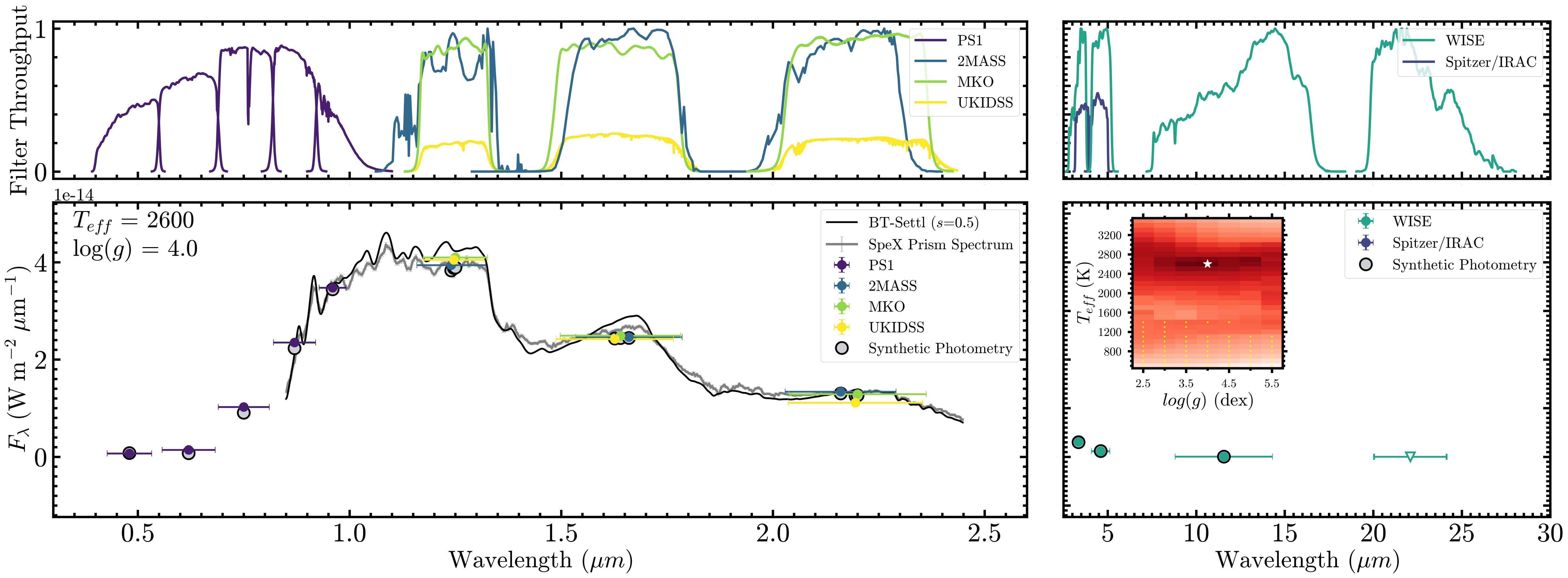}
    \caption{Example flux-calibrated SpeX prism spectrum generated in this work. \emph{Top:} Filter transmission profiles for the PS1, 2MASS, MKO, UKIDSS, WISE, and Spitzer/IRAC filters. \emph{Bottom:} Flux-calibrated SpeX spectrum (slit size = 0.5\arcsec) of 2MASSI J0335020+234235 in gray with the corresponding photometry in the same colors as the filter transmission profiles, if available. Photometry marked with triangles represent upper limits. The gray points represents model synthesized photometry. The black curve correponds to the best-fit atmospheric model. The inset figure shows the $\chi^2$ surface for the atmospheric model fits to the SpeX spectrum in $T_{\mathrm{eff}}$-log $g$ space. The white star marks the location of the model-fit with the smallest $\chi^2$. Yellow plus signs mark the $T_{\mathrm{eff}}$-log $g$ values at which {\sc ATMO} 2020 model spectra were preferred over the {\sc BT-Settl} model spectra based on their lower $\chi^2$.}
    \label{fig:sed}
\end{figure*}

\section{Ultracool Dwarf Spectral Energy Distributions}
\label{sec:SED}
The central goal of the following analysis is to construct and directly integrate the spectral energy distributions of ultracool dwarfs to obtain their bolometric fluxes. Combining this calculated quantity with a distance measurement yields the bolometric luminosity. This then enables the derivation of other fundamental parameters such as masses, radii, effective temperatures, and surface gravities using evolutionary models. In this section, we discuss the spectra used in this work and our spectral flux-calibration procedures. 

\subsection{SpeX Prism Spectra}
\label{sec:spex}
Since most ultracool dwarfs have faint apparent magnitudes, SpeX is often used in its low-resolution ($R\sim$ 75--200) prism mode to acquire 0.8--2.5 $\mu$m spectra using one of the three different slit sizes --- $0\farcs3 \times 15\farcs0$, $0\farcs5 \times 15\farcs0$, and $0\farcs8 \times 15\farcs0$. This wavelength range is sensitive to several well-known absorption features in ultracool dwarf spectra, e.g., $\mathrm{CH_4}$, $\mathrm{H_2O}$, FeH, and VO. SpeX prism spectra are ideal for SED measurements as its broad wavelength coverage captures the NIR region in a single continguous spectrum thereby minimizing systematic uncertainties associated with combining multiple spectra from differing spectral orders or instruments. \citet{2015ApJ...810..158F} find that bolometric luminosities derived from spectral integration are relatively independent of the spectral resolution. Thus, the accuracy of our results will not be hindered by the use of low-resolution spectra.

The large majority of spectra in our study were obtained from the SpeX prism Library \citep{2014ASInC..11....7B}. A comprehensive list of references for our object spectra is provided in the Table of Ultracool Fundamental Properties associated with this paper (see \S\ref{sec:intro}). Flux measurements made at the wavelength extremes of the spectra generally have low S/N and can be unreliable. To mitigate this, we apply cuts to the spectra on the blue end at $0.85\;\mu$m and on the red end at $2.45\;\mu$m. The values are chosen to avoid exclusion of wavelengths included in the filter transmission profiles of the PS1 $y$ band on the blue end and the UKIDSS $K$-band on the red end.

\subsection{Absolute Flux Calibration of SpeX Prism Spectra}
\label{sec:abs-cal}
To derive the bolometric flux from SED integration, we need to first calibrate each object's SpeX spectrum to its observed photometry. We begin with the highest available signal-to-noise ratio (S/N) SpeX spectrum of each object ($f_{\lambda, \mathrm{obs}}$) and the photometry with similar wavelength coverage (PS1 $y$, 2MASS $JH$\Ks, MKO $JHK$, and UKIDSS $JHK$). Each photometric point is converted to a flux density measurement ($f_{\lambda, \mathrm{phot}}$) using the corresponding filter's zero point flux density. For several objects, we find that photometric measurements for some filters have formal uncertainties below the instrumental noise floor \citep[PS1:][]{2020ApJS..251....6M}. To avoid underestimating the uncertainties associated with flux calibration of the spectra, we adopt a uniform noise floor of 0.01 mag for all photometric measurements.

\begin{figure}
    \centering
    \includegraphics[scale=0.4]{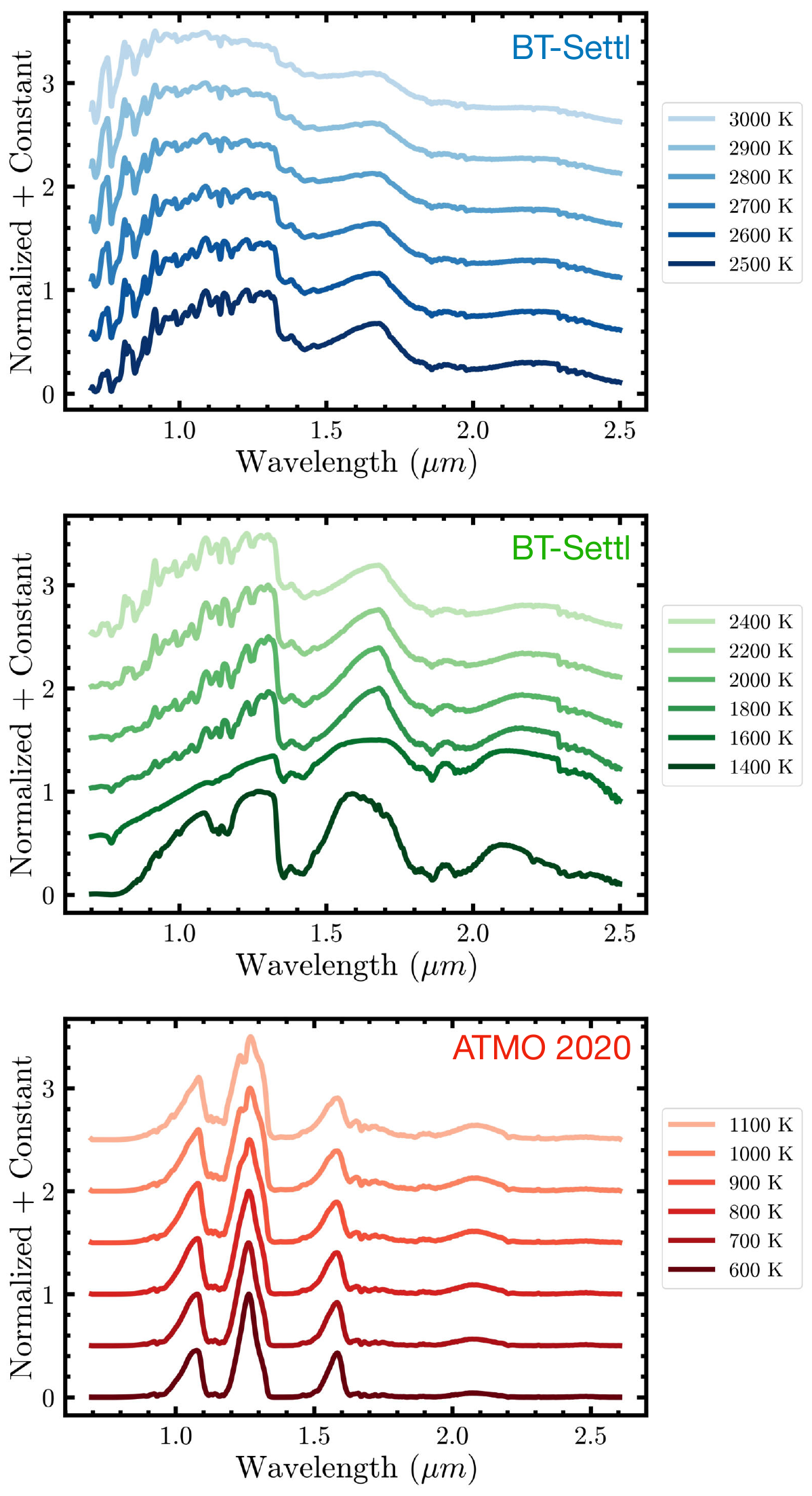}
    \caption{Normalized and vertically offset log $g$ = 5.0 dex {\sc BT-Settl} and {\sc ATMO} 2020 model atmospheric spectra, degraded to the wavelength-dependent resolution of the 0.3\arcsec SpeX prism slit. This sequence spans the effective temperature range from 3000 K to 600 K and approximately corresponds to the ultracool dwarf spectral types M (top), L (middle), and T (bottom).} 
    \label{fig:model}
\end{figure}

Next, we synthesize photometry from the SpeX spectrum of the object for all the filters discussed previously. The filter transmission profile $T_\lambda$ is retrieved from the SVO Filter Profile Service\footnote{\url{http://svo2.cab.inta-csic.es/theory/fps/index.php}} and the SpeX spectrum is interpolated using cubic spline interpolation and mapped to the wavelength grid of the transmission profile. We then calculate the synthetic photometry $f_{\lambda, \mathrm{syn}}$ using,
\begin{equation}
\label{eq:synthesize}
f_{\lambda, \mathrm{syn}} = \frac{\int f_{\lambda, \mathrm{obs}}\;T_\lambda\; d\lambda}{\int T_\lambda\; d\lambda},
\end{equation}
where the limits of the integration are $\lambda_{\mathrm{min}}$ and $\lambda_{\mathrm{max}}$ of the wavelength grid of the filter transmission profile.

Combining the photometric flux densities and the synthetic flux densities derived previously, we can compute a scale factor $k_i$ corresponding to each filter ($i$) as,
\begin{equation}
\label{scale_const:1}
k_i = \left(\frac{f_{\lambda, \mathrm{phot}}}{f_{\lambda, \mathrm{syn}}}\right)_i,
\end{equation}
\begin{equation}
\label{scale_const:2}
\delta k_i = \left(\frac{\delta f_{\lambda, \mathrm{phot}}}{f_{\lambda, \mathrm{syn}}}\right)_i,
\end{equation}
where $\delta k_i$ is the associated uncertainty. We note here that since the synthetic flux density is an integration over the spectrum's many wavelength points corresponding to the filter transmission profile, its uncertainty can be neglected in comparison to the uncertainty in the photometric flux density. We derive the scale factor $k$ for the calibration of the SpeX spectrum by obtaining the weighted average of the individual scale factors $k_i$. A 1\% uncertainty floor is adopted for the weighted average scale factor $k$ to avoid downweighting the photometric noise floor applied earlier. Finally, the calibrated spectrum $f_{\lambda, \mathrm{cal}}$ (Figure \ref{fig:sed}) is found as $f_{\lambda, \mathrm{cal}} = f_{\lambda, \mathrm{obs}} \cdot k$.

\section{Atmospheric Model Fitting}
\label{sec:model_desc}

One of the challenges associated with deriving the bolometric luminosity by SED integration is the lack of optical and mid-infrared (MIR) spectra for most known ultracool dwarfs. However, optical and MIR photometry is well-measured for many of these objects by the Pan-STARRS1 and WISE surveys, respectively. In a few cases, individual objects also have MIR photometry from Spitzer/IRAC Channels 1 and 2 \citep{2004ApJS..154...10F}. We fit the SpeX prism spectrum and the optical and/or MIR photometry with synthetic atmospheric grids produced by the {\sc BT-Settl} \citep[CIFIST2011/2015;][]{2012RSPTA.370.2765A, 2015A&A...577A..42B} and {\sc ATMO} 2020 \citep{2020A&A...637A..38P} models (Figure \ref{fig:model}). Directly integrating the area under the best-fit atmospheric model spectrum in wavelength regions not covered by the SpeX prism spectrum then yields the associated optical and MIR contributions to the bolometric luminosity.

{\sc BT-Settl} (CIFIST2011/2015)\footnote{\url{http://svo2.cab.inta-csic.es/theory/newov2/index.php?models=bt-settl-cifist}} incorporates the role of clouds and dust in its atmospheres, and thus resulting spectra, by self-consistently modeling their formation and sedimentation \citep{2012RSPTA.370.2765A} and accounting for the condensation of nearly 200 liquid and solid species in the atmospheric equation of state. This makes {\sc BT-Settl} appropriate for late-M and L dwarfs, which have cloudy photospheres. Additionally, it includes linelists for several important molecules such as $\mathrm{H_2O}$, CO$_2$, CaH, FeH, CrH, and TiH. Each model is calculated at solar metallicity ($Z = 0$ dex), as defined by \cite{2011SoPh..268..255C}. To ensure a uniformly spaced grid for subsequent uncertainty analysis (Section \ref{sec:lbol}), we omit the 50 K interval model spectra. Given that all our objects have spectral types $\ge$M6, we only consider model spectra upto a temperature of 3500 K to reduce computation time (the grid extends to 7000 K). Thus, the final {\sc BT-Settl} grid adopted in our fitting procedure spans effective temperatures ($T_\mathrm{eff}$) of 1200--3500 K in steps of 100 K and surface gravities ($\log g$) of 2.5--5.5 dex in steps of 0.5 dex.

{\sc ATMO} 2020 is one of the latest cloudless atmospheric model grid geared towards T and Y spectral type ultracool dwarfs \citep{2020A&A...637A..38P}. It brings numerous improvements to the evolutionary and spectroscopic modeling of this coldest subset of brown dwarfs. {\sc ATMO} 2020 makes use of the new hydrogen and helium equation of state from \citet{2019ApJ...872...51C} in its interior structure model. It includes updated line lists for the most prominent molecular opacity sources such CH$_4$ and NH$_3$ from the ExoMol group \citep{2018Atoms...6...26T} and significantly more transitions required to accurately model the atmospheres of brown dwarfs. Furthermore, {\sc ATMO} 2020 provides a more detailed and careful treatment of the Na and K resonant lines, which play an important role in shaping the visible and red-optical spectra of ultracool dwarfs. Similar to {\sc BT-Settl}, each grid point in the model is calculated at solar metallicity ($Z = 0$ dex). In this work, we make use of the weak non-equilibrium chemistry model set (NEQ Weak). Numerous observations have highlighted the importance of non-equilibrium chemistry in the spectra of cool brown dwarfs due to vertical mixing in their atmospheres \citep[e.g.][]{1997ApJ...489L..87N, 2000ApJ...541..374S, 2009ApJ...695..844G, 2015ApJ...799...37L}. \citet{2020A&A...637A..38P} model non-equilibrium chemistry by consistently coupling the relaxation scheme of \citet{2018ApJ...862...31T} to their 1D atmosphere code {\sc ATMO} while considering the non-equilibrium abundances of H$_2$O, CO, CO$_2$, CH$_4$, N$_2$, and NH$_3$. Similar to the approach with {\sc BT-Settl}, to ensure a uniformly spaced rectangular grid for subsequent uncertainty analysis (Section \ref{sec:lbol}), we omit the 50 K interval model spectra. The final {\sc ATMO} 2020 grid in our fitting procedure spans $T_\mathrm{eff}$ of 500--1800 K in steps of 100 K and $\log g$ of 2.5--5.5 dex in steps of 0.5 dex. We note that the NEQ Weak {\sc ATMO} 2020 grid does not provide a spectrum for the $T_\mathrm{eff} =$ 1100 K and $\log g = $ 3.5 dex grid point due to lack of convergence in the model generation. As a substitute, we use an interpolated model spectrum at this grid point (Mark Phillips; private communication). We refer the reader to Section 2 of \citet{2023ApJ...954...22L} for a summary of brown dwarf atmospheric model grids not used in this work.

\begin{figure*}
    \centering
    \includegraphics[scale=0.27]{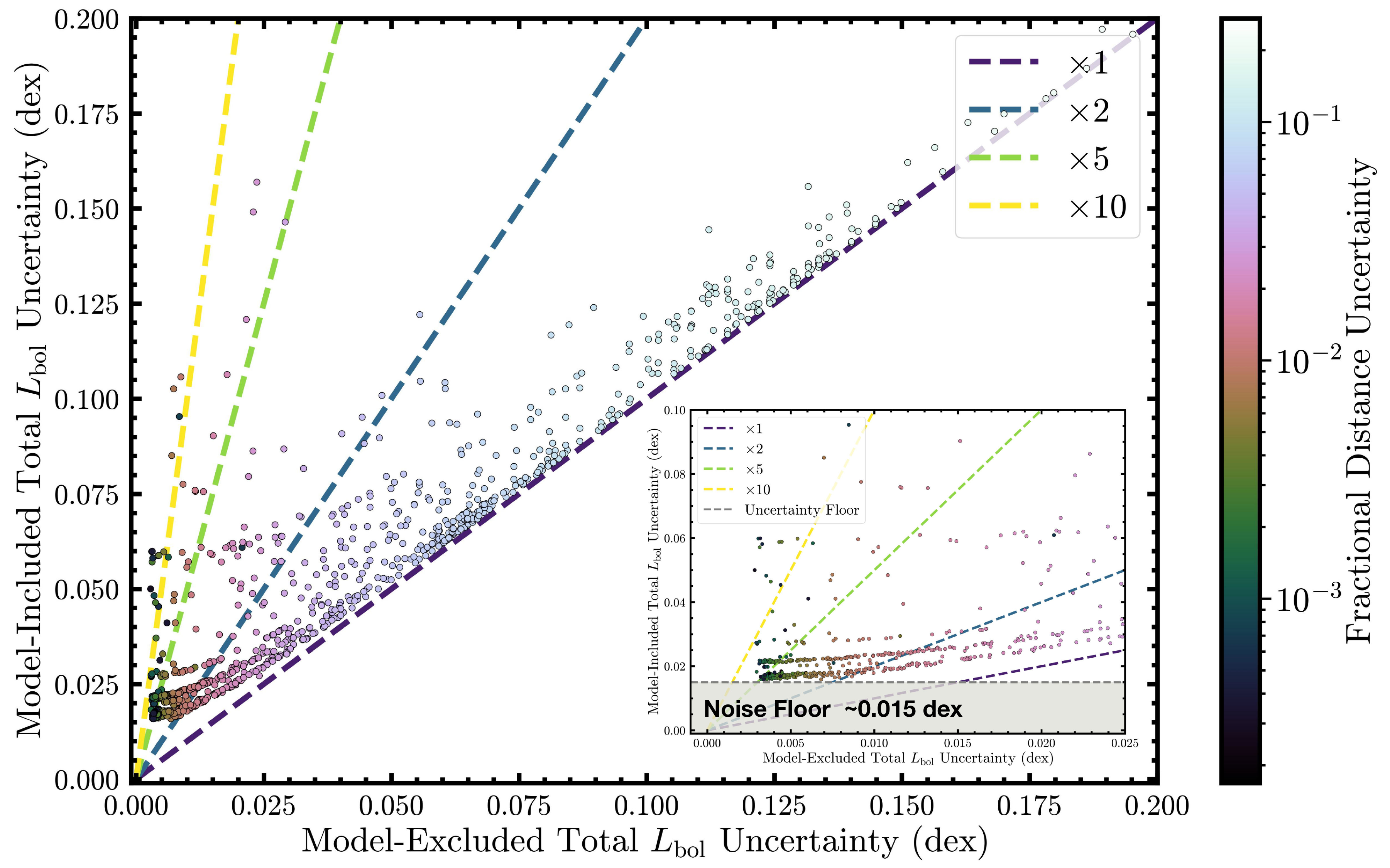}
    \caption{Summary of the bolometric luminosity uncertainty analysis in this work. The model-included total uncertainty (scale factor uncertainty + spectral data points uncertainties + model-contributed uncertainty + distance uncertainty) is plotted on the y-axis against the model-excluded total uncertainty (scale factor uncertainty + spectral data points uncertainties + distance uncertainty). Each point represents one of the \nobject objects in our sample and is colored based on the fractional uncertainty in the distance measurement for that object (color bar at right). Four dashed lines of slopes 1, 2, 5, and 10 plot differing multiplicative relations between the model-included total uncertainty and the model-excluded total uncertainty. The inset figure zooms in near the tail end of the points close to the origin. A noise floor of $\approx$0.015 dex is identified and marked with a dashed gray line. The model-contributed uncertainty component dominates for low fractional distance uncertainty objects, thereby limiting the gains possible with our high precision parallaxes.}
    \label{fig:uncerty}
\end{figure*}

We begin by converting the PS1 optical (excluding $y$ band since it is included in the wavelength range of the SpeX prism spectra) and WISE and Spitzer/IRAC MIR photometry to flux density measurements using the zero points for the corresponding filters. Model spectra are typically provided at high spectral resolution and are first degraded to the non-linear spectral resolution of the SpeX prism data (R$\sim$75--200) corresponding to the three different slit sizes of the SpeX spectrograph (Hurt et al., AAS Journals, submitted). Such a resolution degradation is, however, not required for fitting the optical and/or MIR photometry. Keeping this in mind, we apply the following method to each model spectrum.

First, we map the degraded model spectra to the wavelength grid of the SpeX prism data using cubic spline interpolation. Second, we synthesize the PS1 $griz$, WISE $W1$, $W2$, $W3$, and $W4$, and Spitzer/IRAC Channel 1 and Channel 2 photometric flux densities using the full resolution model spectra, following Equation \ref{eq:synthesize}. Third, we determine the scale factor $k_{\mathrm{fit}}$ for a given model spectrum that would minimize the combined $\chi^2$ of the degraded model spectrum and the synthesized photometriy fit to the observed SED. $k_{\mathrm{fit}}$ is given by 
\begin{equation}
    k_{\mathrm{fit}} = \frac{\sum (f_i F_{k,i} / \sigma_{f, i})}{\sum (f_i^2/ \sigma_{f, i})},
\end{equation}
where $f_i$ represent the individual flux-calibrated SpeX prism flux densities or the individual PS1/WISE/IRAC bandpass photometric flux densities, $F_{k,i}$ represents the model spectrum or the synthesized PS1/WISE/IRAC photometry, and $\sigma_{f, i}$ represent the uncertainty in the observed SpeX spectrum or the PS1/WISE/IRAC photometry. To mitigate telluric contamination due to low atmospheric transmission in the water bands, we exclude spectral data points in the 1.355--1.415 $\mu$m and the 1.830--1.930 $\mu$m wavelength ranges from the above fitting procedure. Additionally, to avoid biasing the model fits, we exclude any spectral data points with unphysical S/N $< -3$. This process is repeated for all available model spectra and the model spectrum that yields the smallest $\chi^2$ value is chosen as the best-fit model spectrum. Figure \ref{fig:sed} shows an example flux-calibrated SED with the best-fit atmospheric model spectrum overlaid, as well as the associated $\chi^2$ surface obtained from the fitting procedures.

\begin{figure}
    \centering
    \includegraphics[scale=0.17]{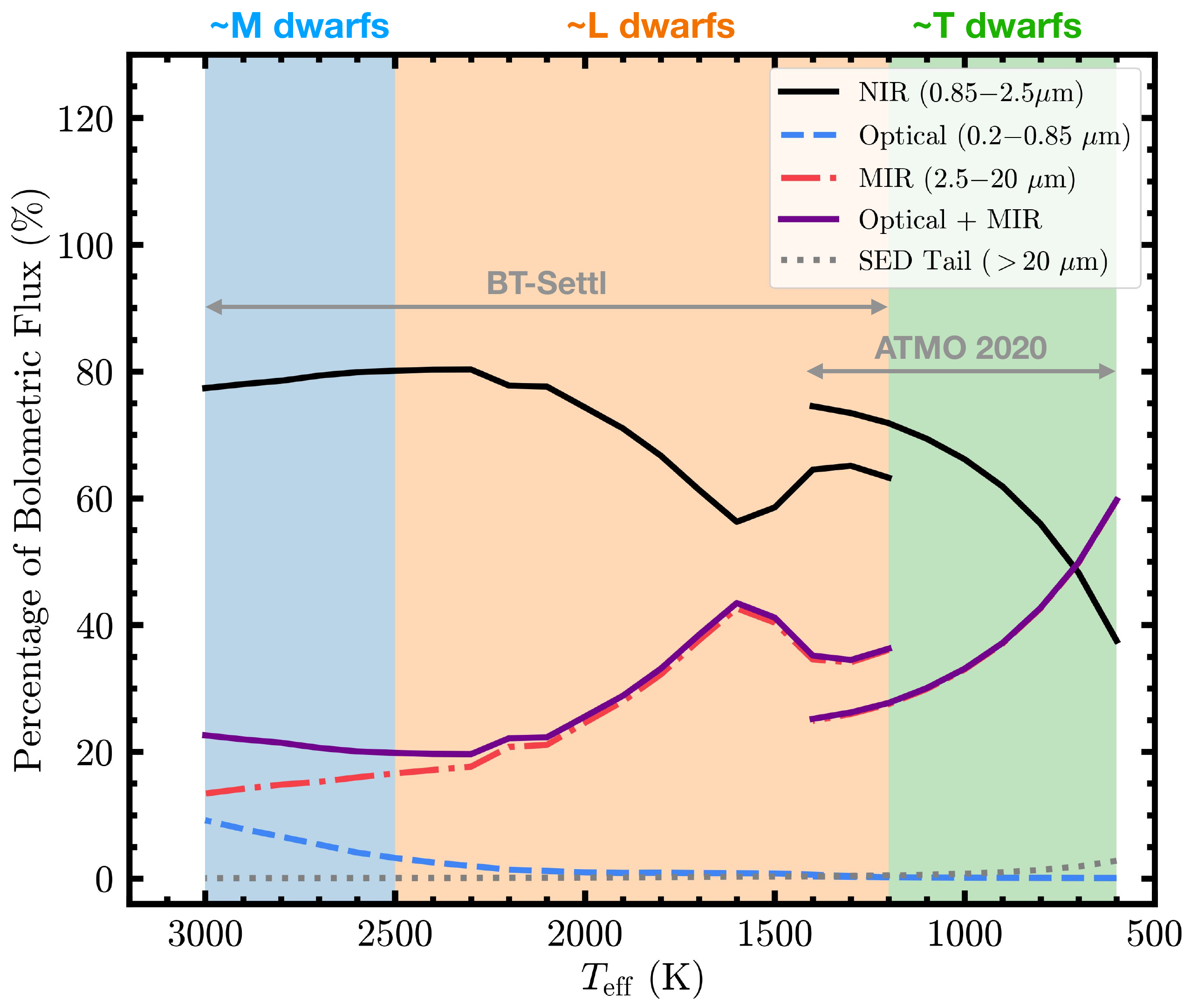}
    \caption{Percentage of bolometric flux contained in different wavelength regions as a function of effective temperature, computed by integrating {\sc BT-Settl} (1200--3000 K) and {\sc ATMO} 2020 (600--1400 K) model spectra with $\mathrm{log}\;g =$ 4.0--5.5. The shaded background areas roughly mark the temperature range corresponding to M, L, and T dwarfs. The NIR contribution dominates for the vast majority of spectral types. The SED tail flux (20--2000 $\mu$m) is a very small percentage of the bolometric flux across all spectral types. The MIR contribution is most important for objects cooler than $\sim$1800 K.}
    \label{fig:flux_cont}
\end{figure}

\begin{figure*}
    \centering
    \includegraphics[scale=0.28]{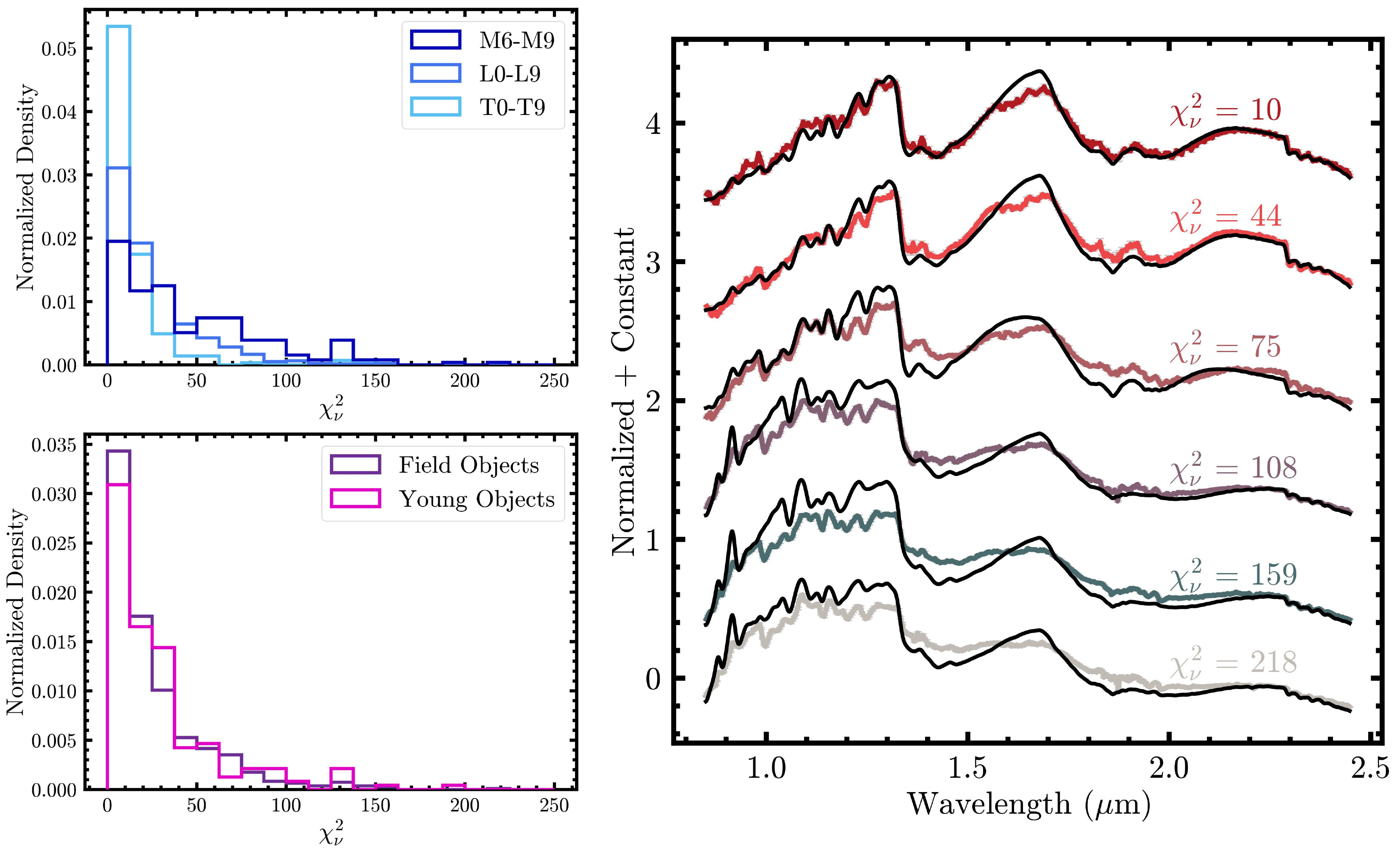}
    \caption{\emph{Left:} Normalized distributions of the best-fit $\chi^2_{\nu}$ values from the model atmosphere fitting for different spectral types (top) and different youth categories (bottom). Young moving group members and low gravity sources are designated as young objects. The bin size is 12.5. Both distributions are skewed significantly towards smaller $\chi^2_{\nu}$ values, indicating good quality model fits for the vast majority of objects in the sample. Model fits to T dwarfs are better than those to L dwarfs, which are in turn better than those to M dwarfs. Field objects yield slightly better fits than young objects. We flag objects with $\chi^2_{\nu}>$ 50 in the Table of Ultracool Fundamental Properties (see \S\ref{sec:fits}). \emph{Right:} Example SpeX prism spectra and the corresponding best-fit model spectra for six objects in the sample with gradually increasing $\chi^2_{\nu}$ values. The S/N of the spectra are similar ($\approx$70--100). This is a visual demonstration of the quality of the fits for different $\chi^2_{\nu}$ values in the distribution.}
    \label{fig:model-fit} 
\end{figure*}

\begin{figure*}[t]
    \centering
    \includegraphics[scale=0.27]{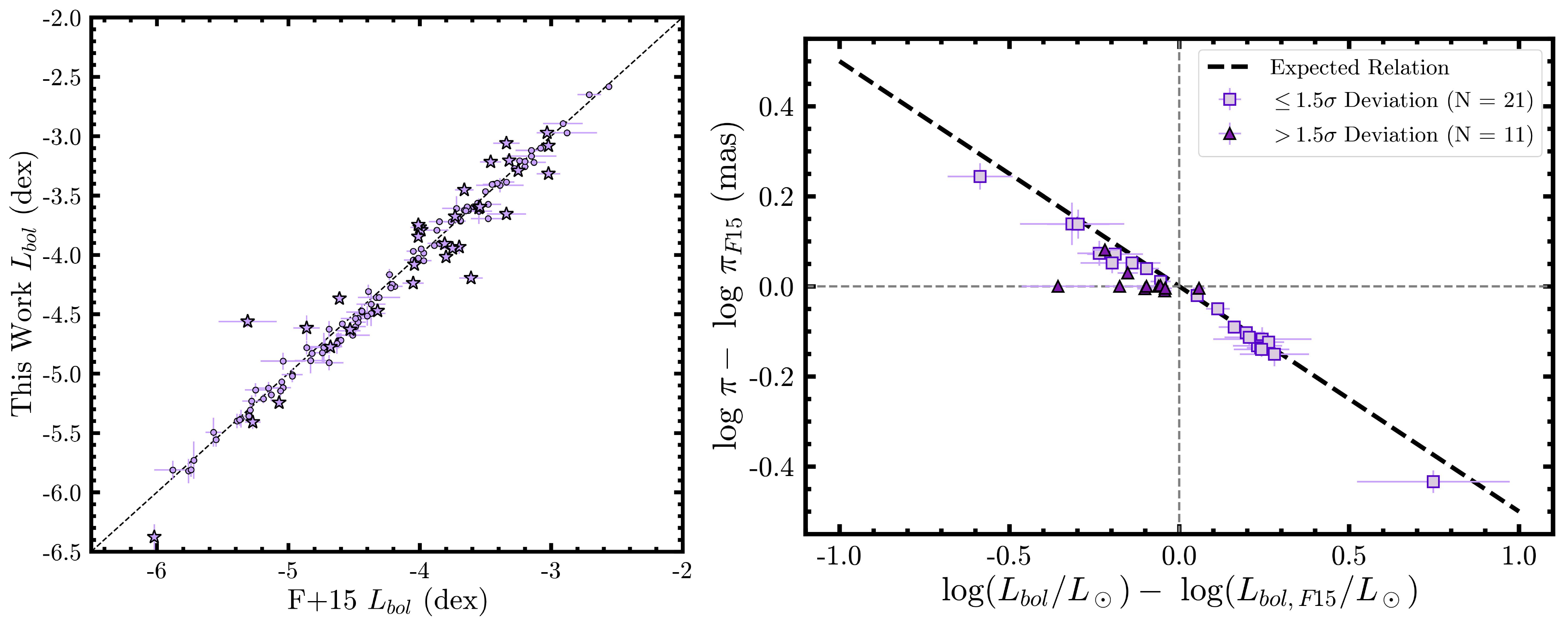}
    \caption{\emph{Left:} Comparison of bolometric luminosities calculated in this work with those in \citet{2015ApJ...810..158F} for objects common to the two samples. The dashed black line represents a 1:1 relation. Objects with disagreements ($>$2$\sigma$ discrepancy) in the two values are marked with a black-bordered star. \emph{Right:} Logarithmic difference in parallax measurements used in the bolometric luminosity calculation vs logarithmic difference in bolometric luminosity measurements between this work and \citet{2015ApJ...810..158F} for the \emph{discrepant objects} in the left panel. The two gray dashed lines correspond to zero luminosity and parallax difference. The black dashed line corresponds to the expected relation between $\Delta$log $\pi$ and $\Delta$log ($L_{\mathrm{bol}}/L_\odot$). Objects marked with a square correspond to the discrepant objects whose luminosity difference is consistent with the theoretically expected parallax difference within 1.5$\sigma$. Objects marked with a triangle correspond to the discrepant objects whose luminosity difference is \emph{not} consistent with the theoretically expected parallax difference within 1.5$\sigma$. Observed differences in bolometric luminosity can thus be explained by changes in parallax measurements for the majority of discrepant objects.}
    \label{fig:lbol-comp-f15}
\end{figure*}

\section{Bolometric Luminosities}
\label{sec:lbol}
Given a flux-calibrated spectrum and best-fit atmospheric model spectrum, we computed the bolometric flux of each object by integration of the SED. We take care to correctly treat correlated and uncorrelated uncertainty contributions to the calculated bolometric flux. 

\subsection{Calculation and Uncertainty Analysis}
We integrate the area under the SpeX spectrum to obtain the NIR contribution to the bolometric flux. The sum of the optical and MIR contribution is obtained by integrating the full resolution best-fit model spectrum at wavelengths outside those of the SpeX spectrum. The full wavelength range of our SED integration is $\approx$ 0.1--2000 $\mu$m. Adding the two components, we obtain a bolometric flux measurement ($f_{\mathrm{bol}}$). The uncertainty in $f_{\mathrm{bol}}$ is a combination of the uncertainty from the NIR contribution and the uncertainty from the summed optical and MIR contributions to the bolometric flux. 

The uncertainty in the NIR flux contribution is defined as the calibration uncertainty and consists of the uncertainty in the scale factor $k_{\mathrm{fit}}$ for the entire SpeX spectrum and the measurement uncertainties for each individual spectral data point. We note that it only comes from the NIR contribution since we use the zero uncertainty model spectrum to estimate the optical and MIR contributions. The calibration uncertainty is determined using Monte Carlo techniques. We generate $10^4$ different SpeX spectra by drawing the scaling factor and the value for each spectral data point from their corresponding posterior distributions. Each generated spectrum is integrated as described previously to obtain the NIR flux contribution. The standard deviation of these measurements yields the calibration uncertainty. 

The uncertainty in the MIR and optical flux contribution is defined as the model-contributed uncertainty ($\sigma_{\mathrm{mod}}$), which arises because the atmospheric model fit to the SpeX spectrum and optical/MIR photometry is not perfect. To calculate the model-contributed uncertainty, we re-compute the sum of the optical and MIR flux contributions to the bolometric flux using the four model spectra adjacent on the model grid in $T_\mathrm{eff}$ and $\mathrm{log}\;g$ values to the best-fit model spectrum. The standard deviation of this set of values gives the model-contributed uncertainty \citep{2008ApJ...678.1372C, 2009ApJ...702..154S}.

The total uncertainty in the bolometric flux ($\delta f_{\mathrm{bol}}$) is then obtained as the quadrature sum of the calibration and model-contributed uncertainty. Given a bolometric flux $f_{\mathrm{bol}} \pm \delta f_{\mathrm{bol}}$ and distance measurement $d \pm \delta d$, we can obtain the bolometric luminosity $L_{\mathrm{bol}}$ and the associated uncertainty $\delta L_{\mathrm{bol}}$ of the object as follows,
\begin{equation}
    L_{\mathrm{bol}} = f_{\mathrm{bol}} \cdot 4\pi d^2,
\end{equation}
\begin{equation}
    \frac{\delta L_{\mathrm{bol}}}{L_{\mathrm{bol}}} = \sqrt{\left(\frac{\delta f_{\mathrm{bol}}}{f_{\mathrm{bol}}}\right)^2 + \left(2 \cdot \frac{\delta d}{d}\right)^2}.
\end{equation}
For 839 objects in our sample, the distance is given by a parallax measurement. For the remaining 215 objects, we use photometric distances computed using absolute magnitude-spectral type relations for 2MASS $J$\Ks, MKO $JK$, and WISE $W2$ bands. We use the polynomials from \citet[][]{2012ApJS..201...19D} for field objects with 2MASS $J$\Ks, MKO $JK$, and ALLWISE $W2$ photometry; \citet[][]{2022RNAAS...6..265F} for field objects with CatWISE $W2$ photometry; or this work (Appendix \ref{sec:abs-spt}) for young objects with 2MASS $J$\Ks, MKO $JK$, and ALLWISE/CatWISE $W2$ photometry. The luminosities are reported as $\mathrm{log}(L_{\mathrm{bol}}/L_\odot)$ where $L_\odot = 3.828 \times 10^{26}$ W \citep{2015arXiv151007674M}.

We examine the effect of the model-contributed uncertainty on the total uncertainty in our $L_{\mathrm{bol}}$ measurements by comparing this total $L_{\mathrm{bol}}$ uncertainty against the total $L_{\mathrm{bol}}$ uncertainty computed assuming zero model-contributed uncertainty. This is presented in Figure \ref{fig:uncerty}, where we have the defined the former quantity as the model-included total uncertainty and the latter quantity as the model-excluded total uncertainty. We find that the model-included total uncertainty is significantly larger than the model-excluded total uncertainty for low fractional distance uncertainty objects. A reduction in the fractional distance uncertainty from high precision parallax programs should have significantly lowered the total uncertainty in the bolometric luminosity (as seen for the $\sigma_{\mathrm{mod}} = 0$ assumption case), but a non-zero $\sigma_{\mathrm{mod}}$ component results in an uncertainty floor at $\approx$0.015 dex. Thus, the central limiting factor for these cases is the presence of the model-contributed uncertainty. Obtaining high S/N spectra of ultracool dwarfs in the optical and MIR wavelengths will be crucial to lowering the identified noise floor.

\begin{deluxetable*}{cccccc}
\label{table:lbol_comp}
\centering
\tabletypesize{\scriptsize}
\tablecaption{Objects with $>$2$\sigma$ Disagreement between This Work's and \citet{2015ApJ...810..158F}'s Bolometric Luminosity Measurements}
\tablehead{\colhead{UCS Name} & \colhead{F+15 Name} & \colhead{$\pi$} & \colhead{$\pi$ (F+15)} & \colhead{log($L_{bol}$/$L_\odot$)} & \colhead{log($L_{bol}$/$L_\odot$) (F+15)} \\ \colhead{} & \colhead{} & \colhead{(mas)} & \colhead{(mas)} & \colhead{(dex)} & \colhead{(dex)}}
\startdata
2MASS J00332386-1521309 & 0033-1521 & $43.50\pm0.80$ & $24.8\pm2.5$ & $-4.20\pm0.03$ & $-3.61\pm0.09$ \\ 
2MASS J00345157+0523050 & 0034+0523 & $118.80\pm2.70$ & $105.4\pm7.5$ & $-5.41\pm0.06$ & $-5.27\pm0.06$ \\ 
2MASS J03231002-4631237 & 0323-4631 & $23.40\pm0.70$ & $17.0\pm3.0$ & $-3.66\pm0.03$ & $-3.34\pm0.15$ \\ 
2MASS J06085283-2753583 & 0608-2753 & $22.62\pm0.16$ & $32.0\pm3.6$ & $-3.06\pm0.02$ & $-3.34\pm0.10$ \\ 
2MASS J09490860-1545485 & 0949-1545 & $42.20\pm2.70$ & $55.3\pm6.6$ & $-4.62\pm0.10$ & $-4.86\pm0.10$ \\ 
2MASS J13595510-4034582 & 1359-4034 & $47.35\pm0.20$ & $64.2\pm5.5$ & $-3.77\pm0.02$ & $-4.0\pm0.07$ \\ 
2MASSI J0439010-235308 & 0439-2353 & $80.70\pm0.40$ & $110.4\pm4.0$ & $-4.37\pm0.03$ & $-4.61\pm0.03$ \\ 
2MASSI J0445538-304820 & 0445-3048 & $61.94\pm0.15$ & $78.5\pm4.9$ & $-3.79\pm0.02$ & $-3.99\pm0.05$ \\ 
2MASSI J0825196+211552$^*$ & 0825+2115 & $92.60\pm0.80$ & $93.8\pm1.0$ & $-4.63\pm0.02$ & $-4.53\pm0.01$ \\ 
2MASSI J0835425-081923 & 0835-0819 & $138.31\pm0.21$ & $117.3\pm11.2$ & $-4.24\pm0.02$ & $-4.05\pm0.08$ \\ 
2MASSI J0847287-153237 & 0847-1532 & $57.51\pm0.21$ & $76.5\pm3.5$ & $-3.75\pm0.02$ & $-4.01\pm0.04$ \\ 
2MASSI J1546291-332511$^*$ & 1546-3325 & $88.00\pm1.90$ & $88.0\pm1.9$ & $-5.25\pm0.04$ & $-5.07\pm0.02$ \\ 
2MASSI J2104149-103736 & 2104-1037 & $58.02\pm0.27$ & $53.0\pm1.71$ & $-3.91\pm0.02$ & $-3.81\pm0.03$ \\ 
2MASSW J1155395-372735 & 1155-3727 & $84.73\pm0.15$ & $104.4\pm4.7$ & $-3.85\pm0.02$ & $-4.01\pm0.04$ \\ 
2MASSW J1506544+132106$^*$ & 1506+1321 & $85.43\pm0.19$ & $70.92\pm2.5$ & $-4.02\pm0.02$ & $-3.8\pm0.03$ \\ 
2MASSW J1515008+484742$^*$ & 1515+4847 & $101.96\pm0.26$ & $95.24\pm2.0$ & $-4.47\pm0.02$ & $-4.32\pm0.02$ \\ 
2MASSW J1552591+294849 & 1552+2948 & $48.90\pm0.16$ & $47.7\pm0.9$ & $-3.60\pm0.02$ & $-3.55\pm0.02$ \\ 
2MASSW J2208136+292121 & 2208+2921 & $25.10\pm1.60$ & $21.2\pm0.7$ & $-3.93\pm0.06$ & $-3.7\pm0.03$ \\ 
BR B0246-1703 & 0248-1651 & $44.62\pm0.13$ & $61.6\pm5.4$ & $-3.22\pm0.02$ & $-3.46\pm0.08$ \\ 
DENIS-P J0751164-253043 & 0751-2530 & $56.42\pm0.12$ & $59.15\pm0.84$ & $-3.68\pm0.02$ & $-3.73\pm0.01$ \\ 
DENIS-P J142527.97-365023.4$^*$ & 1425-3650 & $84.40\pm0.30$ & $86.45\pm0.83$ & $-4.08\pm0.02$ & $-4.04\pm0.01$ \\ 
ESO 207-61 & 0707-4900 & $41.69\pm0.11$ & $54.1\pm4.5$ & $-3.45\pm0.02$ & $-3.66\pm0.07$ \\ 
G 196-3B & 1004+5022 & $46.20\pm0.50$ & $41.0\pm4.1$ & $-3.95\pm0.02$ & $-3.75\pm0.09$ \\ 
LHS 1604$^*$ & 0351-0052 & $68.12\pm0.07$ & $68.1\pm1.9$ & $-3.08\pm0.02$ & $-3.02\pm0.02$ \\ 
LHS 2924$^*$ & 1428+3310 & $91.16\pm0.10$ & $90.8\pm1.3$ & $-3.60\pm0.02$ & $-3.54\pm0.01$ \\ 
LHS 3003 & 1456-2809 & $141.79\pm0.06$ & $159.2\pm5.1$ & $-3.21\pm0.02$ & $-3.32\pm0.03$ \\ 
LP 412-31$^*$ & 0320+1854 & $68.27\pm0.08$ & $68.9\pm0.6$ & $-3.29\pm0.02$ & $-3.25\pm0.01$ \\ 
SDSS J000013.54+255418.6$^*$ & 0000+2554 & $70.80\pm1.90$ & $70.8\pm1.9$ & $-4.78\pm0.04$ & $-4.68\pm0.02$ \\ 
TVLM 831-161058 & 0251+0047 & $28.16\pm0.18$ & $20.5\pm2.2$ & $-3.32\pm0.02$ & $-3.02\pm0.09$ \\ 
UGPS J072227.51-054031.2$^*$ & 0722-0540 & $242.80\pm2.40$ & $242.8\pm2.4$ & $-6.38\pm0.11$ & $-6.02\pm0.02$ \\ 
WISEPA J164715.59+563208.2 & 1647+5632 & $42.70\pm2.10$ & $116.0\pm29.0$ & $-4.56\pm0.05$ & $-5.31\pm0.22$ \\ 
Wolf 359$^*$ & 1056+0700 & $415.18\pm0.07$ & $419.1\pm2.1$ & $-2.97\pm0.02$ & $-3.03\pm0.00$ \\ 
\enddata
\tablecomments{For objects marked with an asterisk ($^*$), the difference in parallax measurements is not consistent with the theoretical $\Delta$log ($L_{\mathrm{bol}}/L_\odot$) = $-2 \cdot \Delta$log $\pi$ scaling relation within 1.5$\sigma$. For the remaining objects, the difference in parallax measurements is consistent with the theoretical scaling relation within 1.5$\sigma$. See \S\ref{sec:comp-lbol-f15} for more details.}
\end{deluxetable*}

\subsection{Quality and Reliability of Atmospheric Model Fits}
\label{sec:fits}
In this work, we estimate the flux contribution from the optical (0.20--0.85 $\mu$m) and MIR (2.5--2000.0 $\mu$m) wavelengths using atmospheric model spectra. Thus, it is important to characterize the quality and reliability of atmospheric model spectrum fits to the objects' NIR SpeX prism spectra and associated PS1 optical and WISE/Spitzer MIR photometry.

The NIR flux contribution to an object's bolometric flux is obtained by directly integrating the observed SpeX prism spectrum across a wavelength range of $\approx$0.85--2.5 $\mu$m. To gauge the percentage of flux missing after this step, we plot the average percentage of the bolometric flux contained in three defined regions --- the optical region (0.20--0.85 $\mu$m), the MIR region (2.5--20.0 $\mu$m), and the SED tail (20.0--2000.0 $\mu$m) --- as a function of effective temperature in Figure \ref{fig:flux_cont}. These values are obtained by integrating the {\sc BT-Settl} and {\sc ATMO} 2020 model spectra for each temperature in the defined region. We average flux percentage values computed across 4.0--5.5 dex $\mathrm{log}\;g$ spectra (most typical range of surface gravities for ultracool dwarfs) at a fixed temperature. It is observed that the SED tail contribution to the bolometric flux is negligible ($<$4\%) across all spectral types and does not affect our luminosity calculations. This quantity is important to check since the longest wavelength WISE photometry ($W4$) point that constrains our fits lies near 20 $\mu$m. The MIR flux contribution to the bolometric flux is between $\approx$14--17\% for M dwarfs, $\approx$16--45\% for L dwarfs, and $\approx$25--60\% for T dwarfs. The optical flux contribution is only significant for M dwarfs ($\approx$3--10\%). Thus, while the NIR contribution dominates for the majority of objects, the MIR contribution does comprise a significant percentage of the bolometric flux, especially for the cooler $\gtrsim$L5 ultracool dwarfs. These are the objects for which the model fits must be examined in greater detail.

We use the reduced chi-square $\chi^2_{\nu}$ as our figure of merit to quantify the goodness of our atmospheric model spectrum fit to the observed SED (SpeX spectrum + optical and MIR photometry). Figure \ref{fig:model-fit} presents the distribution of our best-fit $\chi^2_{\nu}$ values for different spectral classes and youth categories. The distributions are significantly skewed towards smaller $\chi^2_{\nu}$ indicating good model fits to the data. The model fits have better (lower) $\chi^2_{\nu}$ values for T dwarfs, followed by L and M dwarfs respectively. This is reassuring given that good quality model fits to the SEDs of the cooler ultracool dwarfs in our sample are important for accurate estimation of a signficant percentage of their total bolometric flux. The distributions for field and young (young moving group members/low gravity sources) objects are roughly similar with slightly better (lower) $\chi^2_{\nu}$ values for field objects. Figure \ref{fig:model-fit} provides examples of atmospheric model spectra fits to the observed SpeX prism spectra for six objects in our sample at different $\chi^2_{\nu}$ values.

Based on the shape of the $\chi^2_{\nu}$ distributions and visual inspection of the atmospheric model fits, we flag objects with $\chi^2_{\nu}$ values greater than 50 in our Table of Ultracool Fundamental Properties (see \S\ref{sec:intro}) to indicate possibly lower reliability than the other fits. A total of 171 objects ($\sim$16\% of the sample) are flagged: 80 M dwarfs (39\% of M dwarfs in our sample), 83 L dwarfs (13.4\% of the L dwarfs in our sample), and 8 T dwarfs (3.5\% of the T dwarfs in our sample). For these objects, the relatively poorer quality of the model spectrum fit may have impacted the estimated optical and MIR flux contribution, and hence the calculated bolometric luminosity. We note that the discrepancy is likely to be small for the majority of objects in this subset since the MIR contribution to the bolometric flux is significant, compared to the NIR contribution, primarily for $\gtrsim$L5 dwarfs (only 13 objects in the flagged subset).

\subsection{Comparison with \citet{2015ApJ...810..158F}}
\label{sec:comp-lbol-f15}
\begin{figure}
    \centering
    \includegraphics[scale=0.07]{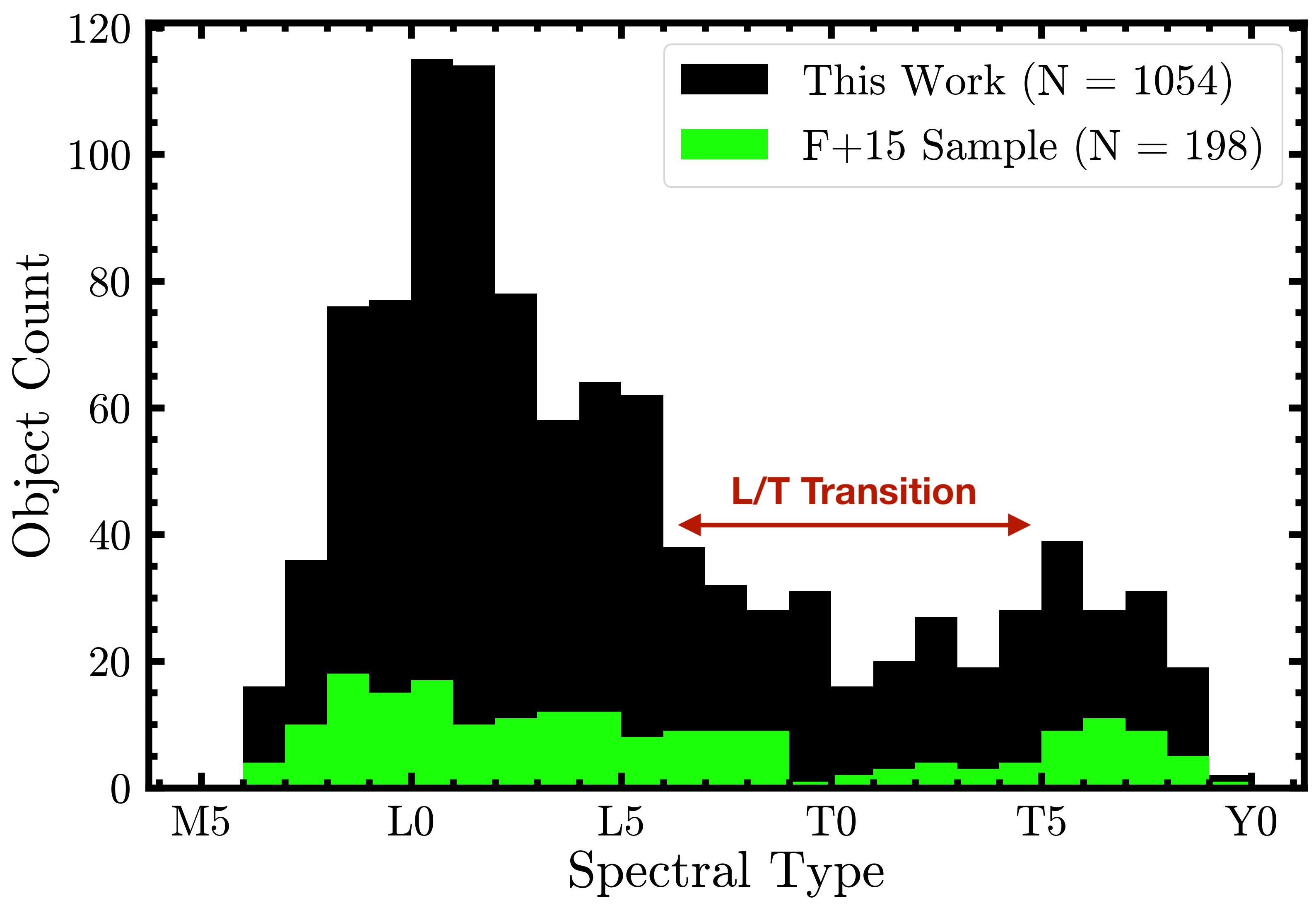}
    \caption{A comparison of the number of SED-integrated $L_{\mathrm{bol}}$ measurements available as a function of spectral type (SpT) before \citep{2015ApJ...810..158F} and after this work. The bin size is 1 SpT. We increase the sample size of $L_{\mathrm{bol}}$ values by a factor of $\sim$5 and contribute a significant number of measurements in the L/T transition region.} 
    \label{fig:sample-comp}
\end{figure}

Figure \ref{fig:lbol-comp-f15} compares our $L_{\mathrm{bol}}$ values with those in \citet{2015ApJ...810..158F} for objects common to the two samples (N = 130) and characterizes measurement differences as a function of the difference in parallaxes used in the calculations. We find that our results are consistent ($<$2$\sigma$ difference) with \citet{2015ApJ...810..158F} for the majority of objects (N = 98). $>$2$\sigma$ discrepancy in the bolometric luminosity is observed for 32 objects (Figure \ref{fig:lbol-comp-f15}, left panel). The logarithmic difference in bolometric luminosity ($\Delta$log [$L_{\mathrm{bol}}/L_\odot$]) is related to the logarithmic difference in parallax ($\Delta$log $\pi$) as $\Delta$log ($L_{\mathrm{bol}}/L_\odot$) = $-2 \cdot \Delta$log $\pi$. We can use this scaling relationship to determine if the source of the discrepancy in bolometric luminosity is a revised parallax measurement. The differences in parallax measurements between our work and \citet{2015ApJ...810..158F} can explain the differences in bolometric luminosity measurements (within 1.5$\sigma$) for 21 of the 32 discrepant objects (Figure \ref{fig:lbol-comp-f15}, right panel). For the 11 discrepant objects (2MASSI J0825196+211552, 2MASSI J1546291-332511, 2MASSW J1506544+132106, 2MASSW J1515008+484742, DENIS-P J142527.97-365023.4, LHS 1604, LHS 2924, LP 412-31, SDSS J000013.54+255418.6, UGPS J072227.51-054031.2, and Wolf 359) where the difference in parallax measurements is not consistent with the theoretical scaling relation within 1.5$\sigma$, the minor differences observed in $L_{\mathrm{bol}}$ values may be attributed to our use of atmospheric models to evaluate the optical and MIR flux contributions to the bolometric flux in contrast to the use of the object's optical and (in some cases) MIR spectra (combined with linear interpolation across SED gaps), as in \citet{2015ApJ...810..158F}. A complete list of the 32 discrepant objects is provided in Table \ref{table:lbol_comp}. Overall, our work increases the sample size of $L_{\mathrm{bol}}$ measurements by a factor of $\sim$5 (Figure \ref{fig:sample-comp}). 

\subsection{Polynomial Relations}
We derive empirical relationships between the bolometric luminosity and spectral type/absolute magnitude for field and young objects by performing inverse variance-weighted polynomial fits to the data. The polynomial fit is carried out for increasing orders and a lack-of-fit test (\emph{F}-test) is conducted for the best-fit coefficients at each order to determine if a higher-order fit is necessary to explain the data. We reject the null-hypothesis (higher-order fit required) for $p > 0.05$. A root-mean-square (rms) value is determined to quantitatively characterize the scatter about the best-fit relation. We note here that the \emph{F}-test does not always yield a definitive answer for the order of the polynomial based on the $p > 0.05$ rejection criterion. For such cases, we adopt the lowest order after which $p$ does not change significantly (factor of $\sim$5--10) as the best-fit polynomial and validate it by visual inspection.

\begin{figure*}
    \centering
    \includegraphics[scale=0.56]{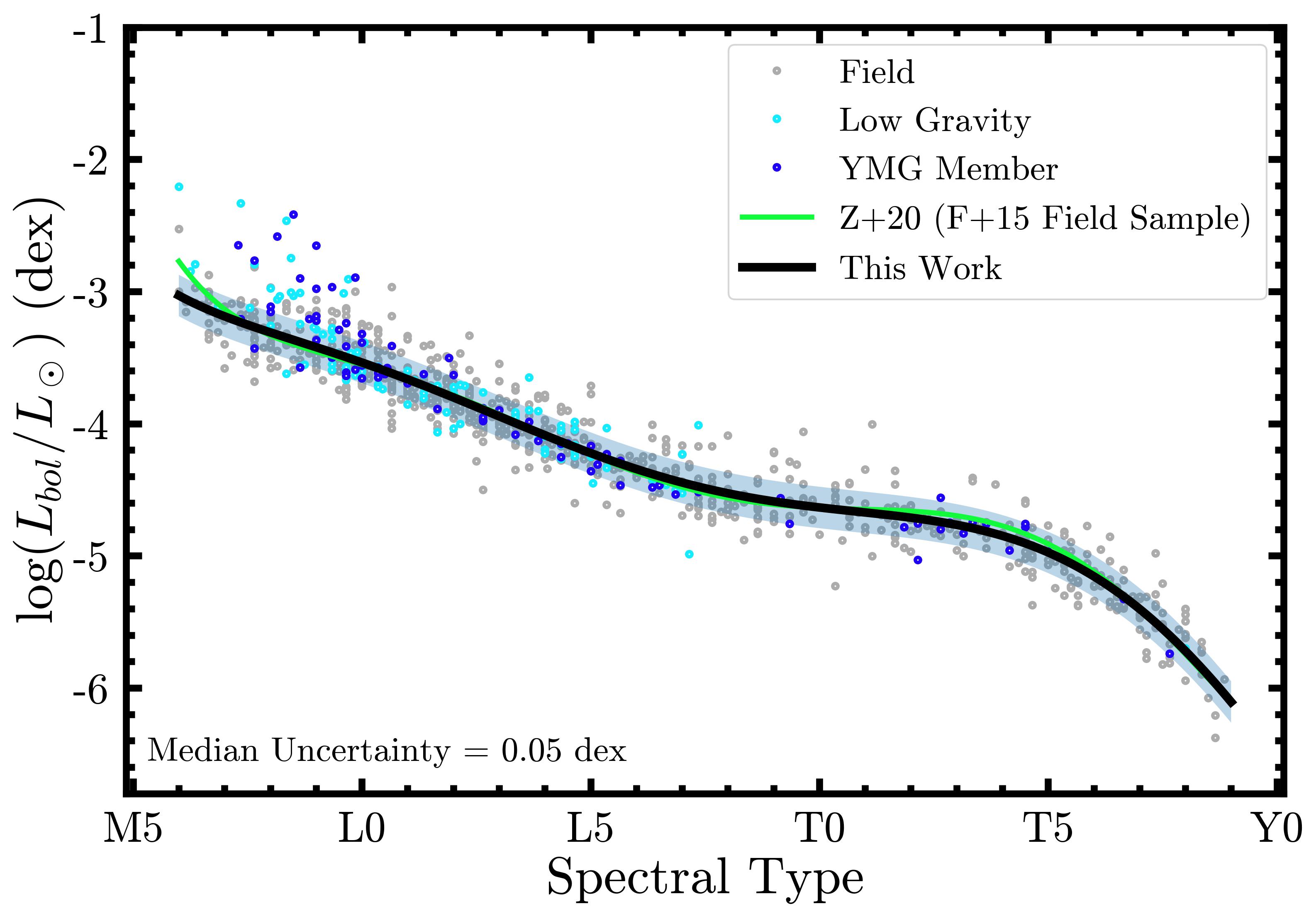}
    \caption{Bolometric luminosities derived for our sample of \nobject ultracool dwarfs as a function of their spectral type (SpT). Field objects, low gravity ultracool dwarfs, and young moving group members are marked by gray, cyan, and blue points respectively. Random scatter in the x direction with amplitude 0.3 SpT is applied to avoid overlapping points. Due to the large number of points, uncertainties in individual measurements are not shown for figure readability. The black line is the best sixth-order weighted polynomial fit to the data and the shaded blue region represents the rms scatter about the fit. The polynomial relation from \citet[][Z+20]{2020ApJ...891..171Z} derived using bolometric luminosities of 198 field-age ultracool dwarfs from \citet[][F+15]{2015ApJ...810..158F} is plotted as a green solid line for comparison.} 
    \label{fig:lbol-spt}
\end{figure*}

Figure \ref{fig:lbol-spt} summarizes our $L_{\mathrm{bol}}$ measurements as a function of spectral type and presents the corresponding sixth-order polynomial relation (Table \ref{table:poly}). For M dwarfs (SpT $= 6-9.5$), young objects show considerable scatter from the relation, as was observed by \citet{2015ApJ...810..158F}. However, for later types, young objects do not scatter significantly different than the field-age objects, an observation not previously confirmed given the small sample size of ultracool dwarf measurements available. While the relation does not exhibit significant differences compared to \citet{2015ApJ...810..158F}, our larger sample size enables a robust determination of the rms scatter in the polynomial relation.

\begin{figure*}
    \centering
    \includegraphics[scale=0.55]{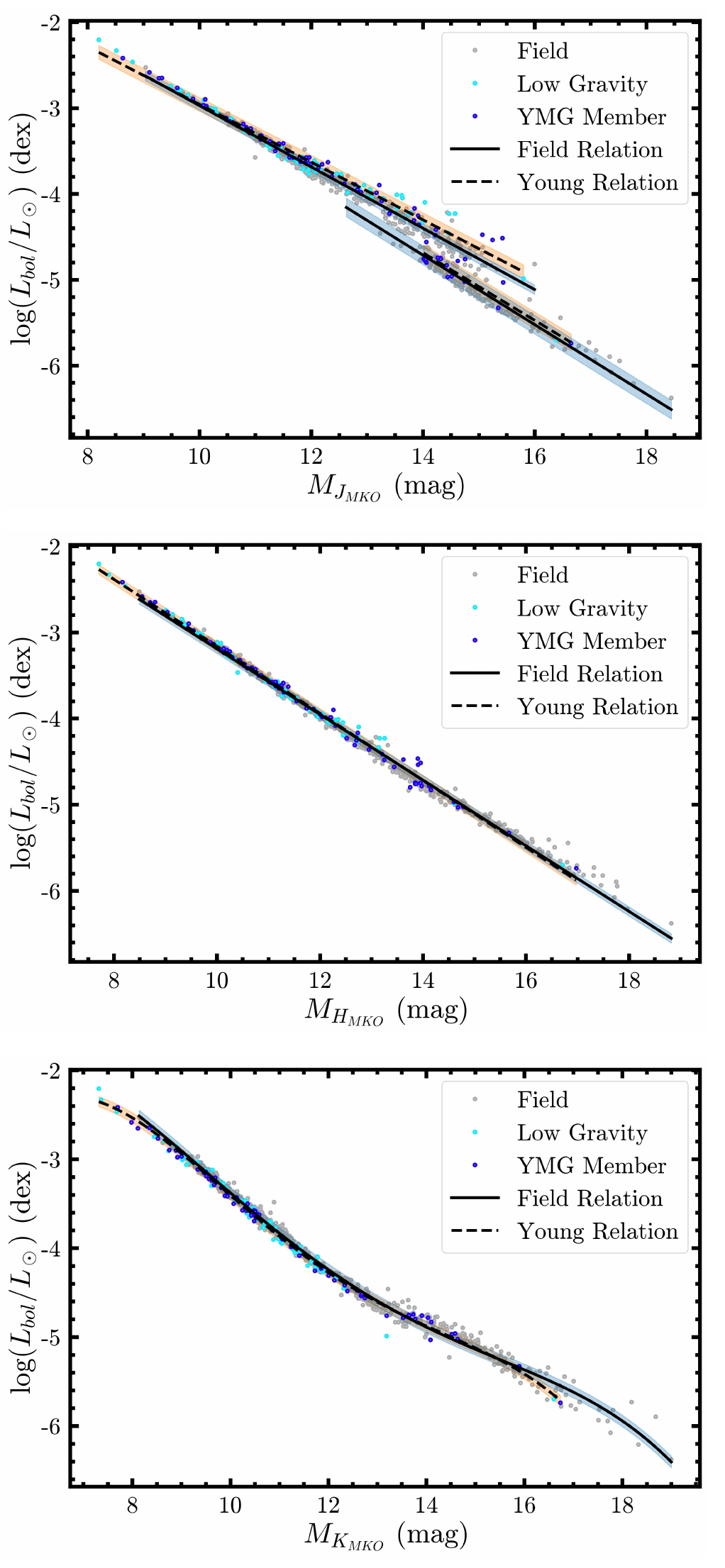}
    \caption{Bolometric luminosity vs absolute magnitude in the MKO $JHK$ bands. Symbols are the same as in Figure \ref{fig:lbol-spt}. The black solid and dashed lines are the best-fit polynomial relations for the field and young objects, respectively. The shaded blue and orange regions represent the rms scatter about the field and young object fits, respectively.}
    \label{fig:lbol_abs_nir}
\end{figure*}

\begin{figure*}[t]
    \centering
    \includegraphics[scale=0.33]{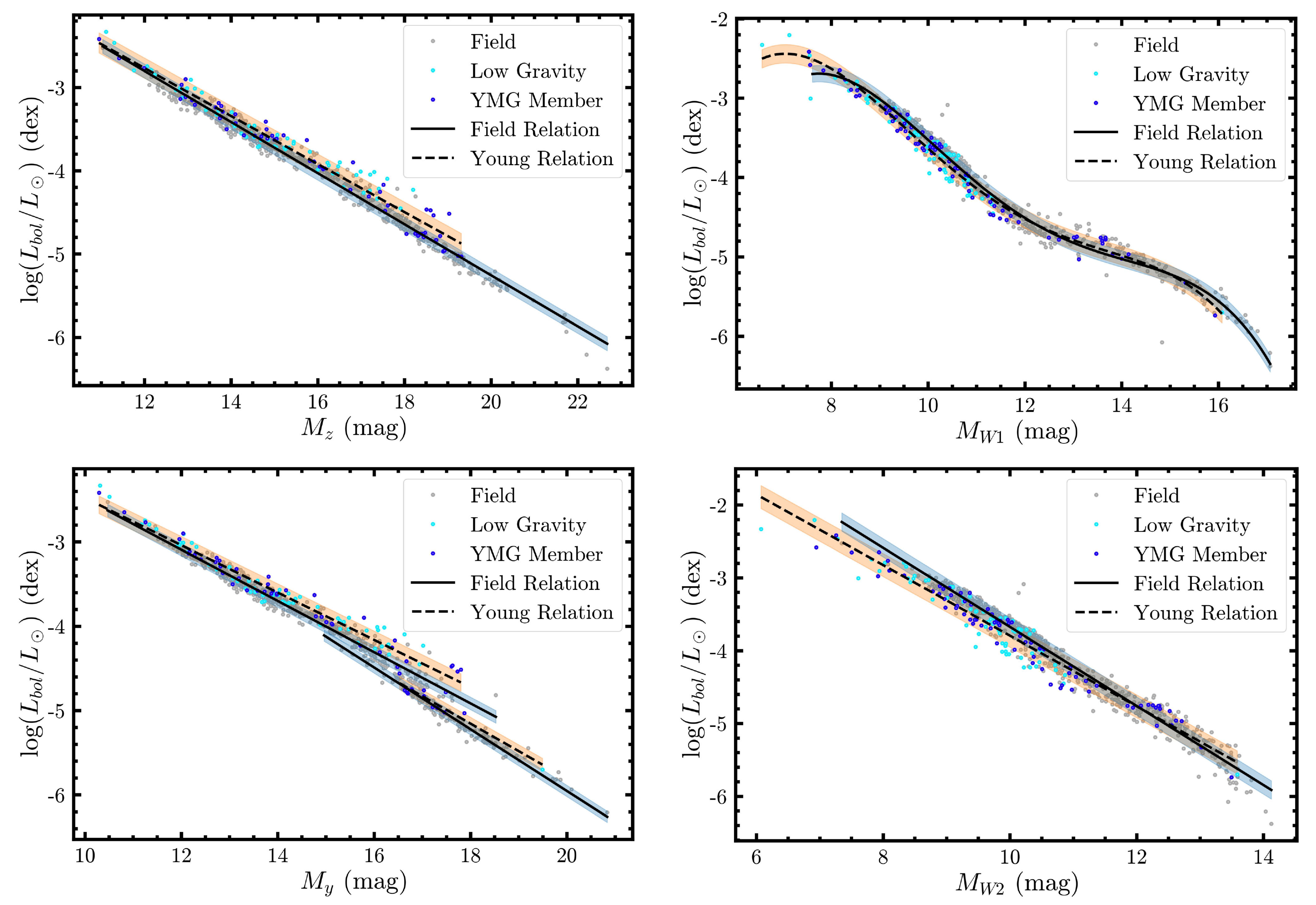}
    \caption{Same as Figure \ref{fig:lbol_abs_nir} for PS1 $zy$ and WISE $W1$ and $W2$ bands.}
    \label{fig:lbol_abs_mir_opt}
\end{figure*}

\begin{figure*}
    \centering
    \includegraphics[scale=0.26]{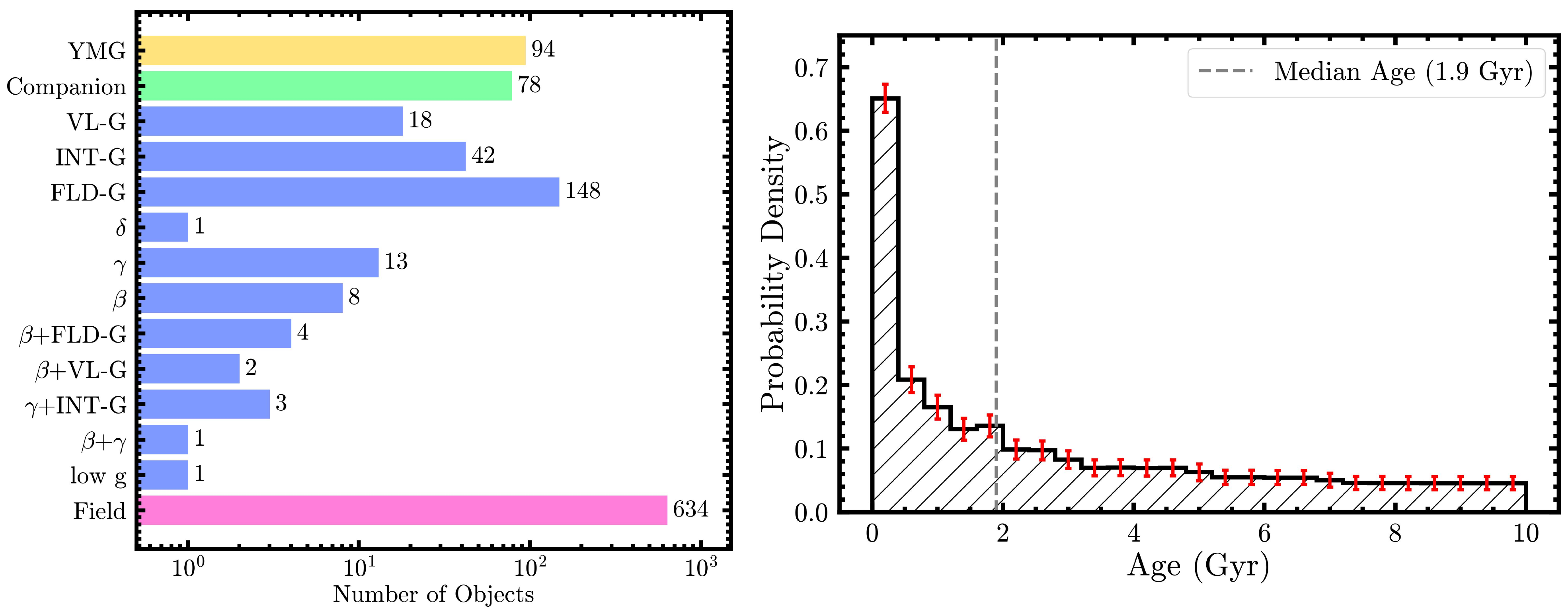}
    \caption{Age properties of the ultracool dwarf sample. \emph{Left:} Number of objects in the sample in each age category as described in \S\ref{sec:ages}. The corresponding count is listed to the right of each bar. \emph{Right:} Mean age distribution of objects in the sample based on Monte Carlo simulations. The bin size is 0.4 Gyr. Uncertainty in the density for each bin is marked with a red errorbar. The median age for the sample is 1.9 Gyr.}
    \label{fig:age-dist}
\end{figure*}

Figures \ref{fig:lbol_abs_nir} and \ref{fig:lbol_abs_mir_opt} summarize our $L_{\mathrm{bol}}$ measurements as a function of absolute magnitude in the MKO $JHK$, PS1 $zy$, and WISE $W1$ and $W2$ bands. The full set of polynomial relations (including those for 2MASS $JH$\Ks) are presented in Table \ref{table:poly}. We find that the MKO $H$-band polynomial relation is the most reliable choice for $L_{\mathrm{bol}}$ determinations using absolute magnitudes, based on its linearity, low rms scatter about the fit, and insensitivity to age and spectral type information. 

\begin{figure*}
    \centering
    \includegraphics[scale=0.069]{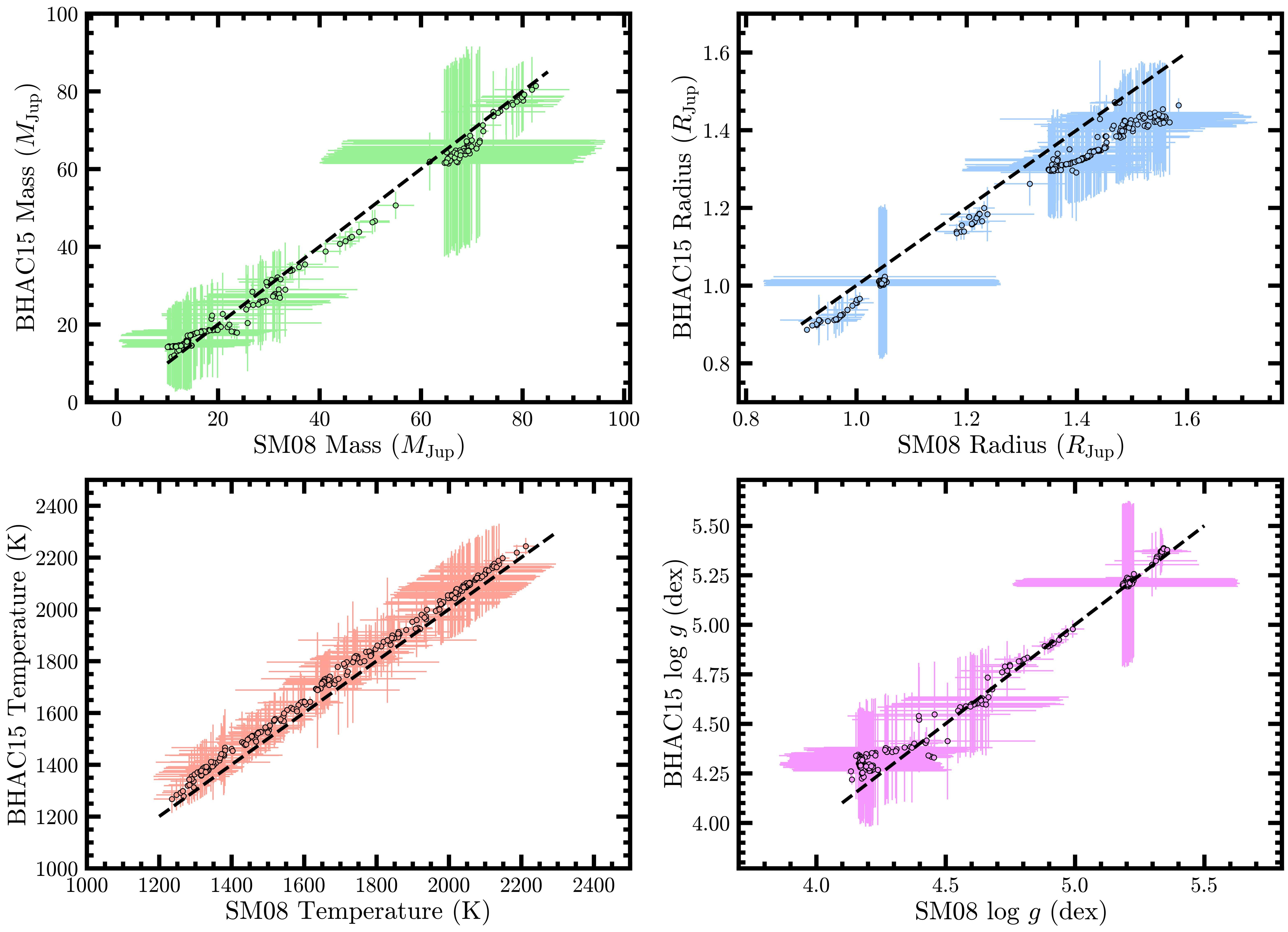}
    \caption{Comparison of masses, radii, effective temperatures, and surface gravities derived with BHAC15 and SM08 evolutionary models. The dashed black line in each panel represents a 1:1 relation between the compared parameters. While the parameters obtained are statistically consistent, we find that BHAC15 generally predicts higher temperatures and smaller radii than SM08.}
    \label{fig:evo}
\end{figure*}

\section{Ages}
\label{sec:ages}
In order to compute properties of ultracool dwarfs from evolutionary models, we must first estimate ages for all the objects. There is no single definitive method for doing this, so we employ multiple methods, making best use of the objects' available measurements and previous assessments in the literature. We follow a tiered approach to assigning ages, with the following priority order:

\begin{enumerate}

\item {\em Membership in a stellar association or moving group:} The BANYAN~$\Sigma$ algorithm \citep{2018ApJ...856...23G} provides a means for uniform probability assessment of objects' memberships in stellar associations or young moving groups (YMGs) within 150~pc based on kinematics ($UVW$) and position ($XYZ$), including handling objects that are missing parallax and/or radial velocity (RV) measurements. We process our entire sample using this algorithm, using the highest S/N astrometry for each object as compiled in {\em The UltracoolSheet}. An object is designated as a YMG member if the BANYAN~$\Sigma$ membership probability is $\ge$90\% and then assigned the corresponding YMG age tabulated in \citet{2018ApJ...856...23G}, assuming a normal age distribution for ages originating from \citet{2015MNRAS.454..593B} and a uniform age distribution for all other ages. (We note that BANYAN~$\Sigma$ is a probabilistic assessment based on an assumed spatial-kinematic model of the solar neighborhood, and individual objects of interest might benefit from greater scrutiny using additional methods.) In order to retain information about objects that might be members, especially with the addition of missing measurements (typically RVs), we assign potential membership for objects with probabilities of 80--90\% and tag these with a ``?'' suffix (e.g. ``ABDMG?'' indicates potential membership in the AB Dor moving group). This suffix is also used for objects whose BANYAN~$\Sigma$ analysis was done with a photometric distance rather than a parallactic one. Finally, for objects where the BANYAN~$\Sigma$ calculation was done with the full 6-d information, we append a ``!'' suffix (e.g., ``ABDMG!'') to indicate the greater robustness of this membership assignment as compared to most objects, where only 5-d information is available since RV measurements are much less common than astrometry and proper motions. In total, 61 objects are designated as YMG members based on the $\ge$90\% probability + parallactic distance criterion (for 22 out of these 61 objects, the full 6-d information is available). 30 additional objects are designated as potential YMG members based on the 80--90\% probability criterion. 3 objects with a $\ge$90\% BANYAN~$\Sigma$ probability are also designated as potential YMG members based on the use of a photometric distance.

\item {\em Companionship to a star}: For ultracool companions with spectral types L6 and later, we use the literature compilation of ages for the primary stars \citep{2001AJ....121.3235K, 2005AN....326..974S, 2005ApJS..159..141V, 2006PASP..118..671R, 2007ApJ...654..570L, 2010ApJ...725.1405B, 2012ApJ...755...94D, 2012ApJ...757..100D, 2012MNRAS.422.1922P, 2014ApJ...792..119D, 2017MNRAS.467.1126D, 2017AJ....154..262M, 2019MNRAS.487.1149G, 2020ApJ...891..171Z, 2021ApJ...911....7Z, 2021ApJ...916...53Z}. For 7~companions in our sample, {\em The UltracoolSheet} does not yet include the literature host star ages, so these objects are not included in the evolutionary model analysis that follows. In total, 78 objects are assigned ages based on the age of the primary star.

\begin{figure*}
    \centering
    \includegraphics[scale=0.6]{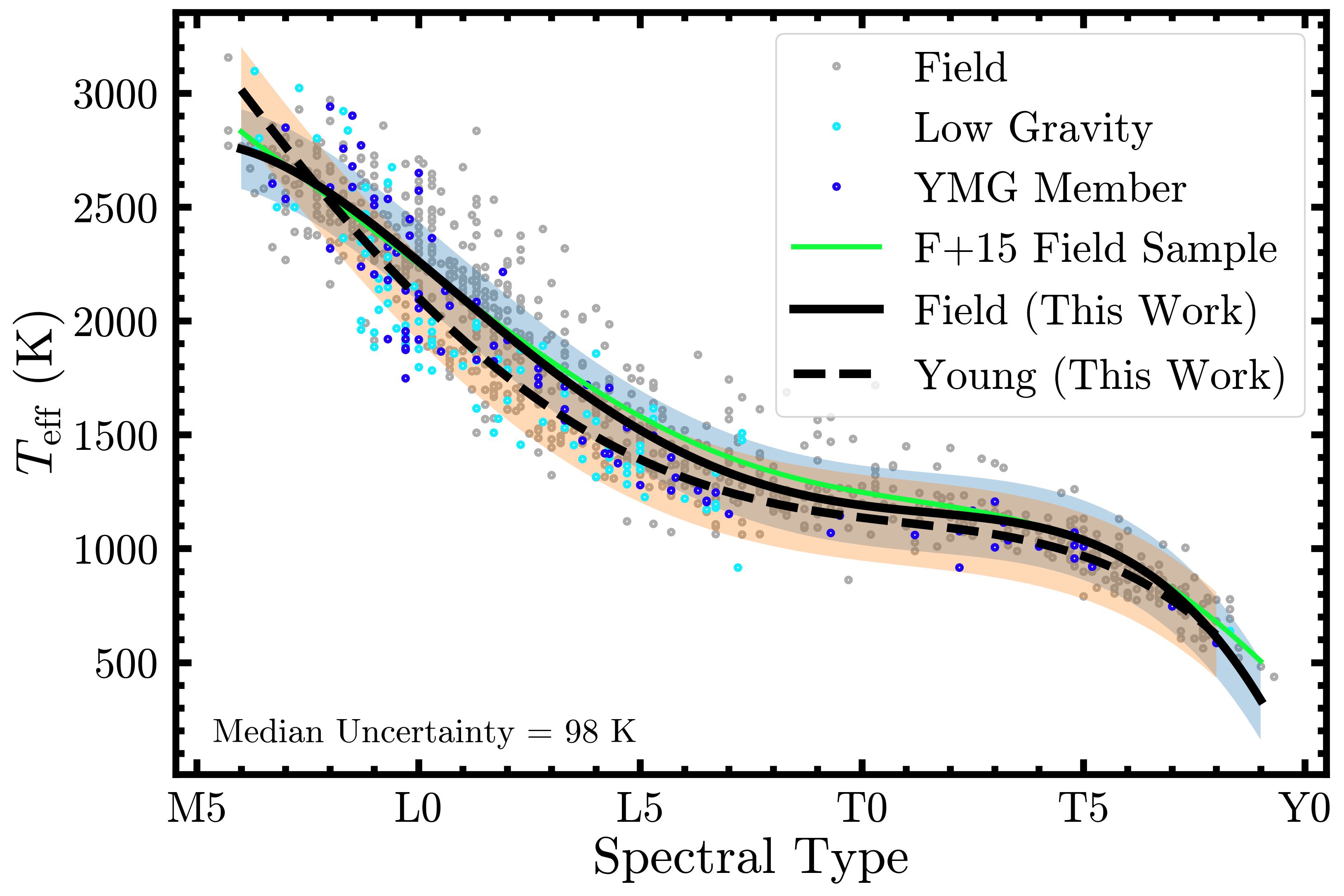}
    \caption{Effective temperatures derived for our sample of \nobject ultracool dwarfs as a function of their spectral type (SpT). Symbols are the same as in Figure \ref{fig:lbol-spt}. Random scatter in the x direction with amplitude 0.3 SpT is applied to avoid overlapping points. The black solid and dashed lines are the weighted best-fit polynomial relations for field and young objects, respectively. The shaded blue and orange regions represent the rms scatter about the field and young object fits, respectively. The field polynomial relation from \citet[][F+15]{2015ApJ...810..158F} is plotted as a green solid line for comparison.}
    \label{fig:teff-spt}
\end{figure*}

\item {\em Spectroscopic gravity:} Spectroscopic signatures of
low-gravity in either optical \citep[e.g.][]{2009AJ....137.3345C} or near-IR \citep[e.g.][]{2013ApJ...772...79A} spectra indicate that an object is young enough ($\lesssim$300~Myr) such that its radius is inflated compared to older (higher-gravity) field objects. We use the set of objects with low-gravity spectra that reside in young moving groups and stellar associations (as determined above with BANYAN~$\Sigma$) to examine the age range corresponding to the different gravity classes. We find objects with \vlg\ near-IR spectra (18 objects) are present in regions with ages of 10--150~Myr (bounded by the TW Hya Association and the AB~Dor moving group\footnote{Younger sources in star-forming regions, e.g., Taurus and Upper Sco, also show \vlg\ spectra \citep[e.g.][]{2018ApJ...858...41Z} but since our sample does not include such regions, we take 10~Myr as the lower age bound for this gravity class.}) while objects with \intg\ near-IR spectra (42 objects) span a range of 20--250 Myr (bounded by the $\beta$~Pic moving group and the cooling age for the \intg\ binary system LP~349-25 [\citealp{2017ApJS..231...15D}]). We assume uniformly distributed age values within each range. Each gravity class does not correspond to a unique age range, as a given moving group may have sources with both gravity classes \citep[e.g.][]{2016ApJ...821..120A}. For objects with optical gravity classifications, we treat the $\beta$ classification (8 objects) as equivalent to \intg\ and the $\delta$ (1 object) and $\gamma$ (13 objects) classifications as equivalent to \vlg. For cases where the optical and near-IR gravity classes are both young but do not agree, or the optical gravity class is uncertain, we adopt an age range that encompasses both gravity classifications: 10--250~Myr for $\beta$+$\vlg$ (2 objects), $\gamma$+$\intg$ (3 objects), and $\beta$+$\gamma$ (1 object), except for the few objects with $\beta$+\fldg\ (4 objects) where we assume 20--250~Myr. For objects with low-gravity spectral signatures but no formal gravity class in the literature (designated as low \emph{g}), we adopt a uniform distribution of 10--300~Myr (1 object). Finally for objects with \fldg\ gravity (148 objects), we assume the same field-age distribution from \citet{2017ApJS..231...15D} described below but with an elevated lower bound of 300~Myr.

\item {\em Field age:} For all remaining objects, i.e., those lacking
evidence of youth or ages based on group membership or companionship (634 objects),
we assign an age drawn from the distribution of
\citet[][hereafter DL17]{2017ApJS..231...15D}. The DL17 age distribution was empirically determined based on high precision dynamical mass measurements of 10 ultracool dwarf binaries and is found to be consistent with the Besançon model of the solar neighborhood \citep{2003A&A...409..523R}. The Besançon model predicts a field age distribution skewed towards younger ages since dynamical heating preferentially excites older objects out of the midplane of the Galaxy. According to DL17 age distribution, the ages are uniformly distributed in the following Besançon model bins, where each bin contains a fraction of the total population as follows: 8.1\% for 0.01--0.15 Gyr, 20.0\% for 0.15--1 Gyr, 16.1\% for 1--2 Gyr, 11.9\% for 2--3 Gyr, 16.6\% for 3--5 Gyr, 12.9\% for 5--7 Gyr, and 14.4\% for 7--10 Gyr. We adopt this as an age probability density function for each field object.

\end{enumerate}

The number of ages obtained with each method is summarized in the left panel of Figure \ref{fig:age-dist}. We derive the age distribution of our ultracool dwarf sample using Monte Carlo simulations. $10^5$ ages are drawn for each object from their assigned age distributions. We construct a sample age distribution spanning 0--10 Gyr with a bin size of 0.4 Gyr from the drawn ages of all objects in each trial. The mean age distribution is derived by averaging the density values across the $10^5$ sample age distributions in each bin. Uncertainty in the density value of each bin is obtained using the standard deviation. The right panel of Figure \ref{fig:age-dist} shows the mean age distribution with uncertainties. The median age of the sample is 1.9 Gyr.

\begin{figure}
    \centering
    \includegraphics[scale=0.33]{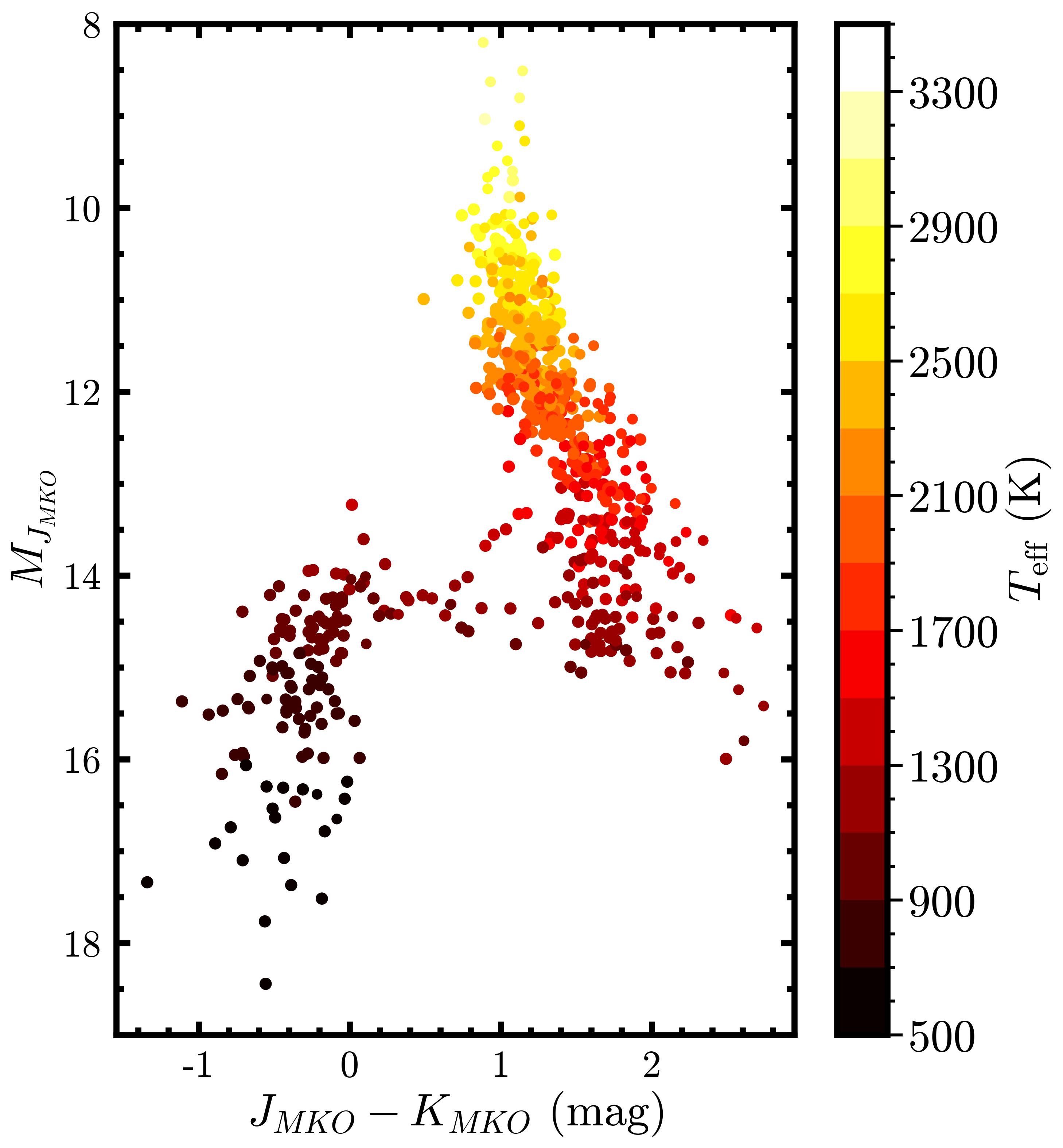}
    \caption{MKO $M_J$ vs $J - K$ color-magnitude diagram for ultracool dwarfs in our sample with individual points colored by their derived effective temperatures. Spectral blends and overluminous objects have been excluded. Due to the large number of points, uncertainties in individual measurements are not shown for figure readability. The median uncertainty in $M_{J_\mathrm{MKO}}$ is 0.06 mag and in $J - K$ is 0.05 mag.}
    \label{fig:cmd}
\end{figure}

\section{Evolutionary Model-Derived Parameters}
\label{sec:evo-sample}
In this section, we use the bolometric luminosities (Section \ref{sec:lbol}) derived for all objects in our sample along with their age estimates (Section \ref{sec:ages}) as inputs to the \citet[][SM08]{2008ApJ...689.1327S} hybrid and \citet[][BHAC15]{2015A&A...577A..42B} evolutionary model grids to estimate their masses, radii, surface gravities, and effective temperatures. Ultracool dwarfs in our sample span a wide range of luminosities and thus, in some cases, are too bright for the SM08 models ($\mathrm{log}(L_{\mathrm{bol}}/L_\odot) \gtrsim -3$) or too faint for the BHAC15 models ($\mathrm{log}(L_{\mathrm{bol}}/L_\odot) \lesssim -4.5$). We use each evolutionary model grid in its respective non-overlapping luminosity-age space. In the region of model overlap, we derive the fundamental properties with both sets of models.

\subsection{Bayesian Rejection Sampling}
\label{sec:brj}
To estimate ultracool dwarf masses, radii, and surface gravities by interpolation of evolutionary model grids, we implement the Bayesian rejection sampling method described in \citet{2017ApJS..231...15D}. The first step is to randomly draw $10^6$ log $L_{\mathrm{bol, model}}$ samples from a uniform distribution spanning the bolometric luminosity range of the evolutionary model grid. The second step involves randomly sampling the object's age distribution, calculating $\chi^2$ values for the sampled ages, and computing a probability for each sample. The exact implementation of this step depends on the adopted age distribution.

\emph{Normal Age Distribution:} $10^6$ $\mathrm{log}\; t_{\mathrm{model}}$ samples are randomly drawn from a uniform distribution spanning the age range of the evolutionary model grid. We compute $\chi^2$ for each sample, 
\begin{equation}
    \chi^2 = \frac{(\mathrm{log}\;L_{\mathrm{\mathrm{bol, model}}} - \mathrm{log}\; L_{\mathrm{bol}})^2}{\sigma_{\mathrm{log}\;L_{\mathrm{bol}}}^2} + \frac{(\mathrm{log}\;t_{\mathrm{model}} - \mathrm{log}\; t)^2}{\sigma_{\mathrm{log}\;t}^2},
\end{equation}
where $\mathrm{log}\; L_{\mathrm{bol}} \pm \sigma_{\mathrm{log}\;L_{\mathrm{bol}}}$ is the bolometric luminosity measurement and $\mathrm{log}\; t \pm \sigma_{\mathrm{log}\;t}$ is the age measurement for the object. The $\chi^2$ for each sample is converted to a probability $p$, normalizing by the sample with the minimum $\chi^2$, as follows, 
\begin{equation}
\label{eq:prob}
    p = \mathrm{exp} \left[-\frac{1}{2}\left(\chi^2 - \chi^2_{\mathrm{min}}\right)\right].
\end{equation}

\emph{Uniform Age Distribution:} $10^6$ $\mathrm{log}\; t_{\mathrm{range}}$ samples are randomly drawn from a uniform distribution spanning the intersection of the quoted age range and the evolutionary model grid age range. We compute the $\chi^2$ for each sample, in this case, as follows, 
\begin{equation}
\label{eq:chi2}
    \chi^2 = \frac{(\mathrm{log}\;L_{\mathrm{bol, model}} - \mathrm{log}\; L_{\mathrm{bol}})^2}{\sigma_{L_{\mathrm{bol}}}^2}.
\end{equation}
The $\chi^2$ for each sample is converted to a probability $p$, normalizing by the sample with the minimum $\chi^2$ using Equation \ref{eq:prob}. 

\begin{figure*}[t]
    \centering
    \includegraphics[scale=0.4]{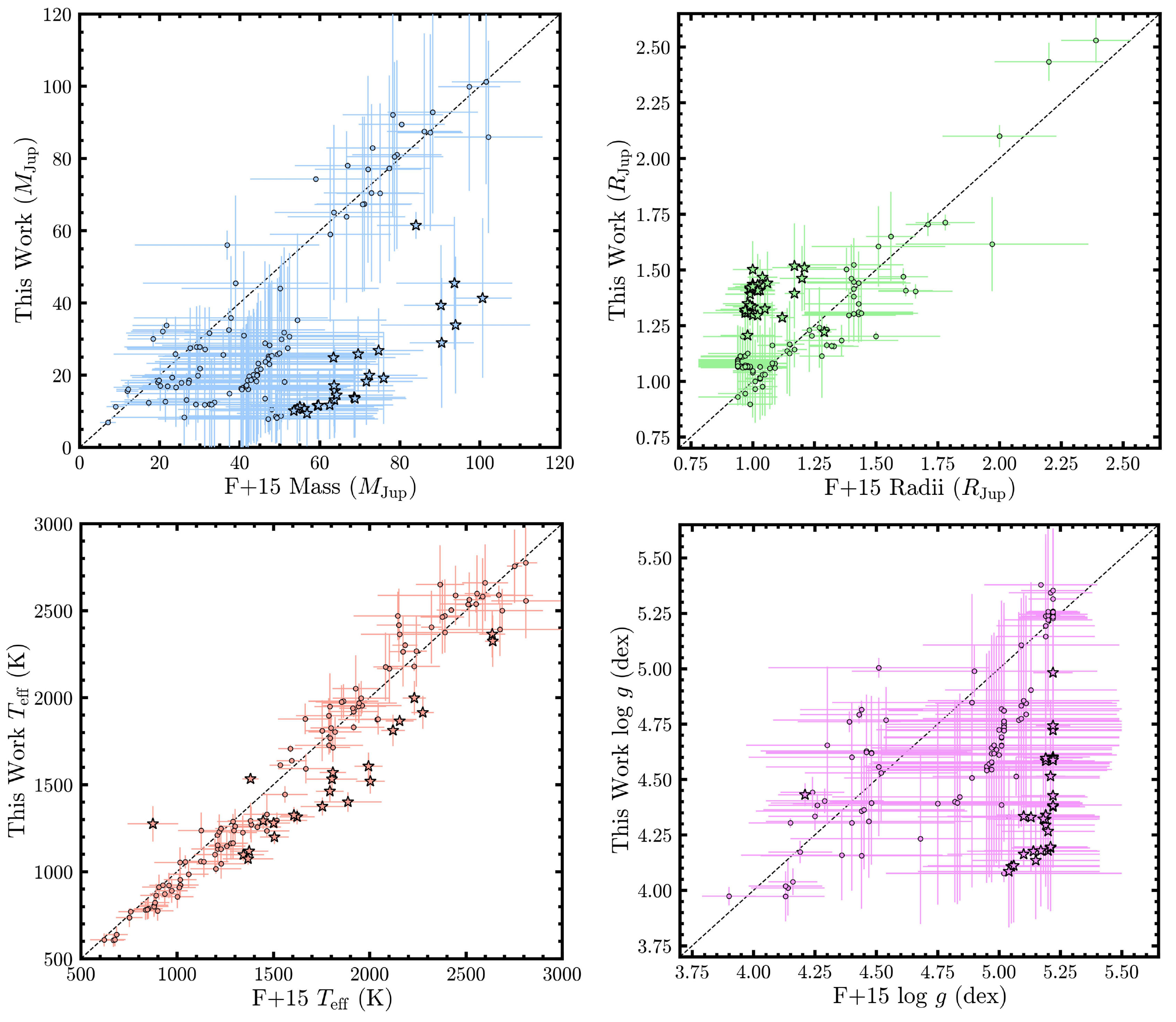}
    \caption{Comparison of mass, radius, effective temperature, and surface gravity derived in this work with those in \citet[][F+15]{2015ApJ...810..158F} for objects common to the two samples. The dashed black line in each panel represents a 1:1 relation between the compared parameters. The parameters obtained are statistically consistent for the large majority of objects. Objects with disagreements ($>$2$\sigma$ discrepancy) in the estimated values are prominently marked with a black-bordered star.}
    \label{fig:phys-comp-f15}
\end{figure*}

\begin{figure*}
    \centering
    \includegraphics[scale=0.45]{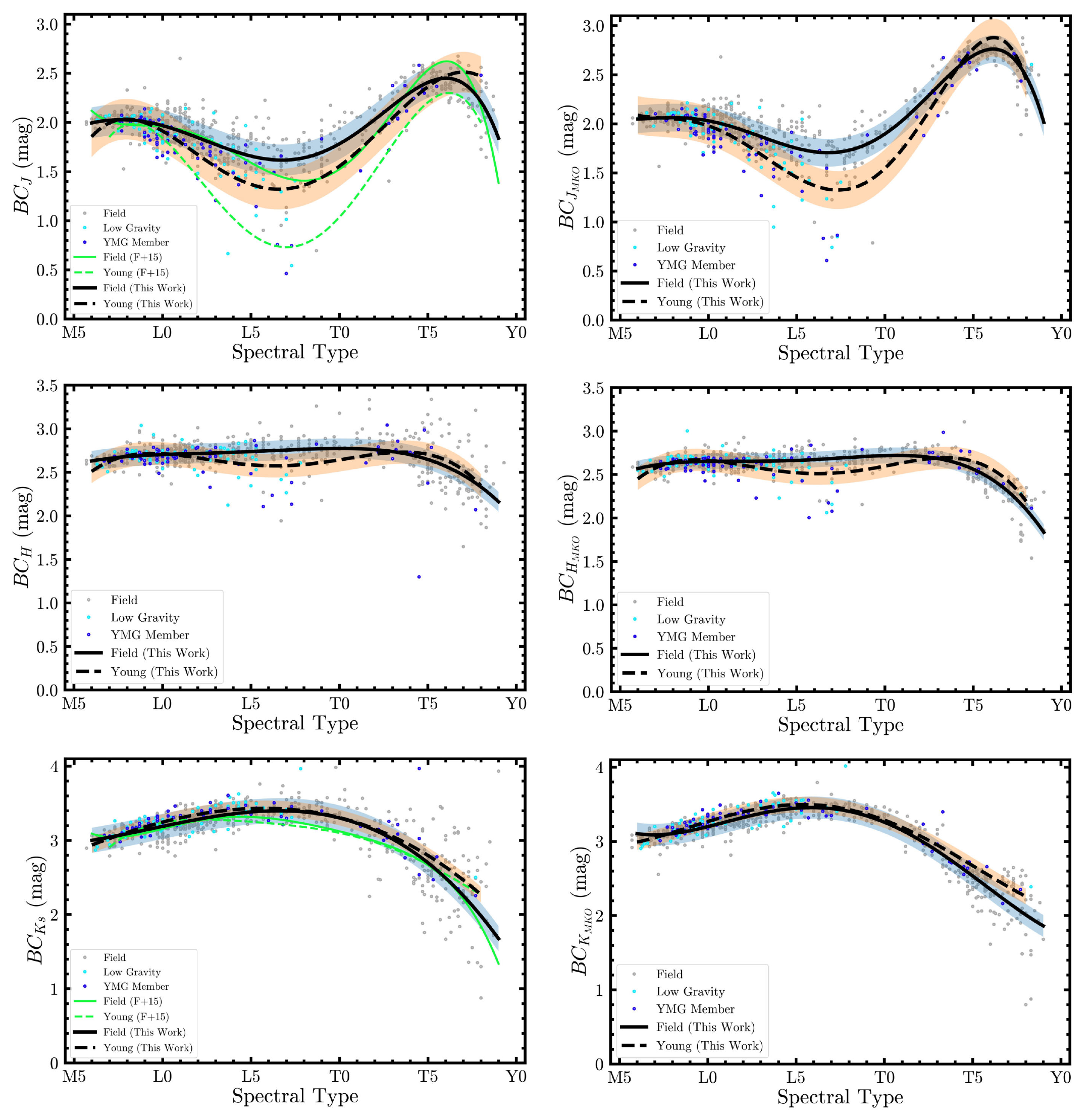}
    \caption{Bolometric corrections in the 2MASS and MKO near-IR bands for ultracool dwarfs in our sample as a function of spectral type (SpT). Random scatter in the x direction with amplitude 0.3 SpT is applied to avoid overlapping points. Symbols are the same as in Figure \ref{fig:lbol-spt}. Due to the large number of points, uncertainties in individual measurements are not shown for figure readability. The median uncertainty of measurements in each panel presented is 0.1 mag. The black solid and dashed lines are the best-fit polynomial relations for field and young objects, respectively. The blue and orange shaded regions are the rms scatter about the field and young object relations, respectively. The field and young object relations from \citet{2015ApJ...810..158F} are plotted as green solid and dashed lines, respectively, for comparison.}
    \label{fig:nir-bcs}
\end{figure*}

\begin{figure*}
    \centering
    \includegraphics[scale=0.32]{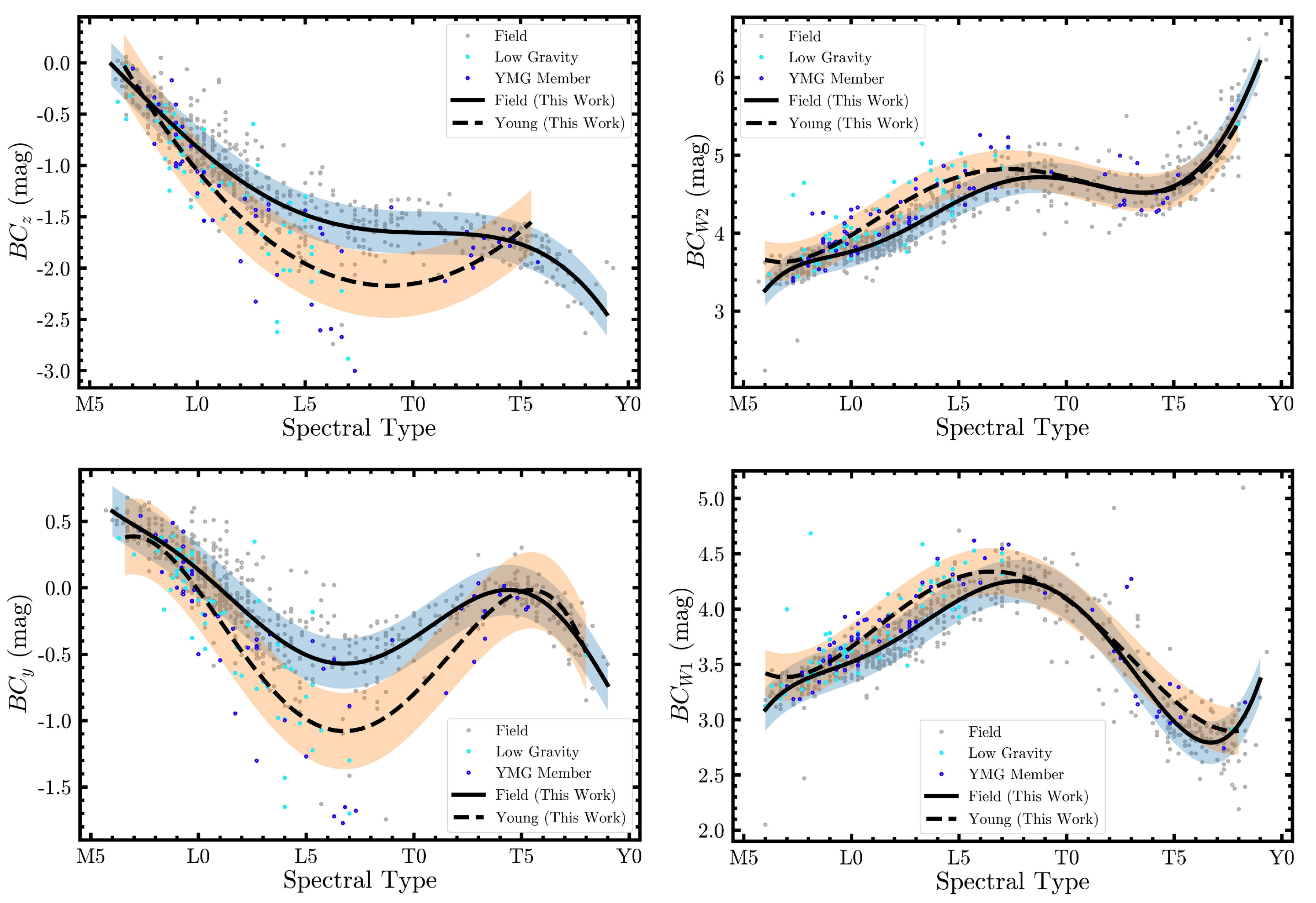}
    \caption{Bolometric corrections in the PS1 optical and WISE mid-infrared bands for ultracool dwarfs in our sample as a function of spectral type (SpT). Figure elements are same as in Figure \ref{fig:nir-bcs}. The median uncertainty of measurements in each panel presented is 0.1 mag.}
    \label{fig:mir-opt-bcs}
\end{figure*}

\emph{DL17 Age Distribution:} $10^6$ $\mathrm{log}\; t_{\mathrm{DL17}}$ samples are randomly drawn from a uniform distribution spanning the intersection of the DL17 age range and the evolutionary model grid age range. In each case, the $\chi^2$ for each sample (Equation \ref{eq:chi2}) is converted to a probability $p_*$, normalizing by the sample with the minimum $\chi^2$ using Equation \ref{eq:prob}. We then compute the likelihood of drawing the sampled age from the DL17 distribution (truncated at 300 Myr) for field-age (\fldg) objects $p_{\mathrm{DL17}} = p(\mathrm{log}\; t_{\mathrm{DL17}} | \mathrm{DL17})$. The final probability for each sample, normalized by the sample with the highest probability, is,
\begin{equation}
\label{eq:prob2}
    p = \frac{p_* \cdot p_{\mathrm{DL17}}}{\mathrm{max}(p_* \cdot p_{\mathrm{DL17}})}
\end{equation}

The third step is to randomly draw $10^6$ uniform variates ($u$) distributed in the range from 0 to 1 and reject any samples where $p < u$. The fourth and final step is to linearly interpolate the evolutionary models at each accepted luminosity-age point to calculate the corresponding mass ($\mathrm{log}\;M$), radius ($\mathrm{log}\;R$), and surface gravity (log $g$). The final mass/radius/gravity measurement and uncertainty are obtained as the median and standard deviation of all the sample's mass/radius/gravity values. Our measurements are summarized in the Table of Ultracool Fundamental Properties associated with this paper (see \S\ref{sec:intro}). A total of 243 sources' parameters were estimated using both BHAC15 and SM08, 339 sources' parameters were estimated using only BHAC15, and 464 sources' parameters were estimated using only SM08.

\begin{figure}[t]
    \centering
    \includegraphics[scale=0.085]{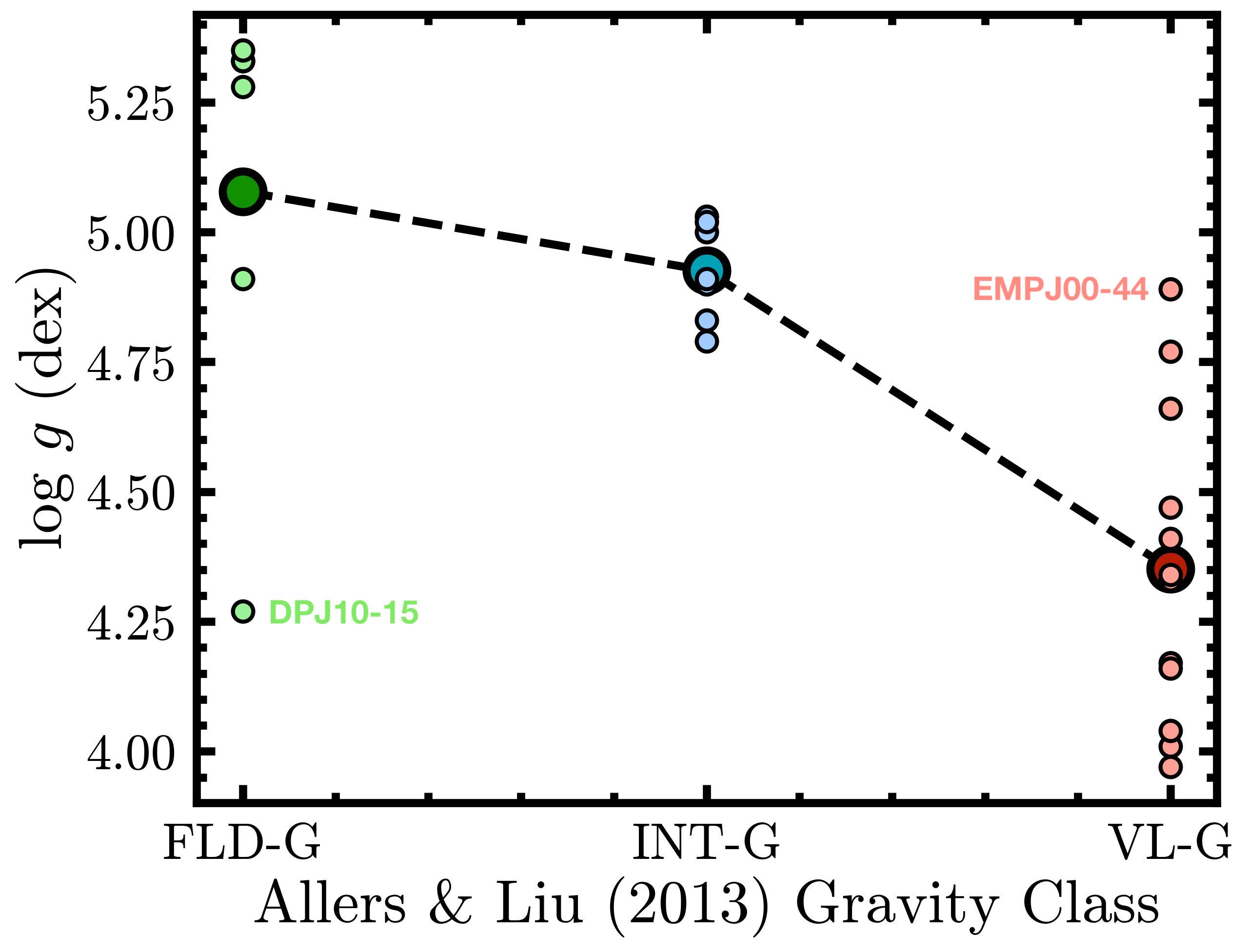}
    \caption{Evolutionary model-derived surface gravity distribution for different near-infrared gravity classifications. The smaller size markers represent individual object measurements and the larger size markers represent the mean surface gravity for a given gravity classification. DENIS-P J1058.7-1548 (DPJ10-15) and EROS-MP J0032-4405 (EMPJ00-44) are significant outliers. The median uncertainty of surface gravity measurements presented here is 0.04 dex. The black dashed line connects the mean surface gravities and serves as a visual aid to identify the trend. While individual groups exhibit varying levels of dispersion in their log $g$ values, the index-based NIR gravity classification scheme accurately groups objects with systematic differences in their evolutionary model-derived surface gravities.}
    \label{fig:surfgrav}
\end{figure}

\begin{figure*}[t]
    \centering
    \includegraphics[scale=0.6]{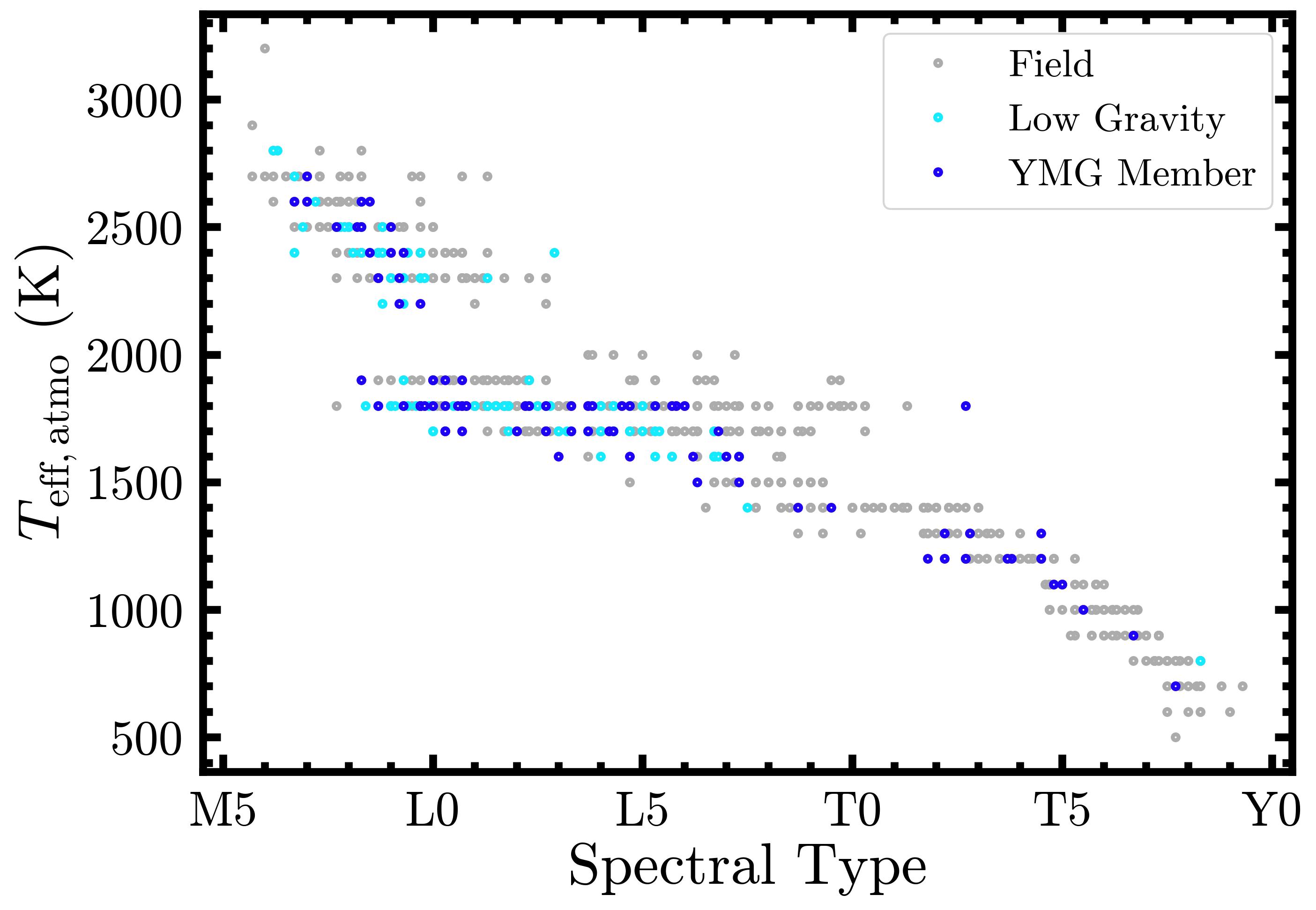}
    \caption{Atmospheric model-derived effective temperatures for objects in our sample as a function of spectral type. The symbols are similar to those described in Figure \ref{fig:teff-spt}. A pile-up of objects at $\approx$1800 K creates two gaps at higher and lower effective temperatures.}
    \label{fig:atmo-temp-spt}
\end{figure*}

The choice of a given model, SM08 or BHAC15, for the adopted measurements is based on the following conditions. BHAC15 is used if the $1\sigma$ lower limit of the measured bolometric luminosity is greater than the luminosity on the lowest mass luminosity-age model track interpolated at the $1\sigma$ lower limit of the measured age (or lower limit of the quoted age range). SM08 is used if the $1\sigma$ upper limit of the measured bolometric luminosity is less than the luminosity on the highest mass luminosity-age model track interpolated at the $1\sigma$ upper limit of the measured age (or upper limit of the quoted age range). When both conditions are satisfied, calculations are made using both evolutionary models. In the Table of Ultracool Fundamental Properties associated with this paper (see \S\ref{sec:intro}), we present the lower uncertainty measurement in such cases. Figure \ref{fig:evo} presents a comparison of the fundamental parameters derived between BHAC15 and SM08. While the parameters obtained with BHAC15 and SM08 are statistically consistent, we find that BHAC15 generally predicts higher temperatures (by $\approx$50 K) and smaller radii (by $\approx$ 0.06 $R_{\mathrm{Jup}}$) than SM08.

This method significantly differs from the process implemented in \citet{2015ApJ...810..158F}. To estimate an object's radius (which is used to derive effective temperature) and mass, \citet{2015ApJ...810..158F} determine a parameter range for each source spanning the minimum and maximum values of all evolutionary model predictions given an object's bolometric luminosity and age. They quote the average of the minimum and maximum value as the final radius/mass and half the radius/mass range as the uncertainty in Table 9. However, this method does not take into account the shape of the underlying radius/mass distribution when estimating the object's properties. Except for the case of a uniform distribution, the average parameter value does not correspond to the distribution's true expectation value. This is further complicated by the use of parameter estimates from several sets of evolutionary models to construct a single range in \citet{2015ApJ...810..158F}. Our Bayesian rejection sampling technique addresses the limitations of the above method by deriving each object's underlying radius and mass distribution independently for each of the two evolutionary model sets in this work. The median value computed from the parameter's cumulative distribution function is then quoted as our final value.

\subsection{Effective Temperature}
Effective temperatures are calculated using the evolutionary model-inferred radius $R$ and SED-integrated bolometric luminosity measurement $L_{\mathrm{bol}}$ using the Stefan-Boltzmann Law,
\begin{equation}
    T_{\mathrm{eff}} = \left(\frac{L_{\mathrm{bol}}}{4\pi R^2\sigma_{\mathrm{SB}}}\right)^{\frac{1}{4}},
\end{equation}
where $\sigma_{\mathrm{SB}}$ is the Stefan-Boltzmann constant. Figure \ref{fig:teff-spt} presents the calculated effective temperatures as a function of spectral type along with our best-fit polynomial relations for field and young (low gravity and YMG member) objects separately (Table \ref{table:poly}). Our field relation is in good agreement with the corresponding field polynomial relation derived in \citet{2015ApJ...810..158F}. Figure \ref{fig:cmd} demonstrates the temperature evolution of ultracool dwarfs as a MKO NIR color-magnitude diagram for objects in our sample with derived fundamental parameters.

\subsection{Comparison with \citet{2015ApJ...810..158F}}
\label{sec:comp-f15-prop}
Figure \ref{fig:phys-comp-f15} compares the masses, radii, effective temperatures, and surface gravities calculated in this work with those in \citet{2015ApJ...810..158F}. We find that our results are consistent ($<$2$\sigma$ difference) with \citet{2015ApJ...810..158F} for the majority of objects (N = 91). $>$2$\sigma$ discrepancy in at least one of the above fundamental properties is observed for 37 objects (Figure \ref{fig:phys-comp-f15}). A complete list of the above objects is provided in Table \ref{table:prop_comp}. 

The primary reason for the observed disagreements is a revision in the age distribution adopted for the objects. For field (\fldg) objects, \citet{2015ApJ...810..158F} adopt a uniform age distribution in the range 0.5--10.0 Gyr. We adopt the (truncated) DL17 age distribution which skews towards younger ages: the greatest probability density in this age distribution lies in the 0.15--1 Gyr bin and the median age is $\sim$2.5 Gyr. Consequently, our derived radii for these objects are generally larger, and our derived masses, temperatures, and gravities are generally smaller. This is the case for 22 out of the 37 discrepant objects. For 11 objects assigned a field age (0.5--10 Gyr) in \citet{2015ApJ...810..158F}, there have been updates to their YMG membership or spectroscopic gravity class. In this work, 2MASS J06244595-4521548, 2MASSW J0030300-145033, 2MASSI J0451009-340214, and DENIS-P J1058.7-1548 are designated as Argus members; 2MASSI J1010148-040649 is designated as a Carina-Near member; and LHS 3003 is designated as an AB Doradus member based on results from BANYAN~$\Sigma$. 2MASS J13595510-4034582, DENIS-P J0652197-253450, and LHS 132 are assigned a $\beta$ optical gravity classification, and 2MASSI J2057540-025230 and Teegarden's Star are assigned an \intg\;NIR gravity classification. Consequently, our derived masses, temperatures, and gravities for these objects are generally smaller, and our derived radii are generally larger. For 2 objects, 2MASS J03264225-2102057 and 2MASSW J2244316+204343, the YMG membership (AB Doradus) does not change but the source of our adopted ages differ. \citet{2015ApJ...810..158F} use an age range of 50--120 Myr from \citet{2013ApJ...762...88M}, whereas we use an older age of $149^{+51}_{-19}$ Myr from \citet{2015MNRAS.454..593B}. Consequently, our derived radii for these objects are generally smaller, and our derived masses, temperatures, and gravities for these objects are generally larger.

Two objects remain to be accounted for: 2MASSW J2208136+292121 and 2MASS J00332386-1521309. 2MASSW J2208136+292121 meets the consistency threshold for mass, radius, and gravity. We derive a lower effective temperature that is marginally inconsistent at the 2.1$\sigma$ level. The age range adopted for this object is the same between \citet{2015ApJ...810..158F} and this work. 2MASS J00332386-1521309 meets the consistency threshold for mass, radius, and gravity. The effective temperature is inconsistent at the 2.8$\sigma$ level. \citet{2015ApJ...810..158F} adopt an age range 100--150 Myr whereas we adopt a truncated DL17 age distribution (lower bound of 300 Myr) based on its \fldg\; classification. This corresponds to an older age assignment compared to \citet{2015ApJ...810..158F}. However, our temperature estimate is lower than that of \citet{2015ApJ...810..158F}, not higher as it would be expected. These two disagreements arise because of the difference in bolometric luminosity measurements (due to changes in parallax measurement). Both objects can be found in Table \ref{table:lbol_comp}. We measure lower bolometric luminosities for 2MASSW J2208136+292121 and 2MASS J00332386-1521309, which explain the lower effective temperatures derived for both objects.

More generally, the differences observed between our evolutionary model-derived parameter estimates and those in \citet{2015ApJ...810..158F} may also be a result of the different estimation methods employed, as described in \S\ref{sec:brj}.

\section{Discussion}
\label{sec:Discussion}

\subsection{Bolometric Corrections}
\label{sec:BC}
We calculate bolometric corrections spanning the optical to MIR filters for both field and young (low gravity and young moving group members) objects and determine empirical relationships as a function of spectral type following the process described in Section \ref{sec:lbol}. For our calculations, we use $M_{bol, \odot} = 4.74$. We note here that the \emph{F}-test does not always yield a definitive answer for the order of the polynomial based on the $p > 0.05$ rejection criterion. For such cases, we adopt the lowest order after which $p$ does not change significantly (factor of $\sim$5-10) as the best-fit polynomial, which is then validated by visual inspection. Figure \ref{fig:nir-bcs} presents our bolometric corrections for the 2MASS and MKO NIR filters as a function of spectral type. Figure \ref{fig:mir-opt-bcs} presents our bolometric corrections for the PS1 $z$ and $y$ optical filters and the WISE $W1$ and $W2$ MIR filters as a function of spectral type. The derived polynomial relations are presented in Table \ref{table:poly}. 

Next, we discuss some key observations. First, based on the rms about the fit, the NIR relations generally show less scatter than the optical and MIR relations and thus are preferable for bolometric magnitude calculations. Second, within the NIR relations, the bolometric corrections generally show smaller rms in the MKO filters than the 2MASS filters, likely due to the higher precision MKO photometry for T-dwarfs in the sample. Finally, we find that both the 2MASS \Ks and MKO $K$-band bolometric corrections are relatively insensitive to the age of the ultracool dwarf and are thus most appropriate for cases when no age information is available.

We find a few significant differences in the 2MASS $J$ and \Ks polynomial relations on comparison with \citet{2015ApJ...810..158F}. Our field relation in the 2MASS $J$ band predicts a higher bolometric correction (by $\approx 0.1-0.3$ mag) in the spectral range $\sim$L5--T2. The change can be attributed to the larger number of L/T transition objects in our work compared to \citet{2015ApJ...810..158F}. This enables a better constraint on the polynomial relation. The difference in the young object relations in the 2MASS $J$ band is even more striking. Our best-fit polynomial predicts a bolometric correction up to $\approx$0.7 mag higher than that in \citet{2015ApJ...810..158F} across spectral types $\sim$L2--T6. A careful comparison of Figure 12 in \citet{2015ApJ...810..158F} with the top left panel of Figure \ref{fig:nir-bcs} leads us to conclude that the larger number of young L2--L5 dwarfs with bolometric corrections $> 1.5$ mag serves to anchor our fit and is responsible for the differing solutions. However, in our measurements, we observe that the bolometric corrections of young objects in the spectral range L5--L7 steeply drop to $\approx$1 mag, a feature not reproduced in our best-fit relation. This is also the case for our MKO $J$ band young object relation. Our results suggest that a more appropriate empirical relationship may lie between the solutions found in \citet{2015ApJ...810..158F} and this work or by adopting the direct averages as a function of spectral type. We note that the key to determining this will be adding more young objects in the range L7--T2, where, presently, there is a noticeable gap in our measurements. Until then, caution should be exercised in using the 2MASS and MKO $J$ band young object relations at the L/T transition. The 2MASS \Ks and MKO $K$-band young relations are more tightly constrained and are the recommended alternatives. Finally, in the 2MASS \Ks band, we observe slight discrepancies between the field relations, within the calculated rms scatter, in the spectral range L5--T2. This can be similarly explained by the larger number of L/T transition object measurements in this work. The \citet{2015ApJ...810..158F} 2MASS \Ks band field relation diverges for T9 dwarfs compared to the one in this work and is likely an edge effect due to the small number of T9 objects in our sample.

\begin{figure*}[t]
    \centering
    \includegraphics[scale=0.25]{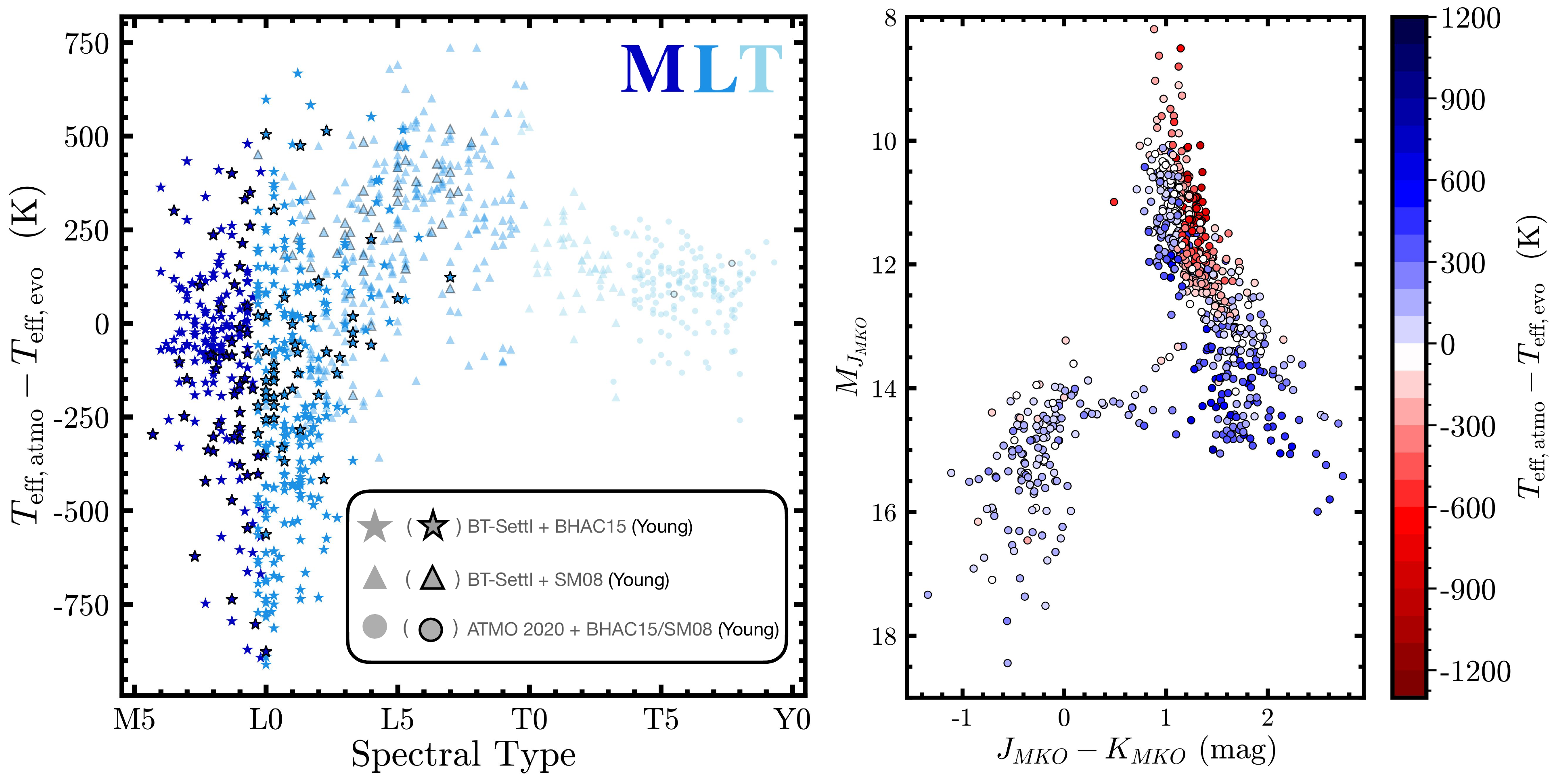}
    \caption{\emph{Left:} Difference between the the atmospheric model-derived effective temperatures and the evolutionary model-derived effective temperatures ($\Delta T_\mathrm{eff}$) as a function of spectral type (SpT). Random scatter in the x direction with amplitude 0.3 SpT is applied to avoid overlapping points. Objects are colored based on their spectral type where the darkest shade corresponds to M-dwarfs, the intermediate shade corresponds to L-dwarfs, and the lightest shade corresponds to T-dwarfs. Objects using the atmospheric-evolutionary model pairings of {\sc BT-Settl}---BHAC15, {\sc BT-Settl}---SM08, and {\sc ATMO} 2020--BHAC15/SM08 are marked with a star, triangle, and circle respectively. {\sc BT-Settl}---BHAC15 objects are presented with a higher color opacity than {\sc BT-Settl}---SM08 and {\sc ATMO} 2020--BHAC15/SM08 objects to emphasize the greater reliability of $\Delta T_\mathrm{eff}$ trends for the former (self-consistently computed models). Symbols with black outlines mark young objects based on signatures of low surface gravity. \emph{Right:} MKO $M_J$ vs $J - K$ color-magnitude diagram for ultracool dwarfs in our sample with each object colored by its corresponding $\Delta T_\mathrm{eff}$ value. Due to the large number of points, uncertainties in individual measurements are not shown for figure readability. The median uncertainty in $M_{J_\mathrm{MKO}}$ is 0.06 mag and in $J - K$ is 0.05 mag.}
    \label{fig:temp-model-diff}
\end{figure*}

\begin{figure*}[t]
    \centering
    \includegraphics[scale=0.27]{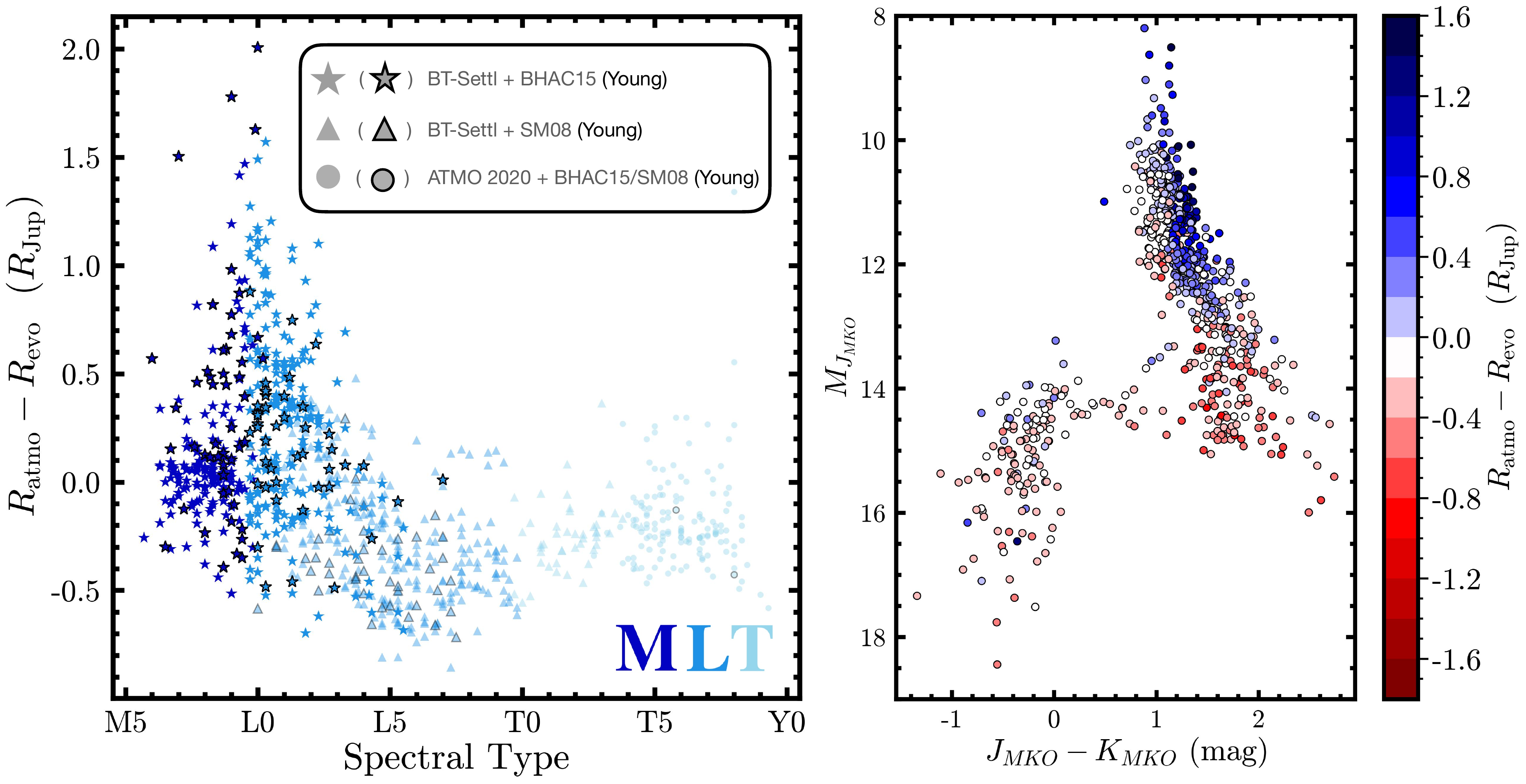}
    \caption{Same as Figure \ref{fig:temp-model-diff} but for difference between atmospheric and evolutionary model-derived radius measurements.}
    \label{fig:radius-model-diff}
\end{figure*}

\subsection{Evolutionary Model-Derived Surface Gravity vs Ultracool Dwarf NIR Gravity Classifications}

\citet{2013ApJ...772...79A} defined a set of near-infrared gravity-sensitive indices based on FeH, VO, K I, Na I, and H-band continuum shape and created a gravity classification scheme consisting of three classes: \fldg, \intg, and \vlg. Here, we examine the empirical distribution of the evolutionary model-derived surface gravity of ultracool dwarfs as a function of their NIR gravity classifications. For an accurate analysis, we exclude ultracool dwarfs with spectra $<$50 S/N in any of the following six bandpasses: $YJHK$ and two blue-half regions in the $H$ and $K$ bands (provided in the Table of Ultracool Fundamental Properties associated with this paper; see \S\ref{sec:intro}). We take care to only include low gravity objects for which the ages were determined based on the ages of their host stars or their membership to a YMG, independent of their NIR gravity classification. This was done to avoid circular logic in the analysis, since the assigned ages directly impact the estimation of surface gravities in our evolutionary model interpolation procedures. This results in 6 \fldg, 7 \intg, and 12 \vlg objects with evolutionary model-derived parameters from this work. These objects are plotted in Figure \ref{fig:surfgrav}. We observe varying levels of scatter in the evolutionary model-derived surface gravity values for each gravity class --- lowest for \intg, followed by \fldg, and highest for \vlg. The \fldg outlier at log $g = 4.27$ dex is DENIS-P J1058.7-1548, and the \vlg outlier at log $g = 4.89$ dex is EROS-MP J0032-4405. DENIS-P J1058.7-1548 is a candidate Argus member (95\% BANYAN-$\Sigma$ probability) and was thus assigned an age of $45 \pm 5$ Myr. Such a young age is uncharacteristic for a \fldg object and would explain the large discrepancy in log $g$ from the \fldg mean value. A comparison of DENIS-P J1058.7-1548's spectrum with L-dwarf \fldg standards does not reveal any discrepancies that could suggest a gravity misclassification and thus it is probable that the YMG assignment is incorrect. EROS-MP J0032-4405 is an AB Doradus (AB Dor) Moving Group member. The \intg classification is characteristic of AB Dor members (all 7 \intg objects in our analysis are AB Dor members). Given that the derived surface gravity for this object is consistent with the mean surface gravity in the \intg class, it may be possible that \intg is a more suitable gravity classification for the object. However, the object also possesses an optical gravity classification of $\gamma$ \citep{2009AJ....137.3345C}, reducing the probability that a NIR gravity misclassification has occurred. We investigated the possibility that this object is a spectral blend using the spectral indices of \citet{2010ApJ...710.1142B} and \citet{2014ApJ...794..143B}, but found that none of the criteria for binarity were met. The relatively higher scatter in the \fldg and \vlg classes may be an edge case artifact due to the limited number of classification groups and possibly highlights the need to add new gravity classes. This can be more reliably determined with a similar analysis consisting of a larger sample of objects. The primary factor limiting the sample size in this work is the lack of high S/N spectra (indicative of reliabilty of the index-based gravity classification) across all the six wavelength regions defined previously. Nevertheless, the mean surface gravities exhibit a trend consistent with the gravity classification scheme --- objects with a \fldg classification, on average, have higher surface gravities than those with an \intg classification which, in turn, have higher surface gravities, on average, than objects with a \vlg classification. 

\subsection{Differences between Atmospheric Model-derived and Evolutionary Model-derived Parameters}

Numerous works in the past have documented discrepant log $g$, $T_{\mathrm{eff}}$, and thus radii values between those derived by fitting atmospheric models to substellar SEDs and those derived by interpolating evolutionary models \citep[e.g.][]{2008ApJ...689..436L, 2009ApJ...692..729D, 2010ApJ...723..850B, 2011ApJ...733...65B, 2012ApJ...754..135M, 2013Sci...341.1492D, 2013ApJ...777L..20L, 2021ApJ...916...53Z, 2022arXiv220814990C}. Here, we take advantage of our large sample to study the noted differences as a function of spectral type and in the near-infrared color-magnitude diagram for ultracool dwarf effective temperatures, radii, and surface gravities.

\subsubsection{Effective Temperature}
\label{sec:teff_diff}
First, we investigate the atmospheric model-derived effective temperatures ($T_{\mathrm{eff, atmo}}$) vs spectral type relation and then systematically characterize the differences with evolutionary model-derived effective temperatures ($T_{\mathrm{eff, evo}}$). 

Figure \ref{fig:atmo-temp-spt} shows $T_{\mathrm{eff, atmo}}$ as a function of spectral type. A pile-up of objects in the temperature range $\approx$1700-1900 K is observed. This pile-up carves out gaps in the temperature sequence at $\approx$2000-2200 K and, less prominently, $\approx$1500-1600 K. Figure \ref{fig:temp-model-diff} investigates the effects of these features in greater detail by presenting the difference in $T_{\mathrm{eff, atmo}}$ and $T_{\mathrm{eff, evo}}$ measurements as a function of spectral type. A comparison between self-consistently computed atmospheric and evolutionary models is the most accurate choice for such an analysis. However, the availability of models only allows such a comparison for objects where we simultaneously employed the {\sc BT-Settl} and BHAC15 models ({\sc BT-Settl}---BHAC15). We find that late-M and early L-dwarfs ($\sim$M6--L2) primarily constitute our sample for this analysis (Figure \ref{fig:temp-model-diff}). Given that we do not observe a significant difference in the temperatures derived with BHAC15 vs SM08 (median difference $\approx$50 K; Figure \ref{fig:evo}), we are also able to perform plausible comparisons for the {\sc BT-Settl}--SM08 and {\sc ATMO} 2020--BHAC15/SM08 pairings. $\sim$L3--T2 dwarfs primarily form the sample for the former and $>$T2 dwarfs form the sample for the latter. 
\begin{enumerate}
    \item {\sc BT-Settl}---BHAC15 objects: In Figure \ref{fig:temp-model-diff}, we find that {\sc BT-Settl} generally underestimates the effective temperature compared to the BHAC15 evolutionary model, with the largest discrepancy (up to $\approx$800 K) at the M/L transition boundary. This discrepancy reflects the $\approx$1800 K pile-up of M/L transition objects in Figure \ref{fig:atmo-temp-spt} and the $\approx$2000-2200 K gap in the temperature sequence. Since cloud opacity increases as the effective temperature decreases across the M/L transition ultracool dwarfs, an underestimated effective temperature highlights that {\sc BT-Settl} does not incorporate sufficient dust in its $\approx$2000-2200 K models. Our results are in agreement with Hurt et al. (AAS Journals, submitted), where a $\approx$1800 K pile-up of objects was also identified in an analysis of the {\sc BT-Settl} model fitting results for 89 late-M and L benchmark dwarfs. They find that M/L transition objects are poorly fit by the {\sc BT-Settl} atmospheric models, which consistently overestimate the $J$ and $H$ band flux. By including an extinction component that follows the interstellar medium (ISM) reddening law, Hurt et al. (AAS Journals, submitted) find a significant improvement in the fits, similarly suggesting a lack of dust opacity in the {\sc BT-Settl} models in this regime.
    
    \item {\sc BT-Settl}--SM08 objects: In Figure \ref{fig:temp-model-diff}, {\sc BT-Settl} generally yields consistent results with the SM08 evolutionary model for spectral types $\sim$L3--L5 but tends to overestimate the effective temperatures of late-L dwarfs. This discrepancy reflects the $\approx$1800 K pile-up of late-L dwarfs in Figure \ref{fig:atmo-temp-spt} and possibly the less prominent $\approx$1500--1600 K gap in the temperature sequence. Since cloud opacity increases from early-L to late-L dwarfs, an overestimated effective temperature for late-L dwarfs may mean that {\sc BT-Settl} over-predicts dust opacity in its $\approx$1700--1800 K models.
    
    \item {\sc ATMO} 2020--BHAC15/SM08 objects: In Figure \ref{fig:temp-model-diff}, we observe that {\sc ATMO} 2020 tends to generally overestimate temperatures at a level of 100--200 K. This may indicate possible incompleteness in incorporating cloud-free opacity sources in the atmospheres of T-dwarfs.
\end{enumerate}

It is important to note that a significant number of objects discussed in the second and third categories are located at the L/T transition. This is a relatively less understood phase of brown dwarf evolution. Since the comparison is not being conducted for self-consistently computed models, the discrepancies seen could be attributed to reasons outside the scope of our investigation. As such, one should be cautious in definitively interpreting {\sc BT-Settl}--SM08 and {\sc ATMO} 2020--BHAC15/SM08 object results. Combining SED-integrated bolometric luminosities with direct radius measurements of ultracool dwarfs would allow strictly empirical tests of the systematics in both atmospheric and evolutionary models. A summary of our results is presented as a MKO $M_J$ vs $J - K$ color-magnitude diagram in the right panel of Figure \ref{fig:temp-model-diff}. 
 
\begin{figure}[t]
    \centering
    \includegraphics[scale=0.34]{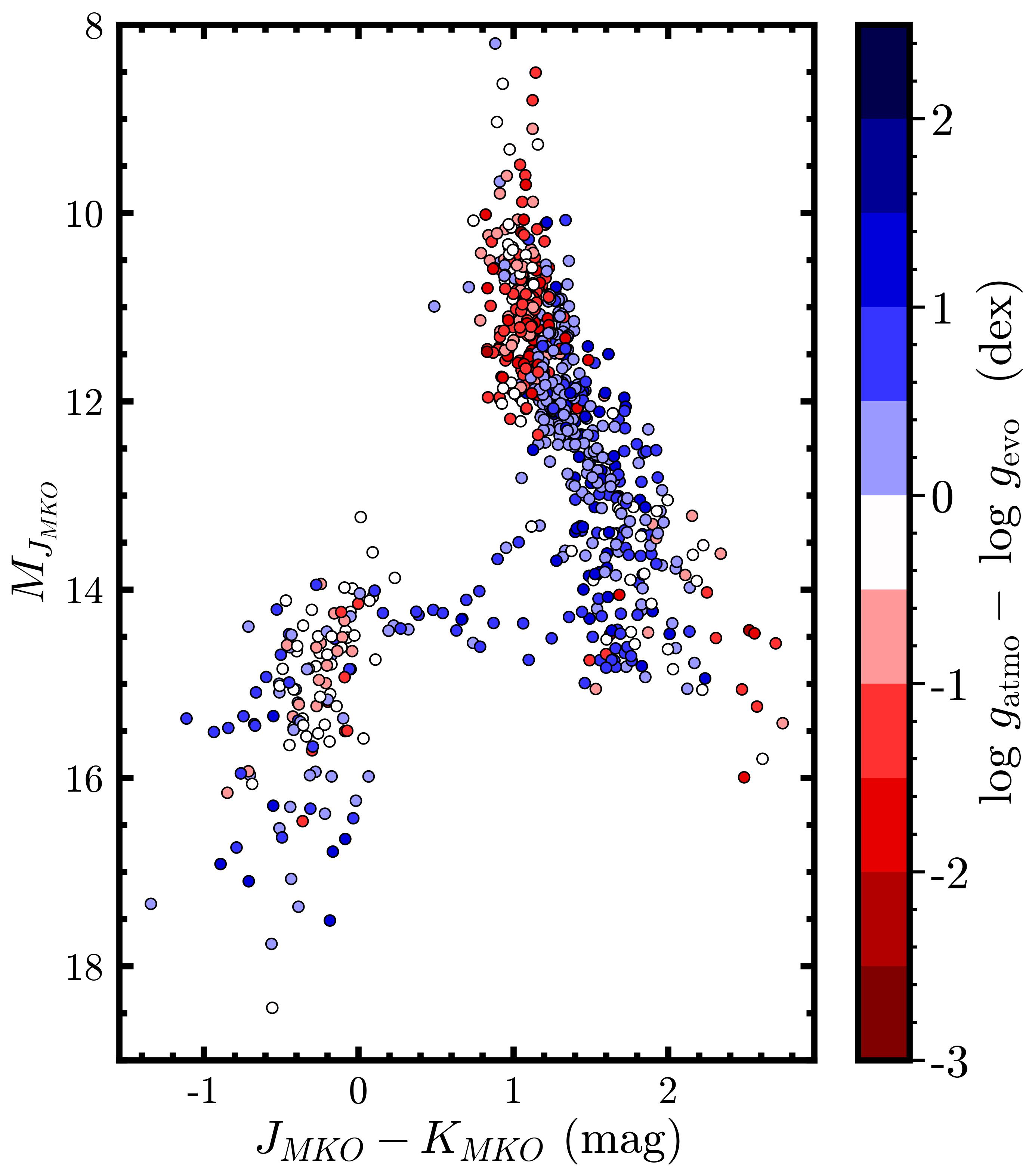}
    \caption{Same as right panel of Figure \ref{fig:temp-model-diff} but for difference between atmospheric and evolutionary model-derived surface gravity measurements.}
    \label{fig:logg-model-diff}
\end{figure}

\begin{figure}[t]
    \centering
    \includegraphics[scale=0.28]{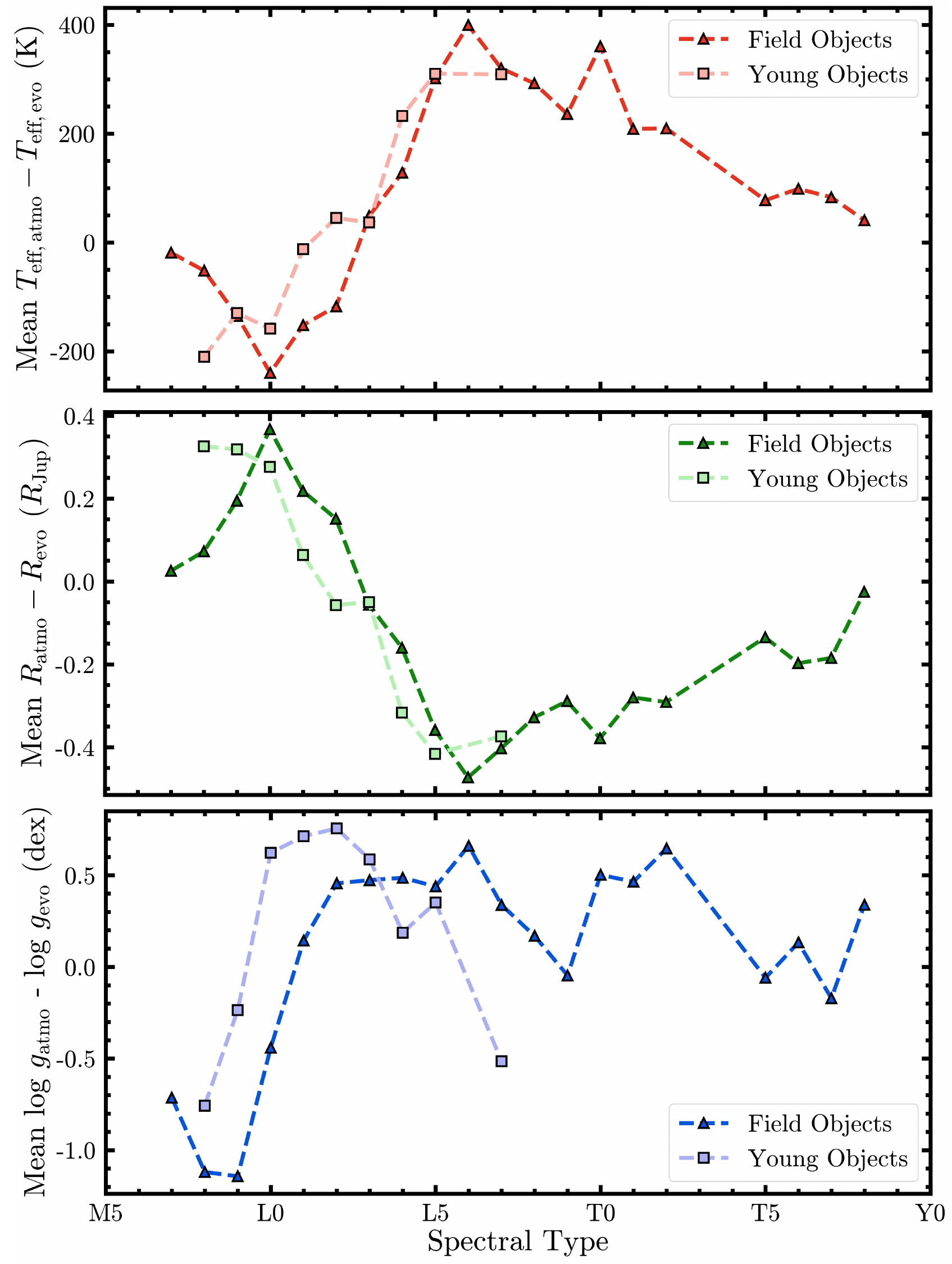}
    \caption{Average difference between the atmospheric and evolutionary model-derived effective temperatures (top), radii (middle), and surface gravities (bottom) as a function of spectral type for field (triangles) and young (squares) objects. Spectral types with $<$10 field objects (M6, T4, T9) and $<$5 young objects (M6, M7, L6, $\ge$L8) in our sample have been skipped. There was no significant difference between the mean and the median parameter difference distributions as a function of spectral type. Such systematic offsets may be adopted when combining fundamental parameter estimates obtained with the two methods.}
    \label{fig:diff-spt}
\end{figure}

\begin{figure}[!t]
    \centering
    \includegraphics[scale=0.28]{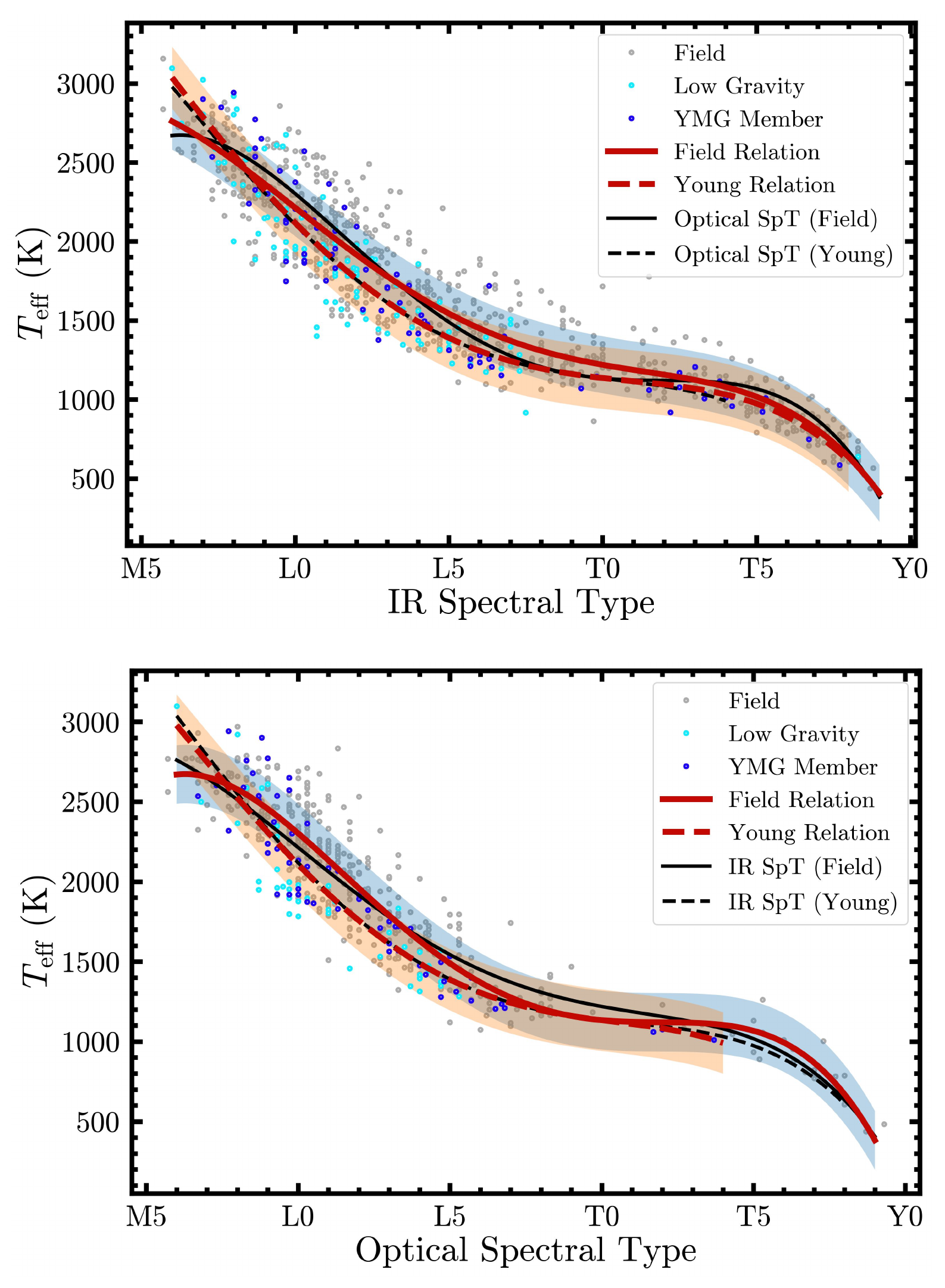}
    \caption{Effective temperatures derived for objects in our sample as a function of infrared (top) and optical (bottom) spectral types (SpTs). Random scatter in the x direction with amplitude 0.3 SpT is applied to avoid overlapping points. The figure elements are similar to those described in Figure \ref{fig:teff-spt}. In each panel, the best-fit polynomial relation for field and young objects from the other panel is presented as a dashed line of the same color as in the other panel.}
    \label{fig:opt-ir-spt-temp}
\end{figure}

\subsubsection{Radius and Surface Gravity}
\label{sec:radius_logg_diff}
Here, we note the observed differences between atmospheric and evolutionary model-derived radii and surface gravities for our sample of ultracool dwarfs. The analysis is conducted similar to \S \ref{sec:teff_diff}.

The atmospheric model-derived radius $R_{\mathrm{atmo}}$ is obtained from the scale factor $\Omega$ of the best-fit model spectrum as,
\begin{equation}
    R_{\mathrm{atmo}} = \sqrt{\Omega \cdot d^2},
\end{equation}
where $d$ is the object's distance. Figure \ref{fig:radius-model-diff} shows the difference between the two sets of radii measurements as a function of spectral type. For {\sc BT-Settl}---BHAC15 objects, which lie at the M/L transition boundary, radius is significantly overestimated by {\sc BT-Settl} (up to 1.5 $R_{\mathrm{Jup}}$). For {\sc BT-Settl}--SM08 and {\sc ATMO} 2020--BHAC15/SM08 objects ($\gtrsim$ L3), atmospheric models underestimate the radius by $\approx$0.1--0.7 $R_{\mathrm{Jup}}$, with maximum discrepancy in the late-L spectral types. A summary of these results is presented as a MKO $M_J$ vs $J - K$ color-magnitude diagram in the right panel of Figure \ref{fig:radius-model-diff}. Finally, Figure \ref{fig:logg-model-diff} presents the trend in log $g$ differences as a MKO $M_J$ vs $J - K$ color-magnitude diagram. Atmospheric models underestimate the gravities of bluer M/L transition dwarfs by up to 2 dex but overestimate the gravities of redder M/L transition dwarfs by $\approx$0.5 dex. No systematic trend is observed for later type objects in the color-magnitude diagram.

\subsection{Characterization of Systematic Offsets}
The significant discrepancies observed in the fundamental parameters obtained from evolutionary and atmospheric models raise questions about the reliability of either method to accurately estimate ultracool dwarf fundamental parameters. Past work on this topic has indicated that atmospheric models are more likely to be the source of the discrepancy: they are sensitive to boundary conditions unlike evolutionary models, involve many uncertain parameters in the modeling inputs and prescriptions, and often yield unphysical parameter estimates \citep[e.g.,][]{2000ApJ...542..464C, 2009ApJ...692..729D, 2010ApJ...721.1725D, 2010ApJ...722..311L, 2011ApJ...733...65B, 2017ApJS..231...15D}. It has been suggested that a systematic offset may exist between atmospheric and evolutionary model-derived parameters. Previously, the characterization of such systematic offsets has been limited to small samples of objects. Here, we take advantage of our large sample size to derive the average systematic offset in effective temperature, radius, and surface gravity as a function of spectral type for field and young objects. We compute the mean of the differences between atmospheric and evolutionary model-derived parameter values ($\Delta P = P_{\mathrm{atmo}} - P_{\mathrm{evo}}$, where $P$ is the relevant fundamental parameter) for each spectral type. We skip spectral types with $<$10 field objects (M6, T4, T9) and spectral types with $<$5 young objects (M6, M7, L6, $\ge$L8) in our sample. Figure \ref{fig:diff-spt} presents our results. The average systematic offsets are as large as 400 K in effective temperature, 0.4 $R_{\mathrm{Jup}}$ in radius, and 1.2 dex in log $g$. Larger systematic offsets are generally found near the M/L and L/T transition regions. Such systematic offsets might be adopted when combining fundamental parameter estimates obtained with the two methods.

\subsection{Optical and Near-Infrared Spectral Type-$T_{\mathrm{eff}}$ Sequences and Comparison}
We present $T_{\mathrm{eff}}$ sequences as a function of optical and IR spectral types in Figure \ref{fig:opt-ir-spt-temp} and the corresponding best-fit polynomial relations for field and young objects in Table \ref{table:poly}. Nearly two times as many objects have IR spectral types in our sample as optical spectral types. We observe that the $T_{\mathrm{eff}}$ sequence has a marginally lower rms about the best-fit as a function of IR spectral type, for field objects, compared to optical spectral types. While the relations for both spectral classifications are consistent, the optical spectral type is observed to predict slightly lower temperatures in the L/T transition for field objects compared to the IR spectral type and marginally higher temperatures for late-T field-age objects. We conclude that both spectral-typing schemes are good indicators of $T_{\mathrm{eff}}$ across the complete ultracool dwarf spectral sequence, given the availability of age information.

\section{Conclusion}
\label{sec:concl}
In this work, we performed a uniform calculation of the fundamental physical properties of \nobject ultracool dwarfs using the most precise photometry and astrometry available to date from \textit{The UltracoolSheet} (version 2.0.0, in preparation). Our results are summarized below:

\begin{enumerate}
    \item We obtained new low-resolution near-IR spectroscopy for 97 objects in our sample using NASA/IRTF. These spectra were collected for the spectroscopic characterization of known objects (57 potential low-gravity sources, benchmark objects, and wide companions) or the discovery of new ultracool dwarfs (40 objects). Additionally, we obtained new parallaxes for 19 objects using CFHT/WIRCam (11 sources) and Pan-STARRS 1 (8 sources).

    \item We have derived the bolometric luminosities ($L_{\mathrm{bol}}$) of 865 field-age and 189 young ultracool dwarfs by assembling and directly integrating flux-calibrated optical to mid-IR SEDs consisting of low-resolution ($R\sim$ 150) near-IR (0.85--2.5 $\mu$m) NASA/IRTF SpeX prism spectra, optical photometry from the Pan-STARRS1 survey, and mid-IR photometry from the CatWISE2020 survey and Spitzer/IRAC. Flux contributions from the optical (0.2--0.85 $\mu$m) and mid-IR (2.5--2000.0 $\mu$m) wavelengths were estimated using the best-fit {\sc BT-Settl} or {\sc ATMO} 2020 atmospheric model spectrum to the SED.
    
    \item A careful treatment of uncertainties revealed the presence of a noise floor in our $L_{\mathrm{bol}}$ measurements at the $\approx$0.015 dex level due to the use of atmospheric models to estimate flux contribution from wavelengths outside our SpeX spectra. This highlighted that high precision parallaxes must be paired with improved spectral coverage in the optical and mid-IR wavelengths to further drive the uncertainty in $L_{\mathrm{bol}}$ lower. Overall, the median uncertainty in our $L_{\mathrm{bol}}$ measurements was 0.05 dex. 
    
    \item Using the derived $L_{\mathrm{bol}}$ values and a new uniform age analysis for all objects in our sample, combined with the BHAC15 and SM08 evolutionary models, we estimated ultracool dwarf masses, radii, and surface gravities by employing Bayesian rejection sampling. Our $L_{\mathrm{bol}}$ measurements and evolutionary model-derived radii enabled the determination of semi-empirical effective temperatures ($T_{\mathrm{eff}}$), with a median uncertainty of 98 K.
    
    \item We showed that our $L_{\mathrm{bol}}$ measurements and evolutionary model-derived fundamental parameter estimates are largely consistent with those from \citet{2015ApJ...810..158F}, for objects common to the two samples (N = 130). For a small subset of these objects, revisions to their parallaxes and age distributions primarily contributed to the discrepancies observed in their $L_{\mathrm{bol}}$ measurements and the evolutionary model-derived fundamental parameter estimates, respectively. 
    
    \item Our large sample enabled us to establish empirical relationships for $L_{\mathrm{bol}}$ as a function of absolute magnitude, and $L_{\mathrm{bol}}$, $T_{\mathrm{eff}}$, and bolometric correction in various photometric bands as a function of spectral type, for both field-age and young objects. Based on the low rms scatter about the fit and relative insensitivity to age, we recommended the use of the MKO $H$-band polynomial relation for $L_{\mathrm{bol}}$ calculations from absolute magnitude. Further, we found that the 2MASS $K_S$-band or MKO $K$-band polynomial relations are the preferred choice for bolometric correction calculations from spectral type.
    
    \item We performed a comparison between the evolutionary model-derived surface gravities and near-infrared gravity classifications of ultracool dwarfs for a carefully constructed sample of young moving group members and companion objects. While the surface gravity measurements exhibited scatter in each of the three categories (\fldg, \intg, and \vlg), we observed a systematic difference in the mean surface gravities between the three gravity classes.
    
    \item A detailed characterization of atmospheric model systematics as a function of spectral type and in the near-infrared color magnitude diagram, based on comparisons with evolutionary model-derived values, revealed significantly large discrepancies in the effective temperatures (up to 800 K), radii (up to 2.0 $R_{\mathrm{Jup}}$), and surface gravities (up to 2 dex) of M/L transition dwarfs. Our results suggested that {\sc BT-Settl} lacks sufficient dust opacity in its models, in this regime. Based on this comparative analysis, we provided average systematic offsets between atmospheric and evolutionary model-derived $T_{\mathrm{eff}}$, radii, and log $g$ as a function of spectral type for field and young objects. These systematic offsets might be adopted when combining fundamental parameter estimates obtained with the two methods.
\end{enumerate}

The large-scale compilation of ultracool dwarf fundamental physical parameters presented in this work represents a factor of $\sim$5 increase in the sample size of such semi-empirical measurements compared to previous work. Determining the ages of ultracool dwarf binaries \citep[e.g.][]{2023MNRAS.519.1688D}, calculating the radii of candidate transiting planets \citep[e.g.][]{2022AJ....164..252T}, and constructing luminosity distributions to constrain the substellar initial mass function \citep[e.g.][]{2021ApJS..253....7K} are just a few of the many exciting applications of this work as we continue to explore the fascinating diversity of substellar objects and companions.

\section*{Acknowledgments}
The authors would like to thank the anonymous referee for many helpful comments which improved the manuscript. We also thank Aaron Meisner and Adam Schneider for helpful discussions on cross-matching ALLWISE sources with CatWISE2020 sources. This work has benefitted from \emph{The UltracoolSheet}, maintained by Will Best, Trent Dupuy, Michael Liu, Aniket Sanghi, Rob Siverd, and Zhoujian Zhang, and developed from compilations by \citet{2012ApJS..201...19D}, \citet{2013Sci...341.1492D}, \citet{2016ApJ...833...96L}, \citet{2018ApJS..234....1B}, \citet{2020AJ....159..257B}, and \citet{2023AJ....166..103S}. The data presented herein were obtained at the W. M. Keck Observatory, which is operated as a scientific partnership among the California Institute of Technology, the University of California, and the National Aeronautics and Space Administration (NASA). The Observatory was made possible by the generous financial support of the W. M. Keck Foundation. The authors wish to recognize and acknowledge the very significant cultural role and reverence that the summit of Maunakea has always had within the indigenous Hawaiian community. We are most fortunate to have the opportunity to conduct observations from this mountain. 

A.~Sanghi acknowledges support from Research Experience for Undergraduate program at the Institute for Astronomy, University of Hawaii-Manoa funded through National Science Foundation (NSF) grant \#2050710. A.~Sanghi would like to thank the Institute for Astronomy for their hospitality during the course of this project. This research was funded in part by the Gordon and Betty Moore Foundation through grant GBMF8550 to M.~Liu. W.M.J.~Best acknowledges support from grant HST-GO-15238 provided by STScI and AURA. T.~Dupuy acknowledges support from UKRI STFC AGP grant ST/W001209/1. For the purpose of open access, the author has applied a Creative Commons Attribution (CC BY) licence to any Author Accepted Manuscript version arising from this submission.

This publication makes use of data products from: the \emph{Two Micron All Sky Survey} (2MASS), which is a joint project of the University of Massachusetts and the Infrared Processing and Analysis Center/California Institute of Technology, funded by the National Aeronautics and Space Administration and the National Science Foundation; the UKIRT Infrared Deep Sky Survey (UKIDSS) project, which is defined in \citet{2007MNRAS.379.1599L}, uses the UKIRT Wide Field Camera \citep[WFCAM:][]{2007A&A...467..777C} and a photometric system described in \citet{2006MNRAS.367..454H}, and the pipeline and science archives described in \citet{2008eic..work..541I} and \citet{2008MNRAS.384..637H}; the \emph{Wide-field Infrared Survey Explorer} (WISE), which is a joint project of the University of California, Los Angeles, and the Jet Propulsion Laboratory/California Institute of Technology, funded by the National Aeronautics and Space Administration; the SIMBAD database, Aladin, and Vizier, operated at CDS, Strasbourg, France; the SpeX prism Spectral Libraries, maintained by Adam Burgasser; the NASA/IPAC Infrared Science Archive, which is operated by the Jet Propulsion Laboratory, California Institute of Technology, under contract with the National Aeronautics and Space Administration; the Spanish Virtual Observatory (https://svo.cab.inta-csic.es) project funded by MCIN/AEI/10.13039/501100011033/ through grant PID2020-112949GB-I00.

\emph{Additional Software/Resources:} {\sc Astropy} \citep{2013A&A...558A..33A, 2018AJ....156..123A, 2022arXiv220614220T}, {\sc Matplotlib} \citep{2007CSE.....9...90H}, {\sc NumPy} \citep{harris2020array}, {\sc SciPy} \citep{2020SciPy-NMeth}, {\sc pandas} \citep{jeff_reback_2022_7037953} and the NASA Astrophysics Data System (ADS). The data presented in this work are made publicly available at \url{https://doi.org/10.5281/zenodo.8315643}. The scripts associated with the methods in this work are made publicly available at \url{https://github.com/cosmicoder/HIPPVI-Code}.

\startlongtable
\begin{longrotatetable}
\begin{deluxetable*}{cccccccccc}
\centering
\tabletypesize{\scriptsize}
\tablecaption{Polynomial Relations}
\tablehead{\colhead{P(x)\tablenotemark{\scriptsize a}} & \colhead{x\tablenotemark{\scriptsize b}} & \colhead{rms} & \colhead{$c_0$} & \colhead{$c_1$} & \colhead{$c_2$} & \colhead{$c_3$} & \colhead{$c_4$} & \colhead{$c_5$} & \colhead{$c_6$}}
\startdata
$L_{\mathrm{bol\;FLD}}$ & 6.0$\le$SpT$\le$29.0 & 0.157 dex & 2.38575339e+00 & -2.34765187e+00 & 4.10403429e-01 & -3.75870300e-02 & 1.79418353e-03 & -4.22643837e-05 & 3.85072198e-07 \\
$T_{\mathrm{eff\;FLD}}$ & 6.0$\le$SpT$\le$29.0 & 174.686 K & 1.58703841e+03 & 5.52980543e+02 & -7.99773680e+01 & 3.71753828e+00 & -5.75357642e-02 & \nodata & \nodata \\
$T_{\mathrm{eff\;YNG}}$ & 6.0$\le$SpT$\le$28.0 & 188.424 K & 4.56313932e+03 & -2.28258451e+02 & -1.11940923e+01 & 1.17847855e+00 & -2.38265571e-02 & \nodata & \nodata \\
$T_{\mathrm{eff\;FLD}}$ & 6.0$\le$Optical SpT$\le$29.0 & 183.423 K & 2.89211204e+02 & 9.22014212e+02 & -1.15315067e+02 & 5.08274865e+00 & -7.58233882e-02 & \nodata & \nodata \\
$T_{\mathrm{eff\;YNG}}$ & 6.0$\le$Optical SpT$\le$24.0 & 191.954 K & 4.07702266e+03 & -8.43172747e+01 & -2.63589282e+01 & 1.85751341e+00 & -3.48367787e-02 & \nodata & \nodata \\
$T_{\mathrm{eff\;FLD}}$ & 6.0$\le$IR SpT$\le$29.0 & 180.226 K & 2.44117451e+03 & 2.80510088e+02 & -5.16856405e+01 & 2.54129573e+00 & -4.05522698e-02 & \nodata & \nodata \\
$T_{\mathrm{eff\;YNG}}$ & 6.0$\le$IR SpT$\le$28.0 & 195.565 K & 4.47499244e+03 & -1.87857056e+02 & -1.60621537e+01 & 1.40051083e+00 & -2.72611817e-02 & \nodata & \nodata \\
$BC_{z\;\mathrm{FLD}}$ & 6.0$\le$SpT$\le$29.0 & 0.208 mag & 8.82748312e-01 & -3.28812471e-02 & -3.01625334e-02 & 2.00785031e-03 & -3.67327336e-05 & \nodata & \nodata \\
$BC_{y\;\mathrm{FLD}}$ & 6.0$\le$SpT$\le$29.0 & 0.189 mag & 4.78290771e+00 & -2.05057405e+00 & 4.09934609e-01 & -4.22427931e-02 & 2.21923901e-03 & -5.63234268e-05 & 5.46529690e-07 \\
$BC_{J\;\mathrm{FLD}}$ & 6.0$\le$SpT$\le$29.0 & 0.157 mag & 1.83270671e+00 & -1.00428244e-01 & 4.91437247e-02 & -6.23850138e-03 & 2.91250402e-04 & -4.49787817e-06 & \nodata \\
$BC_{J_{MKO}\;\mathrm{FLD}}$ & 6.0$\le$SpT$\le$29.0 & 0.141 mag & 3.34867561e+00 & -7.05942778e-01 & 1.41154739e-01 & -1.27017514e-02 & 5.03938151e-04 & -7.12860976e-06 & \nodata \\
$BC_{H\;\mathrm{FLD}}$ & 6.0$\le$SpT$\le$29.0 & 0.114 mag & 2.06836504e+00 & 1.78079042e-01 & -1.90813876e-02 & 9.23336345e-04 & -1.63230829e-05 & \nodata & \nodata \\
$BC_{H_{MKO}\;\mathrm{FLD}}$ & 6.0$\le$SpT$\le$29.0 & 0.091 mag & 1.60830399e+00 & 3.16405418e-01 & -3.51960013e-02 & 1.68671698e-03 & -2.89674748e-05 & \nodata & \nodata \\
$BC_{Ks\;\mathrm{FLD}}$ & 6.0$\le$SpT$\le$29.0 & 0.169 mag & 2.96143878e+00 & -3.73942004e-02 & 9.29897400e-03 & -3.28963542e-04 & \nodata & \nodata & \nodata \\
$BC_{K_{MKO}\;\mathrm{FLD}}$ & 6.0$\le$SpT$\le$29.0 & 0.148 mag & 4.56079546e+00 & -5.30973644e-01 & 6.31017798e-02 & -2.73558092e-03 & 3.72518239e-05 & \nodata & \nodata \\
$BC_{W1\;\mathrm{FLD}}$ & 6.0$\le$SpT$\le$29.0 & 0.193 mag & -5.36228544e+00 & 3.60659733e+00 & -5.96094136e-01 & 4.92274677e-02 & -2.05225834e-03 & 3.99592136e-05 & -2.74443149e-07 \\
$BC_{W2\;\mathrm{FLD}}$ & 6.0$\le$SpT$\le$29.0 & 0.198 mag & -1.07410524e+01 & 6.19843221e+00 & -1.07147549e+00 & 9.35831952e-02 & -4.27523017e-03 & 9.68864595e-05 & -8.53309439e-07 \\
$BC_{z\;\mathrm{YNG}}$ & 6.6$\le$SpT$\le$25.5 & 0.314 mag & 2.88963400e+00 & -5.35800318e-01 & 1.41810437e-02 & \nodata & \nodata & \nodata & \nodata \\
$BC_{y\;\mathrm{YNG}}$ & 6.6$\le$SpT$\le$28.0 & 0.287 mag & -5.21677358e+00 & 2.03998815e+00 & -2.47492191e-01 & 1.12619606e-02 & -1.71769287e-04 & \nodata & \nodata \\
$BC_{J\;\mathrm{YNG}}$ & 6.0$\le$SpT$\le$28.0 & 0.206 mag & -2.92269209e+00 & 1.62463279e+00 & -1.80923490e-01 & 7.81791312e-03 & -1.13687590e-04 & \nodata & \nodata \\
$BC_{J_{MKO}\;\mathrm{YNG}}$ & 6.0$\le$SpT$\le$28.0 & 0.193 mag & 5.27987592e+00 & -1.54348126e+00 & 2.82107387e-01 & -2.39608174e-02 & 9.14238336e-04 & -1.25936127e-05 & \nodata \\
$BC_{H\;\mathrm{YNG}}$ & 6.0$\le$SpT$\le$28.0 & 0.134 mag & -1.33267424e-01 & 8.47837406e-01 & -8.88312926e-02 & 3.83044799e-03 & -5.80866655e-05 & \nodata & \nodata \\
$BC_{H_{MKO}\;\mathrm{YNG}}$ & 6.0$\le$SpT$\le$28.0 & 0.129 mag & -5.08204516e-01 & 9.56683939e-01 & -1.01434926e-01 & 4.42442275e-03 & -6.78222163e-05 & \nodata & \nodata \\
$BC_{Ks\;\mathrm{YNG}}$ & 6.0$\le$SpT$\le$28.0 & 0.106 mag & 2.30496939e+00 & 1.15547026e-01 & -1.02256113e-03 & -1.12479204e-04 & \nodata & \nodata & \nodata \\
$BC_{K_{MKO}\;\mathrm{YNG}}$ & 6.0$\le$SpT$\le$28.0 & 0.106 mag & 3.23630807e+00 & -1.84769176e-01 & 3.30808101e-02 & -1.67472629e-03 & 2.44417318e-05 & \nodata & \nodata \\
$BC_{W1\;\mathrm{YNG}}$ & 6.0$\le$SpT$\le$28.0 & 0.212 mag & 6.86131580e+00 & -1.26576300e+00 & 1.52614866e-01 & -6.80445371e-03 & 9.95645299e-05 & \nodata & \nodata \\
$BC_{W2\;\mathrm{YNG}}$ & 6.0$\le$SpT$\le$28.0 & 0.241 mag & 7.50713316e+00 & -1.44635248e+00 & 1.79292891e-01 & -8.30663097e-03 & 1.30461556e-04 & \nodata & \nodata \\
$L_{\mathrm{bol\;FLD\;ML}}$\tablenotemark{\scriptsize c} & 9.1$\le M_J\le$16.1 & 0.057 dex & 5.53301206e-01 & -3.51344336e-01 & \nodata & \nodata & \nodata & \nodata & \nodata \\
$L_{\mathrm{bol\;FLD\;T}}$\tablenotemark{\scriptsize c} & 12.8$\le M_J\le$18.4 & 0.101 dex & 8.80328596e-01 & -3.93121219e-01 & \nodata & \nodata & \nodata & \nodata & \nodata \\
$L_{\mathrm{bol\;FLD}}$ & 8.5$\le M_H\le$18.1 & 0.066 dex & 6.93323688e-01 & -3.89636901e-01 & \nodata & \nodata & \nodata & \nodata & \nodata \\
$L_{\mathrm{bol\;FLD}}$ & 7.3$\le M_{Ks}\le$18.5 & 0.083 dex & -2.50181488e+01 & 8.16617507e+00 & -1.04402978e+00 & 5.48196484e-02 & -1.04209064e-03 & \nodata & \nodata \\
$L_{\mathrm{bol\;FLD\;ML}}$\tablenotemark{\scriptsize c} & 9.0$\le M_{J_{MKO}}\le$16.0 & 0.055 dex & 6.00343959e-01 & -3.57204309e-01 & \nodata & \nodata & \nodata & \nodata & \nodata \\
$L_{\mathrm{bol\;FLD\;T}}$\tablenotemark{\scriptsize c} & 12.6$\le M_{J_{MKO}}\le$18.4 & 0.108 dex & 9.54537740e-01 & -4.04863108e-01 & \nodata & \nodata & \nodata & \nodata & \nodata \\
$L_{\mathrm{bol\;FLD}}$ & 8.5$\le M_{H_{MKO}}\le$18.8 & 0.053 dex & 6.17668495e-01 & -3.80907000e-01 & \nodata & \nodata & \nodata & \nodata & \nodata \\
$L_{\mathrm{bol\;FLD}}$ & 8.1$\le M_{K_{MKO}}\le$19.0 & 0.064 dex & -8.68942422e+00 & 3.17477734e+00 & -4.83389959e-01 & 2.73739037e-02 & -5.47040525e-04 & \nodata & \nodata \\
$L_{\mathrm{bol\;FLD}}$ & 11.0$\le M_z\le$22.7 & 0.083 dex & 8.78235147e-01 & -3.06670163e-01 & \nodata & \nodata & \nodata & \nodata & \nodata \\
$L_{\mathrm{bol\;FLD\;ML}}$\tablenotemark{\scriptsize c} & 10.5$\le M_y\le$18.5 & 0.074 dex & 5.55740568e-01 & -3.03933139e-01 & \nodata & \nodata & \nodata & \nodata & \nodata \\
$L_{\mathrm{bol\;FLD\;T}}$\tablenotemark{\scriptsize c} & 15.0$\le M_y\le$21.0 & 0.066 dex & 1.37035364e+00 & -3.66205917e-01 & \nodata & \nodata & \nodata & \nodata & \nodata \\
$L_{\mathrm{bol\;FLD}}$ & 7.6$\le M_{W1}\le$17.1 & 0.103 dex & -5.22934625e+01 & 1.85121627e+01 & -2.47203555e+00 & 1.39599009e-01 & -2.87497133e-03 & \nodata & \nodata \\
$L_{\mathrm{bol\;FLD}}$ & 7.3$\le M_{W2}\le$14.1 & 0.125 dex & 1.75395276e+00 & -5.42818335e-01 & \nodata & \nodata & \nodata & \nodata & \nodata \\
$L_{\mathrm{bol\;YNG\;ML}}$\tablenotemark{\scriptsize c} & 8.2$\le M_J\le$15.8 & 0.081 dex & 4.08756088e-01 & -3.34689569e-01 & \nodata & \nodata & \nodata & \nodata & \nodata \\
$L_{\mathrm{bol\;YNG\;T}}$\tablenotemark{\scriptsize c} & 14.1$\le M_J\le$16.7 & 0.108 dex & 9.53221827e-01 & -3.95259421e-01 & \nodata & \nodata & \nodata & \nodata & \nodata \\
$L_{\mathrm{bol\;YNG}}$ & 7.7$\le M_H\le$17.0 & 0.072 dex & 6.58828628e-01 & -3.84611365e-01 & \nodata & \nodata & \nodata & \nodata & \nodata \\
$L_{\mathrm{bol\;YNG}}$ & 7.3$\le M_{Ks}\le$16.8 & 0.057 dex & -1.68245519e+01 & 5.94124964e+00 & -8.33426219e-01 & 4.68755666e-02 & -9.51087115e-04 & \nodata & \nodata \\
$L_{\mathrm{bol\;YNG\;ML}}$\tablenotemark{\scriptsize c} & 8.2$\le M_{J_{MKO}}\le$15.8 & 0.080 dex & 4.06864509e-01 & -3.36361777e-01 & \nodata & \nodata & \nodata & \nodata & \nodata \\
$L_{\mathrm{bol\;YNG\;T}}$\tablenotemark{\scriptsize c} & 14.0$\le M_{J_{MKO}}\le$16.7 & 0.097 dex & 8.31005313e-01 & -3.94047446e-01 & \nodata & \nodata & \nodata & \nodata & \nodata \\
$L_{\mathrm{bol\;YNG}}$ & 7.7$\le M_{H_{MKO}}\le$17.0 & 0.057 dex & 7.25956089e-01 & -3.88884874e-01 & \nodata & \nodata & \nodata & \nodata & \nodata \\
$L_{\mathrm{bol\;YNG}}$ & 7.3$\le M_{K_{MKO}}\le$16.7 & 0.057 dex & -2.06743904e+01 & 7.40114284e+00 & -1.03611814e+00 & 5.90347236e-02 & -1.21640941e-03 & \nodata & \nodata \\
$L_{\mathrm{bol\;YNG}}$ & 11.0$\le M_z\le$19.3 & 0.121 dex & 6.92111297e-01 & -2.88226403e-01 & \nodata & \nodata & \nodata & \nodata & \nodata \\
$L_{\mathrm{bol\;YNG\;ML}}$\tablenotemark{\scriptsize c} & 10.3$\le M_y\le$17.8 & 0.104 dex & 3.28920351e-01 & -2.80644117e-01 & \nodata & \nodata & \nodata & \nodata & \nodata \\
$L_{\mathrm{bol\;YNG\;T}}$\tablenotemark{\scriptsize c} & 16.5$\le M_y\le$19.5 & 0.075 dex & 7.09044657e-01 & -3.25773704e-01 & \nodata & \nodata & \nodata & \nodata & \nodata \\
$L_{\mathrm{bol\;YNG}}$ & 6.6$\le M_{W1}\le$16.1 & 0.116 dex & -3.70291470e+01 & 1.38286472e+01 & -1.95883111e+00 & 1.15867819e-01 & -2.48769429e-03 & \nodata & \nodata \\
$L_{\mathrm{bol\;YNG}}$ & 6.1$\le M_{W2}\le$13.6 & 0.158 dex & 1.05360011e+00 & -4.84834949e-01 & \nodata & \nodata & \nodata & \nodata & \nodata \\
\enddata
\tablecomments{Caution should be exercised in using the 2MASS and MKO $J$ band young object relations at the L/T transition as discussed in \S\ref{sec:BC}. Since we exclude objects in star-forming regions from our sample, the YNG relations (calculated for YMG members and low gravity sources) are not appropriate for objects with ages $\le$10 Myr.}
\tablenotetext{a}{$P(x) = \sum_{i=0}^{n} c_i x^i$}
\tablenotetext{b}{Polynomials expressed as a function of spectral type (SpT) accept inputs in the range 6–29 corresponding to spectral types M6-T9.}
\tablenotetext{c}{The subscripts $\mathrm{ML}$ and $\mathrm{T}$ designate distinct polynomial relations for objects with numerical spectral types $[6,19]$ and $[20,29]$ respectively.}
\label{table:poly}
\end{deluxetable*}
\end{longrotatetable}
\startlongtable
\centerwidetable
\movetabledown=3mm
\begin{longrotatetable}
\begin{deluxetable*}{cccccccccccc}
\centering
\tabletypesize{\scriptsize}
\tablecaption{Objects with $>$2$\sigma$ Disagreement between This Work's and \citet{2015ApJ...810..158F}'s Evolutionary Model-derived Parameters}
\tablehead{\colhead{UCS Name} & \colhead{F+15 Name} & \colhead{UCS Age\tablenotemark{\scriptsize{a}}} & \colhead{F+15 Age} & \colhead{Mass} & \colhead{Mass (F+15)} & \colhead{Radius} & \colhead{Radius (F+15)} & \colhead{$T_{\mathrm{eff}}$} & \colhead{$T_{\mathrm{eff}}$ (F+15)} & \colhead{log $g$} & \colhead{log $g$ (F+15)} \\ \colhead{} & \colhead{} & \colhead{(Gyr)} & \colhead{(Gyr)} & \colhead{($M_\mathrm{{Jup}}$)} & \colhead{($M_\mathrm{{Jup}}$)} & \colhead{($R_\mathrm{{Jup}}$)} & \colhead{($R_\mathrm{{Jup}}$)} & \colhead{(K)} & \colhead{(K)} & \colhead{(dex)} & \colhead{(dex)}}
\startdata
2MASS J00332386-1521309$^*$ & 0033-1521 & Truncated DL17 & 0.010-0.150 & $11.93\pm9.98$ & $29.05\pm18.53$ & $1.35\pm0.11$ & $1.43\pm0.22$ & $1401.0\pm56.0$ & $1887.0\pm175.0$ & $4.31\pm0.25$ & $4.47\pm0.42$ \\ 
2MASS J03185403-3421292 & 0318-3421 & Truncated DL17 & 0.500-10.000 & $8.59\pm7.57$ & $48.98\pm24.02$ & $1.32\pm0.08$ & $0.97\pm0.12$ & $1097.0\pm37.0$ & $1344.0\pm107.0$ & $4.11\pm0.26$ & $5.05\pm0.36$ \\ 
2MASS J03264225-2102057 & 0326-2102 & 0.130-0.200 Gyr & 0.050-0.120 & $30.03\pm2.30$ & $18.4\pm6.46$ & $1.16\pm0.02$ & $1.3\pm0.07$ & $1534.0\pm22.0$ & $1381.0\pm42.0$ & $4.76\pm0.05$ & $4.39\pm0.21$ \\ 
2MASS J05002100+0330501 & 0500+0330 & Truncated DL17 & 0.500-10.000 & $13.14\pm10.51$ & $63.68\pm14.44$ & $1.50\pm0.13$ & $1.0\pm0.08$ & $1462.0\pm62.0$ & $1793.0\pm72.0$ & $4.18\pm0.26$ & $5.2\pm0.19$ \\ 
2MASS J06244595-4521548 & 0624-4521 & 0.040-0.050 Gyr & 0.500-10.000 & $10.78\pm0.27$ & $56.04\pm18.52$ & $1.34\pm0.01$ & $0.99\pm0.1$ & $1279.0\pm6.0$ & $1501.0\pm85.0$ & $4.18\pm0.01$ & $5.14\pm0.27$ \\ 
2MASS J13595510-4034582 & 1359-4034 & 0.020-0.250 Gyr & 0.500-10.000 & $24.86\pm10.22$ & $63.44\pm15.48$ & $1.30\pm0.12$ & $1.0\pm0.08$ & $1826.0\pm88.0$ & $1799.0\pm106.0$ & $4.58\pm0.25$ & $5.19\pm0.2$ \\ 
2MASS J21481628+4003593 & 2148+4003 & Truncated DL17 & 0.500-10.000 & $11.26\pm9.54$ & $55.03\pm18.55$ & $1.39\pm0.10$ & $0.99\pm0.1$ & $1292.0\pm48.0$ & $1446.0\pm72.0$ & $4.33\pm0.25$ & $5.13\pm0.28$ \\ 
2MASSI J0439010-235308 & 0439-2353 & Truncated DL17 & 0.500-10.000 & $10.50\pm9.03$ & $48.01\pm23.41$ & $1.31\pm0.09$ & $0.97\pm0.12$ & $1287.0\pm44.0$ & $1290.0\pm82.0$ & $4.33\pm0.23$ & $5.04\pm0.37$ \\ 
2MASSI J0445538-304820 & 0445-3048 & Truncated DL17 & 0.500-10.000 & $15.71\pm11.56$ & $63.8\pm15.0$ & $1.44\pm0.14$ & $1.0\pm0.08$ & $1713.0\pm83.0$ & $1809.0\pm90.0$ & $4.29\pm0.28$ & $5.19\pm0.2$ \\ 
2MASSI J0451009-340214 & 0451-3402 & 0.040-0.050 Gyr & 0.500-10.000 & $19.95\pm1.87$ & $72.38\pm12.0$ & $1.47\pm0.03$ & $1.04\pm0.06$ & $1865.0\pm23.0$ & $2155.0\pm72.0$ & $4.43\pm0.06$ & $5.22\pm0.13$ \\ 
2MASSI J0835425-081923 & 0835-0819 & Truncated DL17 & 0.500-10.000 & $11.75\pm9.45$ & $62.47\pm15.86$ & $1.33\pm0.10$ & $1.0\pm0.08$ & $1374.0\pm52.0$ & $1754.0\pm112.0$ & $4.32\pm0.24$ & $5.19\pm0.21$ \\ 
2MASSI J0847287-153237 & 0847-1532 & Truncated DL17 & 0.500-10.000 & $17.12\pm11.29$ & $63.55\pm14.79$ & $1.42\pm0.14$ & $1.0\pm0.08$ & $1768.0\pm85.0$ & $1794.0\pm81.0$ & $4.34\pm0.27$ & $5.19\pm0.2$ \\ 
2MASSI J0859254-194926 & 0859-1949 & Truncated DL17 & 0.500-10.000 & $8.73\pm7.71$ & $50.29\pm23.1$ & $1.32\pm0.08$ & $0.98\pm0.11$ & $1117.0\pm41.0$ & $1374.0\pm100.0$ & $4.11\pm0.26$ & $5.06\pm0.35$ \\ 
2MASSI J1010148-040649 & 1010-0406 & 0.150-0.250 Gyr & 0.500-10.000 & $18.14\pm3.23$ & $51.16\pm23.06$ & $1.21\pm0.03$ & $0.98\pm0.11$ & $1257.0\pm24.0$ & $1416.0\pm123.0$ & $4.51\pm0.09$ & $5.07\pm0.34$ \\ 
2MASSI J1526140+204341 & 1526+2043 & Truncated DL17 & 0.500-10.000 & $10.62\pm9.57$ & $54.39\pm21.4$ & $1.31\pm0.09$ & $0.98\pm0.1$ & $1295.0\pm45.0$ & $1518.0\pm157.0$ & $4.33\pm0.23$ & $5.1\pm0.31$ \\ 
2MASSI J2057540-025230 & 2057-0252 & 0.020-0.250 Gyr & 0.500-10.000 & $25.86\pm10.41$ & $69.56\pm13.04$ & $1.30\pm0.12$ & $1.02\pm0.07$ & $1877.0\pm89.0$ & $2044.0\pm87.0$ & $4.60\pm0.25$ & $5.22\pm0.15$ \\ 
2MASSI J2104149-103736 & 2104-1037 & Truncated DL17 & 0.500-10.000 & $13.91\pm11.24$ & $68.59\pm12.82$ & $1.43\pm0.14$ & $1.02\pm0.07$ & $1605.0\pm79.0$ & $1994.0\pm73.0$ & $4.19\pm0.29$ & $5.21\pm0.16$ \\ 
2MASSW J0030300-145033 & 0030-1450 & 0.040-0.050 Gyr & 0.500-10.000 & $10.14\pm1.06$ & $53.55\pm21.14$ & $1.35\pm0.04$ & $0.98\pm0.1$ & $1235.0\pm39.0$ & $1466.0\pm117.0$ & $4.16\pm0.03$ & $5.1\pm0.31$ \\ 
2MASSW J1036530-344138 & 1036-3441 & Truncated DL17 & 0.500-10.000 & $8.18\pm6.88$ & $49.24\pm24.45$ & $1.32\pm0.08$ & $0.97\pm0.12$ & $1074.0\pm34.0$ & $1368.0\pm131.0$ & $4.08\pm0.25$ & $5.04\pm0.37$ \\ 
2MASSW J1439284+192915 & 1439+1929 & Truncated DL17 & 0.500-10.000 & $18.28\pm12.01$ & $71.64\pm11.6$ & $1.41\pm0.14$ & $1.03\pm0.06$ & $1812.0\pm90.0$ & $2121.0\pm61.0$ & $4.38\pm0.28$ & $5.22\pm0.13$ \\ 
2MASSW J1448256+103159 & 1448+1031 & Truncated DL17 & 0.500-10.000 & $11.58\pm9.27$ & $59.43\pm16.66$ & $1.41\pm0.09$ & $0.99\pm0.09$ & $1315.0\pm43.0$ & $1623.0\pm91.0$ & $4.18\pm0.25$ & $5.17\pm0.23$ \\ 
2MASSW J1506544+132106 & 1506+1321 & Truncated DL17 & 0.500-10.000 & $13.55\pm10.82$ & $68.77\pm12.85$ & $1.41\pm0.13$ & $1.02\pm0.07$ & $1519.0\pm70.0$ & $2004.0\pm75.0$ & $4.20\pm0.27$ & $5.21\pm0.15$ \\ 
2MASSW J1507476-162738 & 1507-1627 & Truncated DL17 & 0.500-10.000 & $11.68\pm9.56$ & $59.66\pm15.62$ & $1.42\pm0.10$ & $0.99\pm0.09$ & $1325.0\pm46.0$ & $1607.0\pm70.0$ & $4.32\pm0.25$ & $5.18\pm0.23$ \\ 
2MASSW J1515008+484742 & 1515+4847 & Truncated DL17 & 0.500-10.000 & $9.45\pm9.12$ & $56.83\pm17.52$ & $1.34\pm0.09$ & $0.99\pm0.1$ & $1199.0\pm40.0$ & $1505.0\pm74.0$ & $4.13\pm0.26$ & $5.15\pm0.26$ \\ 
2MASSW J2208136+292121$^*$ & 2208+2921 & 0.010-0.150 Gyr & 0.010-0.150 & $13.15\pm7.39$ & $26.71\pm15.54$ & $1.52\pm0.12$ & $1.41\pm0.2$ & $1532.0\pm64.0$ & $1804.0\pm132.0$ & $4.16\pm0.24$ & $4.44\pm0.41$ \\ 
2MASSW J2244316+204343 & 2244+2043 & 0.130-0.200 Gyr & 0.050-0.120 & $15.58\pm1.60$ & $11.91\pm2.79$ & $1.22\pm0.02$ & $1.29\pm0.03$ & $1211.0\pm19.0$ & $1209.0\pm17.0$ & $4.43\pm0.05$ & $4.21\pm0.11$ \\ 
DENIS-P J0652197-253450 & 0652-2534 & 0.020-0.250 Gyr & 0.500-10.000 & $26.85\pm11.67$ & $74.67\pm11.18$ & $1.33\pm0.13$ & $1.05\pm0.05$ & $1998.0\pm97.0$ & $2231.0\pm60.0$ & $4.60\pm0.26$ & $5.22\pm0.12$ \\ 
DENIS-P J1058.7-1548 & 1058-1548 & 0.040-0.050 Gyr & 0.500-10.000 & $14.52\pm0.34$ & $64.24\pm14.01$ & $1.42\pm0.01$ & $1.0\pm0.07$ & $1570.0\pm11.0$ & $1809.0\pm68.0$ & $4.27\pm0.01$ & $5.2\pm0.19$ \\ 
LHS 132 & 0102-3737 & 0.020-0.250 Gyr & 0.500-10.000 & $39.32\pm16.69$ & $90.28\pm9.22$ & $1.39\pm0.16$ & $1.17\pm0.04$ & $2364.0\pm133.0$ & $2636.0\pm67.0$ & $4.72\pm0.30$ & $5.22\pm0.07$ \\ 
LHS 2924 & 1428+3310 & Truncated DL17 & 0.500-10.000 & $19.21\pm12.72$ & $75.98\pm10.89$ & $1.44\pm0.14$ & $1.06\pm0.05$ & $1915.0\pm94.0$ & $2275.0\pm58.0$ & $4.38\pm0.28$ & $5.22\pm0.11$ \\ 
LHS 3003 & 1456-2809 & 0.130-0.200 Gyr & 0.500-10.000 & $61.46\pm3.74$ & $84.02\pm9.72$ & $1.29\pm0.02$ & $1.12\pm0.05$ & $2535.0\pm24.0$ & $2511.0\pm65.0$ & $4.98\pm0.04$ & $5.22\pm0.08$ \\ 
Teegarden's Star & 0253+1652 & 0.020-0.250 Gyr & 0.500-10.000 & $45.48\pm18.35$ & $93.64\pm14.47$ & $1.46\pm0.15$ & $1.2\pm0.09$ & $2500.0\pm126.0$ & $2688.0\pm212.0$ & $4.74\pm0.27$ & $5.22\pm0.07$ \\ 
UGPS J072227.51-054031.2 & 0722-0540 & DL17 & 0.500-10.000 & $8.34\pm7.68$ & $26.11\pm18.14$ & $1.12\pm0.10$ & $0.98\pm0.16$ & $437.0\pm20.0$ & $569.0\pm45.0$ & $4.23\pm0.38$ & $4.68\pm0.53$ \\ 
WISEPA J164715.59+563208.2 & 1647+5632 & DL17 & 0.500-10.000 & $30.98\pm23.35$ & $41.07\pm28.95$ & $1.07\pm0.17$ & $0.94\pm0.16$ & $1275.0\pm102.0$ & $875.0\pm132.0$ & $4.85\pm0.49$ & $4.89\pm0.57$ \\ 
Wolf 359 & 1056+0700 & Truncated DL17 & 0.500-10.000 & $41.34\pm22.09$ & $100.7\pm7.22$ & $1.65\pm0.20$ & $1.56\pm0.1$ & $2562.0\pm156.0$ & $2517.0\pm81.0$ & $4.60\pm0.33$ & $5.19\pm0.05$ \\ 
$\left[\mathrm{HB88}\right]$ M19 & 2127-4215 & Truncated DL17 & 0.500-10.000 & $33.91\pm18.97$ & $93.91\pm18.63$ & $1.51\pm0.19$ & $1.21\pm0.13$ & $2391.0\pm152.0$ & $2676.0\pm320.0$ & $4.59\pm0.34$ & $5.22\pm0.08$ \\ 
vB 8 & 1655-0823 & Truncated DL17 & 0.500-10.000 & $28.94\pm17.95$ & $90.47\pm8.0$ & $1.52\pm0.19$ & $1.17\pm0.03$ & $2324.0\pm147.0$ & $2639.0\pm40.0$ & $4.51\pm0.34$ & $5.21\pm0.06$ \\ 
\enddata
\label{table:prop_comp}
\tablecomments{For objects marked with an asterisk ($^*$), the difference in bolometric luminosity measurements is responsible for the discrepancies in fundamental parameters. For the remaining objects, revisions to their ages are responsible for the discrepancies in fundamental parameters. See \S\ref{sec:comp-f15-prop} for more details.}
\tablenotetext{a}{According to \citet[][DL17]{2017ApJS..231...15D} age distribution, the ages are uniformly distributed in the following age bins, where each bin contains a fraction of the total population as follows: 8.1\% for 0.01-0.15 Gyr, 20.0\% for 0.15–1 Gyr, 16.1\% for 1–2 Gyr, 11.9\% for 2–3 Gyr, 16.6\% for 3–5 Gyr, 12.9\% for 5–7 Gyr, and 14.4\% for 7–10 Gyr. The Truncated DL17 distribution is identical to the above distribution but with an elevated lower bound of 300~Myr.}
\end{deluxetable*}
\end{longrotatetable}
\appendix
\restartappendixnumbering 

\section{NIR and MIR Young Object Absolute Magnitude-Spectral Type Relations}
\label{sec:abs-spt}

\begin{figure*}
    \centering
    \includegraphics[scale=0.58]{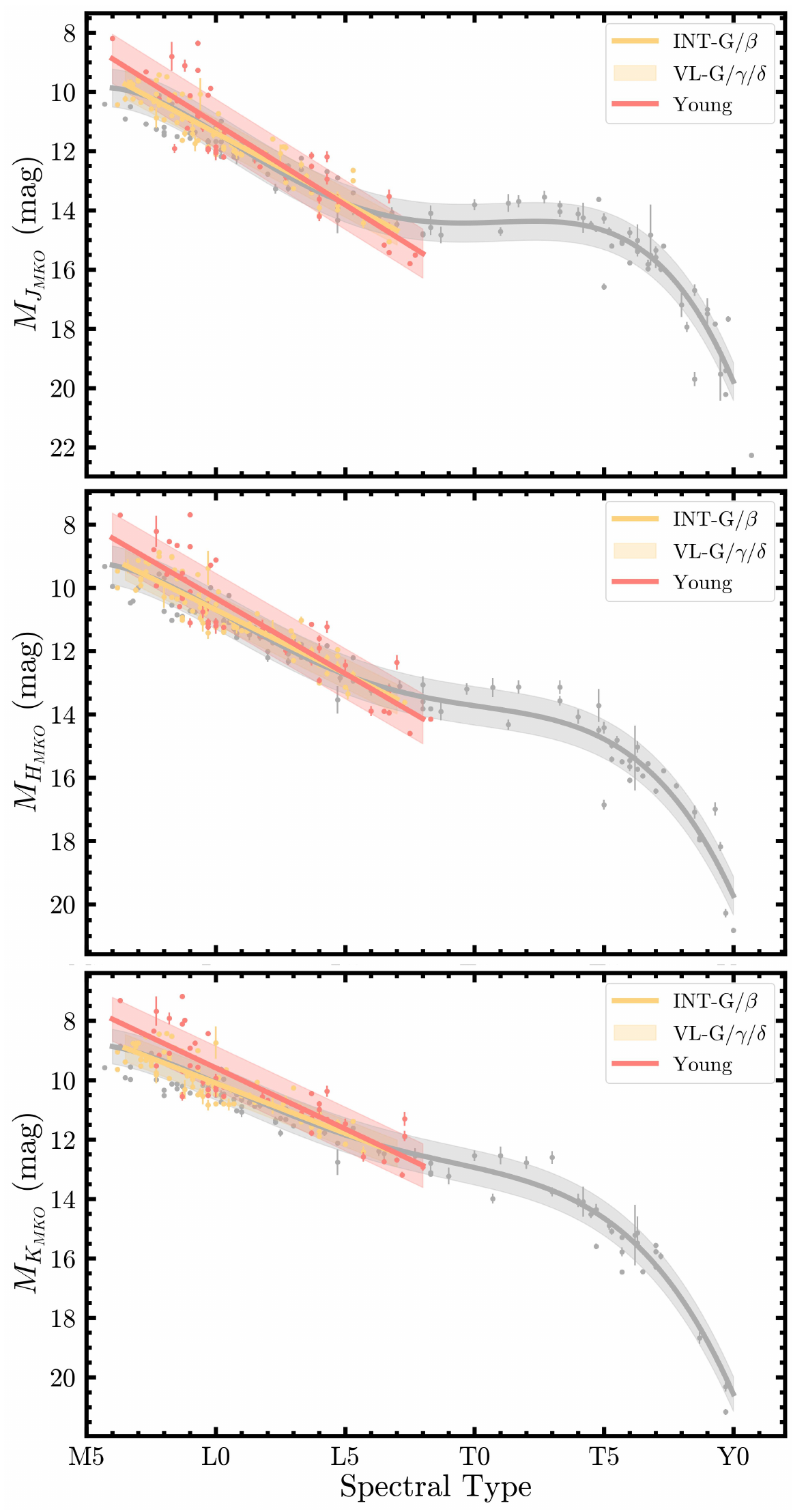}
    \caption{Absolute magnitude vs spectral type relations for intermediate gravity (yellow), low gravity (red), and young (gray) objects in the MKO $JHK$ bands. Random scatter in the x direction with amplitude 0.3 SpT is applied to avoid overlapping points.}
    \label{fig:mko_rel}
\end{figure*}

\begin{figure*}
    \centering
    \includegraphics[scale=0.58]{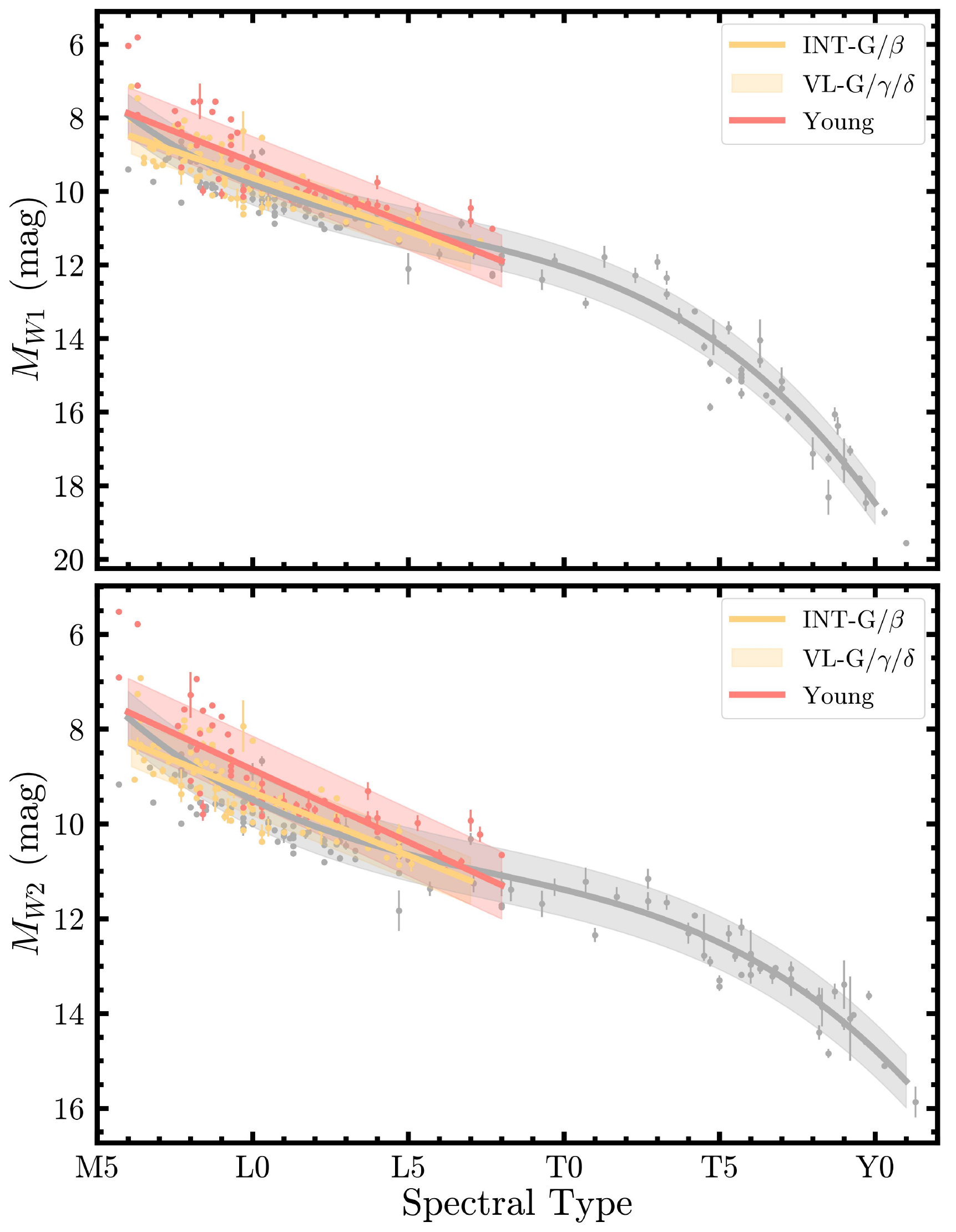}
    \caption{Same as Figure \ref{fig:mko_rel} for the WISE $W1$ and $W2$ bands.}
    \label{fig:wise_rel}
\end{figure*}

Polynomial relations for absolute magnitude as a function of spectral type are routinely used in literature to compute photometric distances to objects in the absence of a parallax measurement \citep[e.g.][]{2013AJ....145....2F,  2016ApJ...822L...1S, 2017AJ....153..196S, 2021AJ....162..102S, 2022ApJ...935...15Z}. Such relations are well-determined for field and young objects but based on fewer measurements in the latter case \citep{2012ApJS..201...19D, 2015ApJ...810..158F, 2016ApJS..225...10F, 2016ApJ...833...96L}. Here, we revisit the polynomial relations for young ultracool dwarfs (spectral type $\ge$M6) using the most-up-to-date information available from \emph{The UltracoolSheet} (version 2.0.0, in preparation). Our analysis excludes unresolved multiples, subdwarfs, field objects, objects in active star-forming regions, accreting objects, objects with disks, and objects in moving groups with an age $>300$ Myr (Hyades, Ursa Major, Rhea 468). Additionally, we excluded 2MASS J04221413+1530525 from the analysis due to its outlier NIR and MIR absolute magnitude. For remaining objects in \emph{The UltracoolSheet}, we define three groups: intermediate surface gravity objects (classification of \intg or $\beta$), low surface gravity objects (classification of \vlg, $\gamma$, or $\delta$), and a combined group of the above spectroscopically young (low gravity classification) objects and young moving group members (as determined by BANYAN-$\Sigma$ with a $\ge$90\% probability). A least-squares polynomial fit is peformed for each of the above groups individually, for the absolute magnitudes in the 2MASS $JH$\Ks, MKO $JHK$, and WISE $W1$ and $W2$ bands as a function of spectral type (photometry quoted without uncertainties, i.e., upper limits, are ignored in the fitting procedures). We could not use the \emph{F}-test to determine the appropriate degree of the polynomial relation due to the low number of data points in several cases. In such a situation, a combination of visual inspection and changes in the rms of the fit was used to determine the degree of the best-fit polynomial relation. The coefficients of the best-fit polynomials are reported in Table \ref{table:poly-appendix} and the MKO and WISE absolute magnitude relations are plotted in Figures \ref{fig:mko_rel} and \ref{fig:wise_rel}, respectively.

\begin{deluxetable*}{cccccccc}
\centering
\tabletypesize{\footnotesize}
\tablecaption{Absolute Magnitude-Spectral Type Polynomial Relations}
\tablehead{\colhead{P(x)\tablenotemark{\scriptsize a}} & \colhead{x\tablenotemark{\scriptsize b}} & \colhead{rms\tablenotemark{\scriptsize c}} & \colhead{$c_0$} & \colhead{$c_1$} & \colhead{$c_2$} & \colhead{$c_3$} & \colhead{$c_4$}}
\startdata
$M_{J\;\mathrm{INT-G}}$ & 6.0$\le$SpT$\le$17.0 & 0.509 & 6.49686469e+00 & 4.89593866e-01 & \nodata & \nodata & \nodata \\
$M_{J\;\mathrm{VL-G}}$ & 6.0$\le$SpT$\le$18.0 & 0.867 & 5.30775821e+00 & 5.85807060e-01 & \nodata & \nodata & \nodata \\
$M_{J\;\mathrm{YNG}}$ & 6.0$\le$SpT$\le$29.0 & 0.639 & 9.28655107e+00 & -7.67241260e-01 & 1.79452020e-01 & -9.69277254e-03 & 1.64408961e-04 \\
$M_{H\;\mathrm{INT-G}}$ & 6.0$\le$SpT$\le$17.0 & 0.490 & 6.44068903e+00 & 4.14482184e-01 & \nodata & \nodata & \nodata \\
$M_{H\;\mathrm{VL-G}}$ & 6.0$\le$SpT$\le$18.0 & 0.803 & 5.18182206e+00 & 5.12068029e-01 & \nodata & \nodata & \nodata \\
$M_{H\;\mathrm{YNG}}$ & 6.0$\le$SpT$\le$29.0 & 0.604 & 7.39739920e+00 & -2.46457596e-01 & 1.12382175e-01 & -6.67377104e-03 & 1.21697109e-04 \\
$M_{Ks\;\mathrm{INT-G}}$ & 6.0$\le$SpT$\le$17.0 & 0.479 & 6.45136804e+00 & 3.60752292e-01 & \nodata & \nodata & \nodata \\
$M_{Ks\;\mathrm{VL-G}}$ & 6.0$\le$SpT$\le$18.0 & 0.768 & 5.23321049e+00 & 4.46473667e-01 & \nodata & \nodata & \nodata \\
$M_{Ks\;\mathrm{YNG}}$ & 6.0$\le$SpT$\le$29.0 & 0.597 & 5.69859407e+00 & 2.48832335e-01 & 5.12670883e-02 & -3.96997960e-03 & 8.40328391e-05 \\
$M_{J\;\mathrm{MKO, INT-G}}$ & 6.5$\le$SpT$\le$17.0 & 0.492 & 6.68728621e+00 & 4.70045485e-01 & \nodata & \nodata & \nodata \\
$M_{J\;\mathrm{MKO, VL-G}}$ & 6.0$\le$SpT$\le$18.0 & 0.829 & 5.59034869e+00 & 5.48250755e-01 & \nodata & \nodata & \nodata \\
$M_{J\;\mathrm{MKO, YNG}}$ & 6.0$\le$SpT$\le$30.0 & 0.638 & 1.59439463e+01 & -2.59148613e+00 & 3.53603963e-01 & -1.66425599e-02 & 2.62569032e-04 \\
$M_{H\;\mathrm{MKO, INT-G}}$ & 6.5$\le$SpT$\le$17.0 & 0.469 & 6.66895264e+00 & 4.02842646e-01 & \nodata & \nodata & \nodata \\
$M_{H\;\mathrm{MKO, VL-G}}$ & 6.0$\le$SpT$\le$18.0 & 0.789 & 5.57417854e+00 & 4.75801078e-01 & \nodata & \nodata & \nodata \\
$M_{H\;\mathrm{MKO, YNG}}$ & 6.0$\le$SpT$\le$30.0 & 0.608 & 1.25445838e+01 & -1.57813057e+00 & 2.32989880e-01 & -1.11747660e-02 & 1.80928732e-04 \\
$M_{K\;\mathrm{MKO,INT-G}}$ & 6.5$\le$SpT$\le$17.0 & 0.455 & 6.72066085e+00 & 3.37867787e-01 & \nodata & \nodata & \nodata \\
$M_{K\;\mathrm{MKO, VL-G}}$ & 6.0$\le$SpT$\le$18.0 & 0.736 & 5.48128477e+00 & 4.10589460e-01 & \nodata & \nodata & \nodata \\
$M_{K\;\mathrm{MKO, YNG}}$ & 6.0$\le$SpT$\le$30.0 & 0.581 & 1.13550558e+01 & -1.25394819e+00 & 1.90091338e-01 & -9.30942538e-03 & 1.56902337e-04 \\
$M_{W1\;\mathrm{INT-G}}$ & 6.0$\le$SpT$\le$17.0 & 0.483 & 6.73106434e+00 & 2.90360683e-01 & \nodata & \nodata & \nodata \\
$M_{W1\;\mathrm{VL-G}}$ & 6.0$\le$SpT$\le$18.0 & 0.707 & 5.86952106e+00 & 3.34558077e-01 & \nodata & \nodata & \nodata \\
$M_{W1\;\mathrm{YNG}}$ & 6.0$\le$SpT$\le$30.0 & 0.569 & 2.21271512e+00 & 1.33939334e+00 & -7.38075759e-02 & 1.57439376e-03 & \nodata \\
$M_{W2\;\mathrm{INT-G}}$ & 6.0$\le$SpT$\le$17.0 & 0.496 & 6.66789839e+00 & 2.66624181e-01 & \nodata & \nodata & \nodata \\
$M_{W2\;\mathrm{VL-G}}$ & 6.0$\le$SpT$\le$18.0 & 0.711 & 5.81406449e+00 & 3.04483215e-01 & \nodata & \nodata & \nodata \\
$M_{W2\;\mathrm{YNG}}$ & 6.0$\le$SpT$\le$31.0 & 0.563 & 2.89919460e+00 & 1.10045304e+00 & -5.44095936e-02 & 1.03057887e-03 & \nodata \\
\enddata
\tablecomments{Since we exclude objects in star-forming regions from our sample, the YNG relations (calculated for YMG members and low gravity sources) are not appropriate for objects with ages $<$10 Myr.}
\tablenotetext{a}{$P(x) = \sum_{i=0}^{n} c_i x^i$}
\tablenotetext{b}{Polynomials expressed as a function of spectral type (SpT) accept inputs in the range 6--29 corresponding to spectral types M6--T9.}
\tablenotetext{c}{Units are in magnitudes except where noted.}
\label{table:poly-appendix}
\end{deluxetable*} 

\clearpage
\bibliography{\string references}

\end{document}